\documentclass[12pt,twoside]{book}
\usepackage{amssymb}
\usepackage{amsmath}
\usepackage[dvips]{graphicx}
\usepackage{fancyhdr}
\fancypagestyle{plain}{%
\fancyhead{} %
 
}
\usepackage[retainorgcmds]{IEEEtrantools}
\begin{document}

\setcounter{page}{1}
%\newcounter{equation}[section]
\newtheorem{t1}{Theorem}[section]
\newtheorem{d1}{Definition}[section]
\newtheorem{c1}{Corollary}[section]
\newtheorem{l1}{Lemma}[section]
\newtheorem{r1}{Remark}[section]
\newcommand{\cA}{{\cal A}}
\newcommand{\cB}{{\cal B}}
\newcommand{\cC}{{\cal C}}
\newcommand{\cD}{{\cal D}}
\newcommand{\cE}{{\cal E}}
\newcommand{\cF}{{\cal F}}
\newcommand{\cG}{{\cal G}}
\newcommand{\cH}{{\cal H}}
\newcommand{\cI}{{\cal I}}
\newcommand{\cJ}{{\cal J}}
\newcommand{\cK}{{\cal K}}
\newcommand{\cL}{{\cal L}}
\newcommand{\cM}{{\cal M}}
\newcommand{\cN}{{\cal N}}
\newcommand{\cO}{{\cal O}}
\newcommand{\cP}{{\cal P}}
\newcommand{\cQ}{{\cal Q}}
\newcommand{\cR}{{\cal R}}
\newcommand{\cS}{{\cal S}}
\newcommand{\cT}{{\cal T}}
\newcommand{\cU}{{\cal U}}
\newcommand{\cV}{{\cal V}}
\newcommand{\cX}{{\cal X}}
\newcommand{\cW}{{\cal W}}
\newcommand{\cY}{{\cal Y}}
\newcommand{\cZ}{{\cal Z}}
%==============================================================

\def\cl{\centerline}
\def\bd{\begin{description}}
\def\be{\begin{enumerate}}
\def\ben{\begin{equation}}
\def\benn{\begin{equation*}}
\def\een{\end{equation}}
\def\eenn{\end{equation*}}
\def\benr{\begin{eqnarray}}
\def\eenr{\end{eqnarray}}
\def\benrr{\begin{eqnarray*}}
\def\eenrr{\end{eqnarray*}}
\def\ed{\end{description}}
\def\ee{\end{enumerate}}
\def\al{\alpha}
\def\b{\beta}
\def\bR{\bar\R}
\def\bc{\begin{center}}
\def\ec{\end{center}}
\def\dg{\dagger}
\def\d{\dot}
\def\D{\Delta}
\def\del{\delta}
\def\ep{\epsilon}
\def\g{\gamma}
\def\G{\Gamma}
\def\h{\hat}
\def\iny{\infty}
\def\La{\Longrightarrow}
\def\la{\lambda}
\def\m{\mu}
\def\n{\nu}
\def\noi{\noindent}
\def\Om{\Omega}
\def\om{\omega}
\def\p{\psi}
\def\pr{\prime}
\def\r{\ref}
\def\R{{\bf R}}
\def\ra{\rightarrow}
\def\up{\uparrow}
\def\dn{\downarrow}
\def\lr{\leftrightarrow}
\def\s{\sum_{i=1}^n}
\def\si{\sigma}
\def\Si{\Sigma}
\def\t{\tau}
\def\th{\theta}
\def\Th{\Theta}

\def\vep{\varepsilon}
\def\vp{\varphi}
\def\pa{\partial}
\def\un{\underline}
\def\ov{\overline}
\def\fr{\frac}
\def\sq{\sqrt}
\def\ot{\otimes}
\def\tf{\textbf}
\def\WW{\begin{stack}{\circle \\ W}\end{stack}}
\def\ww{\begin{stack}{\circle \\ w}\end{stack}}
\def\st{\stackrel}
\def\Ra{\Rightarrow}
\def\R{{\mathbb R}}
\def\mf{\mathbf }
\def\bi{\begin{itemize}}
\def\ei{\end{itemize}}
\def\i{\item}
\def\bt{\begin{tabular}}
\def\et{\end{tabular}}
\def\lf{\leftarrow}
\def\nn{\nonumber}
\def\va{\vartheta}
\def\wh{\widehat}
\def\vs{\vspace}
\def\Lam{\Lambda}
\def\sm{\setminus}
\def\ba{\begin{array}}
\def\ea{\end{array}}
\def\bd{\begin{description}}
\def\ed{\end{description}}
\def\lan{\langle}
\def\ran{\rangle}
%======================================================================================from chap-2====================
\def\cl{\centerline}
\def\bd{\begin{description}}
\def\be{\begin{enumerate}}
\def\ben{\begin{equation}}
\def\benn{\begin{equation*}}
\def\een{\end{equation}}
\def\eenn{\end{equation*}}
\def\benr{\begin{eqnarray}}
\def\eenr{\end{eqnarray}}
\def\benrr{\begin{eqnarray*}}
\def\eenrr{\end{eqnarray*}}
\def\ed{\end{description}}
\def\ee{\end{enumerate}}
\def\al{\alpha}
\def\b{\beta}
\def\bR{\bar\R}
\def\bc{\begin{center}}
\def\ec{\end{center}}
\def\d{\dot}
\def\D{\Delta}
\def\del{\delta}
\def\ep{\epsilon}
\def\g{\gamma}
\def\G{\Gamma}
\def\h{\hat}
\def\iny{\infty}
\def\La{\Longrightarrow}
\def\la{\lambda}
\def\m{\mu}
\def\n{\nu}
\def\noi{\noindent}
\def\Om{\Omega}
\def\om{\omega}
\def\p{\psi}
\def\pr{\prime}
\def\r{\ref}
\def\R{{\bf R}}
\def\ra{\rightarrow}
\def\s{\sum_{i=1}^n}
\def\si{\sigma}
\def\Si{\Sigma}
\def\t{\tau}
\def\th{\theta}
\def\Th{\Theta}

\def\vep{\varepsilon}
\def\vp{\varphi}
\def\pa{\partial}
\def\un{\underline}
\def\ov{\overline}
\def\fr{\frac}
\def\sq{\sqrt}

\def\WW{\begin{stack}{\circle \\ W}\end{stack}}
\def\ww{\begin{stack}{\circle \\ w}\end{stack}}
\def\st{\stackrel}
\def\Ra{\Rightarrow}
\def\R{{\mathbb R}}
\def\bi{\begin{itemize}}
\def\ei{\end{itemize}}
\def\i{\item}
\def\bt{\begin{tabular}}
\def\et{\end{tabular}}
\def\lf{\leftarrow}
\def\nn{\nonumber}
\def\va{\vartheta}
\def\wh{\widehat}
\def\vs{\vspace}
\def\Lam{\Lambda}
\def\sm{\setminus}
\def\ba{\begin{array}}
\def\ea{\end{array}}
\def\bd{\begin{description}}
\def\ed{\end{description}}
\def\lan{\langle}
\def\ran{\rangle}

%=========================================chapter 4 definition===========

\def\cl{\centerline}
\def\bd{\begin{description}}
\def\be{\begin{enumerate}}
\def\ben{\begin{equation}}
\def\benn{\begin{equation*}}
\def\een{\end{equation}}
\def\eenn{\end{equation*}}
\def\benr{\begin{eqnarray}}
\def\eenr{\end{eqnarray}}
\def\benrr{\begin{eqnarray*}}
\def\eenrr{\end{eqnarray*}}
\def\ed{\end{description}}
\def\ee{\end{enumerate}}
\def\al{\alpha}
\def\b{\beta}
\def\bR{\bar\R}
\def\bc{\begin{center}}
\def\ec{\end{center}}
\def\dg{\dagger}
\def\d{\dot}
\def\D{\Delta}
\def\del{\delta}
\def\ep{\epsilon}
\def\g{\gamma}
\def\G{\Gamma}
\def\h{\hat}
\def\iny{\infty}
\def\La{\Longrightarrow}
\def\la{\lambda}
\def\m{\mu}
\def\n{\nu}
\def\noi{\noindent}
\def\Om{\Omega}
\def\om{\omega}
\def\p{\psi}
\def\pr{\prime}
\def\r{\ref}
\def\R{{\bf R}}
\def\ra{\rightarrow}
\def\up{\uparrow}
\def\dn{\downarrow}
\def\lr{\leftrightarrow}
\def\s{\sum_{i=1}^n}
\def\si{\sigma}
\def\Si{\Sigma}
\def\t{\tau}
\def\th{\theta}
\def\Th{\Theta}

\def\vep{\varepsilon}
\def\vp{\varphi}
\def\pa{\partial}
\def\un{\underline}
\def\ov{\overline}
\def\fr{\frac}
\def\sq{\sqrt}
\def\ot{\otimes}
\def\tf{\textbf}
\def\WW{\begin{stack}{\circle \\ W}\end{stack}}
\def\ww{\begin{stack}{\circle \\ w}\end{stack}}
\def\st{\stackrel}
\def\Ra{\Rightarrow}
\def\R{{\mathbb R}}
\def\mf{\mathbf }
\def\bi{\begin{itemize}}
\def\ei{\end{itemize}}
\def\i{\item}
\def\bt{\begin{tabular}}
\def\et{\end{tabular}}
\def\lf{\leftarrow}
\def\nn{\nonumber}
\def\va{\vartheta}
\def\wh{\widehat}
\def\vs{\vspace}
\def\Lam{\Lambda}
\def\sm{\setminus}
\def\ba{\begin{array}}
\def\ea{\end{array}}
\def\bd{\begin{description}}
\def\ed{\end{description}}
\def\lan{\langle}
\def\ran{\rangle}
\def\l{\label}
\def\mb{\mathbb}
\def\ti{\times}

%===============================================================
\def\cl{\centerline}
\def\bd{\begin{description}}
\def\be{\begin{enumerate}}
\def\ben{\begin{equation}}
\def\benn{\begin{equation*}}
\def\een{\end{equation}}
\def\eenn{\end{equation*}}
\def\benr{\begin{eqnarray}}
\def\eenr{\end{eqnarray}}
\def\benrr{\begin{eqnarray*}}
\def\eenrr{\end{eqnarray*}}
\def\ed{\end{description}}
\def\ee{\end{enumerate}}
\def\al{\alpha}
\def\b{\beta}
\def\bR{\bar\R}
\def\bc{\begin{center}}
\def\ec{\end{center}}
\def\d{\dot}
\def\D{\Delta}
\def\del{\delta}
\def\ep{\epsilon}
\def\g{\gamma}
\def\G{\Gamma}
\def\h{\hat}
\def\iny{\infty}
\def\La{\Longrightarrow}
\def\la{\lambda}
\def\m{\mu}
\def\n{\nu}
\def\noi{\noindent}
\def\Om{\Omega}
\def\om{\omega}
\def\p{\psi}
\def\pr{\prime}
\def\r{\ref}
\def\R{{\bf R}}
\def\ra{\rightarrow}
\def\s{\sum_{i=1}^n}
\def\si{\sigma}
\def\Si{\Sigma}
\def\t{\tau}
\def\th{\theta}
\def\Th{\Theta}

\def\vep{\varepsilon}
\def\vp{\varphi}
\def\pa{\partial}
\def\un{\underline}
\def\ov{\overline}
\def\fr{\frac}
\def\sq{\sqrt}

\def\WW{\begin{stack}{\circle \\ W}\end{stack}}
\def\ww{\begin{stack}{\circle \\ w}\end{stack}}
\def\st{\stackrel}
\def\Ra{\Rightarrow}
\def\R{{\mathbb R}}
\def\bi{\begin{itemize}}
\def\ei{\end{itemize}}
\def\i{\item}
\def\bt{\begin{tabular}}
\def\et{\end{tabular}}
\def\lf{\leftarrow}
\def\nn{\nonumber}
\def\va{\vartheta}
\def\wh{\widehat}
\def\vs{\vspace}
\def\Lam{\Lambda}
\def\sm{\setminus}
\def\ba{\begin{array}}
\def\ea{\end{array}}
\def\ds{\displaystyle}
\def\lan{\langle}
\def\ran{\rangle}
\baselineskip 15truept
\large

\thispagestyle{empty}

\frontmatter

\begin{figure}[!ht]
\begin{center}\includegraphics[width=10cm,height=12cm]{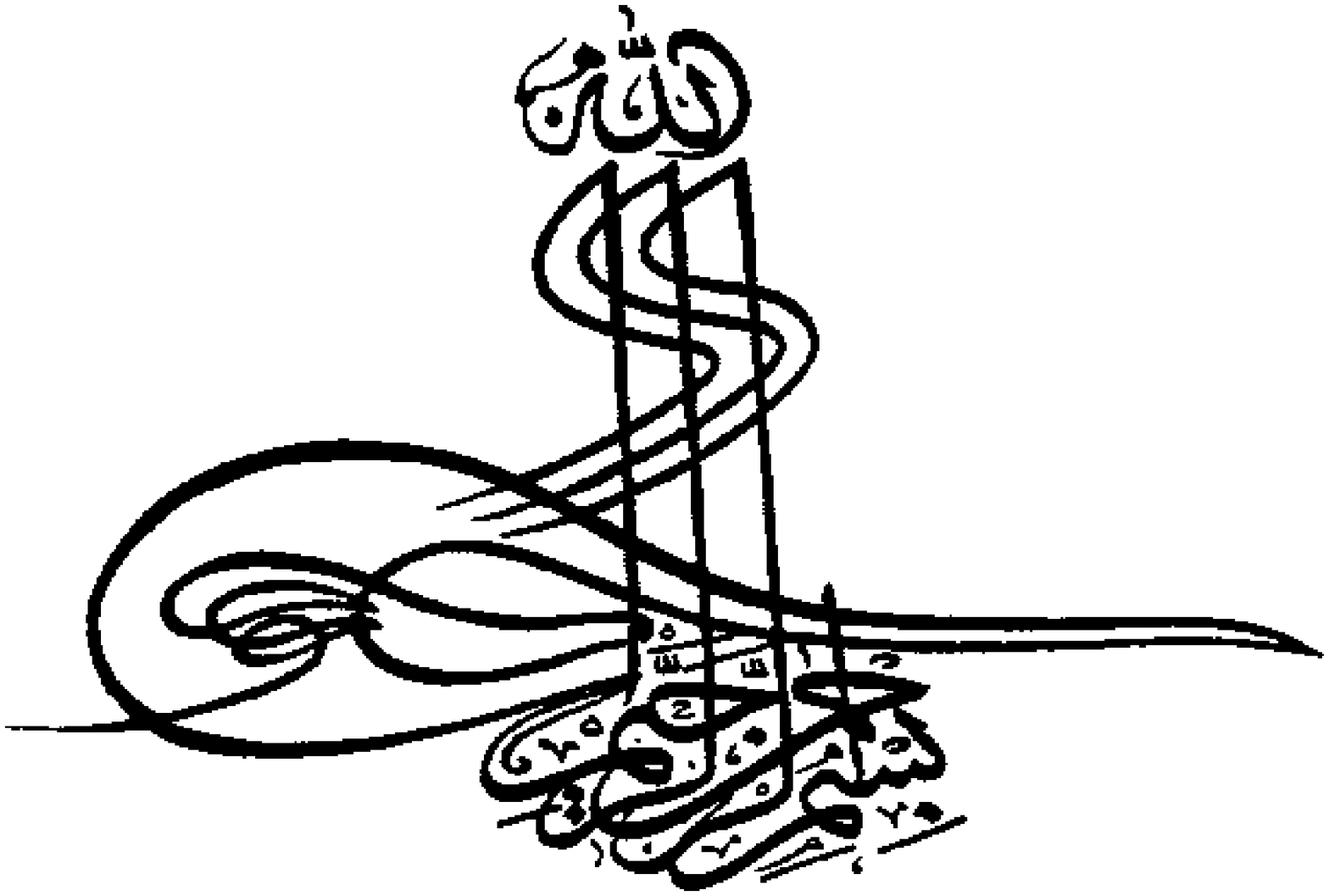}
\end{center}
\end{figure}

\title{GENERATION OF ENTANGLEMENT, MEASURE OF MULTIPARTITE ENTANGLEMENT IN FERMIONIC SYSTEMS AND QUANTUM DISCORD IN BIPARTITE SYSTEMS AND HEISENBERG CHAINS.},\author{By\\\\Behzad Lari\\Department of Physics,\\University of Pune, Ganeshkhind, Pune-411007\\\\\\Under the Supervision of \\Prof. Pramod. S. Joag \\Department of Physics \\University of Pune,\\ Pune 411007.\\\\\\Dissertation submitted in partial fulfillment of the requirements\\for the degree of\\Doctor of Philosophy in Physics}
\maketitle
\newpage
\textbf{
\begin{center}
CERTIFICATE
\end{center}
}

CERTIFIED that the work incorporated in the thesis \\\textbf{``Generation of entanglement, measure of multipartite entanglement in fermionic systems and quantum discord in bipartite systems and Heisenberg chains."}\\ submitted by \textbf{Behzad Lari} was carried out by the candidate under my guidance. Such material as has been obtained from other sources has been duly acknowledged in the thesis.\\\\\\
\begin{flushright}
Prof. P. S. Joag\\(supervisor)
\end{flushright}
\newpage
\textbf{
\begin{center}
DECLARATION
\end{center}
}
I declare that the thesis entitled \\\textbf{``Generation of entanglement, measure of multipartite entanglement in fermionic systems and quantum discord in bipartite systems and Heisenberg chains."}\\ submitted by me for the degree of Doctor of Philosophy, is the record of work carried out by me during the period time \date{Oct. /2007} to \date{Jan. /2011} under the guidance of \textbf{Prof. Pramod. S. Joag} and this has not formed the basis for the award of any degree, diploma, associateship, fellowship, title in this or any other university or other institution of higher learning.\\I further declare that the material obtained from other sources has been duly acknowledged in the thesis.\\
\begin{flushright}
Behzad Lari
\end{flushright}
\newpage
\begin{center}
\textbf{LIST OF PUBLISHED / COMMUNICATED PAPERS.}
\end{center}
1. \textbf{Entanglement Capacity of Nonlocal Hamiltonians : A Geometric
Approach}\\Behzad lari, Ali Saif M. Hassan, and Pramod S. Joag, {\it Physical Review A} \textbf{80}, 062305 (2009).\\\\
2. \textbf{Multipartite entanglement in fermionic systems via a geometric
measure, A Geometric Approach}\\ Behzad lari, P. Durganandini, and Pramod S. Joag, {\it Physical Review A} \textbf{82}, 062302 (2010).\\\\
3. \textbf{Thermal quantum and classical correlations in two qubit XX
model in a nonuniform external magnetic field}\\ Ali Saif M. Hassan, Behzad lari, and Pramod S. Joag, {\it J. Phys. A: Math. Theor.} \textbf{43}, (2010) 485302.\\\\
4. \textbf{Geometric measure of quantum discord for an arbitrary state
of a bipartite quantum system}\\ Ali Saif M. Hassan, Behzad lari, and Pramod S. Joag, (With journal).

\newpage
\
\\
\
\\
\
\\
\
\\
\
\\
\
\\
\
\\
\
\\

\begin{center}
{\it To my dear wife POOPAK}\\
{\it\ and}\\
{\it To my Mother and Father, wife's Dad and Mom}\\
{\it and }\\
{\it To all my Family}\\
\end{center}

\newpage
\begin{center}
\large {\bf Acknowledgments}
\end{center}

\begin{flushleft}
\begin{sloppypar}
It gives me great pleasure to express my gratitude to my supervisor Prof. Pramod S. Joag for his constant guidance, encouragement and support. He has been always great {\it teacher in physics and guide } to me and remains an ideal in my life as a physicist and as a unrivalled friend and human being.\\
\end{sloppypar}
\
\\

\begin{sloppypar}
I thank Dr. P. Durganandini for several illuminating discussions and for her guidence and encouragement at various stages of my work.\\
\
\\
I thank my friend and the senior member of our group, Dr. Ali Saif M. Hassan for a very fruitful collaboration on various parts of this thesis. I really enjoyed discussing various research problems with him, which have enriched my expertise and experience.\\
\end{sloppypar}
\
\\

\begin{sloppypar}
I wish to thank Prof. S. Ghosh, Prof. R. Shankar, Prof. V. Subramanian, Dr. S. Goyal for thier useful discussions and Dr. F. Jamali, Dr. M. kazemian Abyaneh,
Ali. R. Khalili, S. M. Ghazi for thier encouragement.\\
\end{sloppypar}
\
\\

\begin{sloppypar}
I would like to thank the Department of Physics, University of Pune for providing necessary facilities.
\end{sloppypar}

\begin{flushright}
Behzad Lari
\end{flushright}
\begin{sloppypar}
January 2011
\end{sloppypar}
\end{flushleft}

\tableofcontents

\mainmatter

\chapter{Introduction and Overview}
\begin{figure}[!ht]
\begin{center}
\includegraphics[width=8cm,height=.5cm]{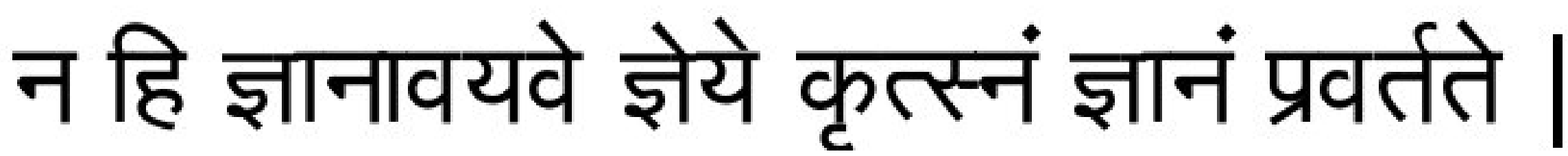}

Charak Samhita (First book of Aurveda)\\
\scriptsize\textsc{ Even if we separately know all the organs making up a human body, we still fall short of knowing the whole body.}
\end{center}
\end{figure}
\begin{center}
%\scriptsize\textsc{$\cdots$ But I can safely say that nobody understands Quantum Mechanics$\cdots$\\ Richard Feynmann}
\end{center}

In this chapter we introduce the research areas dealt with in this thesis and give some background material.

\section{ Quantum entanglement}

 Entanglement is a subtle and eluding property of quantum systems comprising many parts. Entanglement induces correlations between the measurable properties of different parts of a quantum system which cannot be reproduced by any procedure involving only the local operations (LO) and classical communication (CC) between various parts of the system \cite{pv07}.  In consonance with this, entanglement in a quantum system cannot increase (or be created) via LOCC. This principle is connected to another intriguing property of entanglement: a multipartite quantum system can get entangled in various inequivalent ways, which cannot be transformed into each other via LOCC. However, the most challenging aspect of entanglement is that it cannot be `built in parts', that is, the entanglement of N parts is not a sum or a simple function of the entanglement of $M ( < N)$ partite subsystems \cite{eg05}.

The concept of entanglement has played a crucial role in the development of quantum physics. In the early days entanglement was mainly perceived as the qualitative
feature of quantum theory that most strikingly distinguishes it from our classical intuition. The subsequent development of Bell inequalities made this distinction quantitative, and therefore rendered the nonlocal features of quantum theory accessible to experimental verification \cite{bel64,bel66,perb93}. Bell inequalities may indeed
be viewed as an early attempt to quantify quantum correlations that are responsible for the counterintuitive features of quantum mechanically entangled states. At the time it was almost unimaginable that such quantum correlations between distinct quantum systems could be created in well controlled environments. However, the
technological progress of the last few decades means that we are now able to coherently prepare, manipulate, and measure individual quantum systems, as well as create controllable quantum correlations. In parallel with these developments, quantum correlations have come to be recognized as a novel resource that may be used to perform
tasks that are either impossible or very inefficient in the classical realm. These developments have provided the seed for the development of modern quantum information
science.

 Given the status of entanglement as a resource it is quite natural and important to discover the mathematical structures underlying its theoretical description. We will see that such a description aims to provide answers to three questions about entanglement, namely (1) its detection and classification, (2) its creation and manipulation and, (3) its quantification.

In this thesis, we deal with the second and the third problem. We have used the geometric (Bloch representation) approach for studying the creation and quantification of entanglement we give a geometric measure for quantifying the entanglement of multipartite pure states of fermionic systems. Our measure satisfies all properties for a good measure of entanglement.

We have loosely described entanglement as the quantum correlations that can occur in many-party quantum states. This leads to the question  what differentiates quantum correlations from as implied by entanglement classical correlations? In the context of quantum information one of the precise way to define classical correlations is via LOCC operations. Classical correlations can be defined as those that can be generated by LOCC operations. If we observe a quantum system and find correlations that cannot be simulated classically, then we usually attribute them to quantum effects, and hence label them quantum correlations. The entanglement is a resource because it lifts the so-called LOCC constraint, i.e. entanglement and LOCC together can perform tasks that cannot be accomplished by LOCC alone. Using LOCC-operations as the only other tool, the inherent quantum correlations of entanglement are required to implement general, and therefore nonlocal, quantum operations on two or more parts \cite{ejpp,clp01}. As LOCC-operations alone are insufficient to achieve these transformations, we conclude that entanglement may be defined as the sort of correlations that may not be created by LOCC alone.

Entanglement has proved to be a vital physical resource for various kinds of quantum-information processing, including quantum state teleportation \cite{bbcjpw,yc06}, cryptographic key distribution \cite{bbdcjmps}, classical communication over quantum channels \cite{bw92,bfsb97,bsst}, quantum error correction \cite{sho95}, quantum computational speedups \cite{deu85}, and distributed computation \cite{gro97,cb97}. Further, entanglement is expected to play a crucial role in the many particle phenomena such as quantum phase transitions, transfer of information across a spin chain \cite{on02,pv07a} etc. Therefore, quantification of entanglement of multipartite quantum  states is fundamental to the whole field of quantum information and in general, to the physics of multicomponent quantum systems.

Whereas the entanglement in pure bipartite state is well understood, the understanding of entanglement in mixed bipartite state is far from complete. In section 1.1, we review the entanglement of bipartite quantum system. We will state the available measures and criteria for detecting entanglement for both bipartite pure and mixed states. In section 1.2, we deal with multipartite entangled states. In section 1.3, we explain the capability of creating entanglement for a general physical interaction acting on two qubits. In section 1.4, we discuss entanglement in indistinguishable particle systems. In section 1.5, we summarize the quantum discord as a measure of {\it ``quantumness''} of the system. The material in section 1.3, 1.4, 1.5 and 1.6 forms a background for chapters 2, 3, 4, and 5. Section 1.7 is chapterwise summary.

\section{ Bipartite Entanglement}

In this section, we define the entanglement in bipartite quantum states. We review the work that has been done in the bipartite systems.
Consider a system consisting of two subsystems. Quantum mechanics associates to each subsystem a Hilbert space. Let $\mathcal{H}_A$ and $\mathcal{H}_B$ denote these two Hilbert spaces, let $|i\rangle_A$ (where $i=1,2,3,\cdots$) represent a complete orthonormal basis for $\mathcal{H}_A$ and  $|j\rangle_B$ (where $j=1,2,3,\cdots$) a complete orthonormal basis for $\mathcal{H}_B$. Quantum mechanics associates with the system, i.e. the two subsystems taken together, the Hilbert space spanned by the states $|i\rangle_A \otimes |j\rangle_B$.  In  the following, we will drop the tensor product symbol $\otimes$ and write $|i\rangle_A \otimes |j\rangle_B$ as $|i\rangle_A |j\rangle_B$, and so on. Any linear combination of the basis states $|i\rangle_A  |j\rangle_B$ is a state of the system, and any state $|\psi\rangle_{AB}$ of the system can be written \cite{prb98}

\ben \label{ee1}
|\psi\rangle_{AB}=\sum_{ij} c_{ij}|i\rangle_A  |j\rangle_B,	
\een

where the $c_{ij}$ are complex coefficients, we take  $|\psi\rangle_{AB}$ to be normalized, hence $\sum_{ij} |c_{ij}|^2=1$.

If we can write $|\psi\rangle_{AB}= |\psi^{(A)}\rangle_A |\psi^{(B)}\rangle_B$, we say the $|\psi\rangle_{AB}$ is product state (separable state).
If $|\psi\rangle_{AB}$ is not a product state, we say that it is entangled.

By using local operators and classical communication (LOCC) any state $|\psi\rangle_{AB}$ of two subsystems $A$ and $B$ can be transformed to the form \cite{abhh,ncb00}

\ben \label{ee2}
|\psi\rangle_{AB}=\sum_{i=1}^k d_i |\phi_i\rangle_A |\phi'_i\rangle_B; \;\; k\le dim(\mathcal{H}_A \otimes \mathcal{H}_B),	
\een

where the positive coefficients $d_i$ are called Schimdt coefficients. The state is entangled if at least two coefficients do not vanish. Pure entangled state contains quantum correlation which can not be simulated  by any classical tools. A fundamental Theorem was proved by Bell \cite{bel64}, who showed that if the constraint of locality was imposed on the hidden variables, then there was an upper bound on the correlations of results of measurements that could be performed on the two distant systems. That upper bound, mathematically expressed by Bell's inequality \cite{bel64}, is violated by some state in quantum mechanics, thus the state contains quantum correlation which is Non-local property of quantum state \cite{per96a}.

However, in real conditions, owing to interaction with the environment, called decoherence, we encounter mixed states rather than pure ones. A mixed state is a classical mixture of pure quantum states \cite{prb98}. These mixed states can still possess some residual entanglement. A mixed state is considered to be entangled if it is not a mixture of product states \cite{w89}. In mixed states the quantum correlations are weakened and hence the manifestations of mixed state entanglement can be very subtle \cite{pop95}. From the definition of entanglement of mixed state it is difficult to apply this definition directly to know the quantum state is entangled or not, because the mixed state contains both classical and quantum correlations, and can be prepared using infinite possible ensembles.

For pure states, it is easily shown that the CHSH inequality is violated by any nonfactorable state \cite{cfs73,gp92}, while on the other hand a factorable state trivially admits a (contextual) LHV model \cite{bel66}.

For mixed states, Werner \cite{w89} constructed a density matrix $\rho_w$ for a pair of spin-j particles. Werner's state $\rho_w$ can not be written as a sum of direct products of density matrices, $\sum_j c_j \rho^A_j\otimes \rho^B_j$, where $A$ and $B$ refer to the two distant particles and $j$ runs over the states in the ensemble. Therefore, genuinely quantum correlations are involved in $\rho_w$. Nevertheless, for any pair of ideal local measurements performed on the two particles, the correlations derived from $\rho_w$ not only satisfy the CHSH inequality, but, as Werner showed \cite{w89}, it is possible to introduce an explicit LHV model that correctly reproduces all the observable correlations for these ideal measurements \cite{per96a}. Thus for mixed states entanglement and nonlocality are two different resources.

\section{ Multipartite Entanglement}

Multiparticle entanglement is genuinely different from entanglement in quantum systems consisting of two parts. To understand what is so different consider, say, a quantum system that is composed of three qubits. Each of the qubits is to be held by one of three laboratories distantly separated. It may come as quite a surprise that states of such composite quantum systems may contain tripartite entanglement, while at the same time showing no bi-partite entanglement at all. In contrast to the bipartite setting, there is no longer a natural ``unit'' of entanglement, the role that was taken by the maximally entangled state  of a system of two qubits. Quite strikingly, the very concept of being maximally entangled becomes void. Instead, we will see that in two ways there are ``inequivalent kinds of entanglement''. Consider multipartite entanglement of pure quantum states. A theory of entanglement should not discriminate states that differ only by a local operation. Here, ``local operation'' can mean merely a change of local bases (LU operations) or, else, general local quantum operations assisted by classical communication, that are either required to be successful at each instance (LOCC) or just stochastically  (SLOCC). For each notion of locality, local unitary operation (LU) or Local operations and classical communication (LOCC) or just stochastic-LOCC (SLOCC), the questions that have to be addressed are how many equivalence classes exist, how are they parameterized and how can one decide whether two given states belong to the same class?

For the case of bi-partite qubit states, two quantum states are LU-equivalent if
and only if their respective Schmidt normal forms coincide. All classes are parameterized by only one real parameter. Some simple parameter counting arguments show that in the case of N-qubit systems the situation must be vastly more complex. Indeed, disregarding a global phase, it takes $2^{N+1}-2$ real parameters to fix a normalized quantum state in $H = (C^2)^{\otimes N}$. The group of local unitary
transformations $SU(2)\times\cdots \times SU(2)$ on the other hand has 3N real parameters \cite{eg05}. Therefore, one needs at least $2^{N+1}-3N-2$ real numbers to parameterize the sets of inequivalent pure quantum states \cite{lp98}. This lower bound turns out to be tight \cite{chs00}. It is a striking result that the ratio of non-local to local parameters grows exponentially in the number of systems. In particular, the finding rules out all hopes of a generalization of the Schmidt normal form. A general pure tripartite qubit state, say, cannot be cast into the form
$sin\theta |000\rangle + cos\theta|111\rangle$ by the action of local unitaries \cite{perb93}. Considerable effort has been undertaken to describe the structure of LU-equivalence classes by the use of invariants or normal forms \cite{lp98,chs00,grb98,rain97b,aajt}. Ac$\acute{i}$n {\it et al.} \cite{aacjlt} have proved for any pure three-qubit state the existence of local bases which allow one to build a set of five orthogonal product states in terms of which the state can be written in a unique form. This leads to a canonical form which generalizes the two-qubit Schmidt decomposition. It is uniquely characterized by the five entanglement parameters.  When one deals with SLOCC operations, the group of SLOCC,  $SL(C^2)\times\cdots \times SL(C^2)$ has $2^{N+1}-6N-2$ parameters that are necessary to label SLOCC equivalence classes of qubit systems. It turns out that the three-qubit pure states are partitioned into a total of Six SLOCC-equivalent classes \cite{dvc00}. The picture is complete for three-qubit : any fully entangled state is SLOCC-equivalent to either $|GHZ\rangle$ or $|W\rangle$ \cite{dvc00}. Three-qubit W-states and GHZ-states have already been experimentally realized, both purely optically using postselection  \cite{ekbkw,bpdwz} and in ion traps \cite{rrhhblbsb}. The two states behave differently, however, if a system is traced out. Specifically, tracing
out the first qubit of the GHZ state will leave the remaining systems in a complete mixture. For $|W\rangle$ will leave the remaining systems in a mixed entangled bipartite state. Thus, the entanglement of $|W\rangle$ is more robust under particle loss than the one of $|GHZ\rangle$ \cite{dvc00}. From point of view of asymptotic manipulation of multipartite quantum states, there is no longer a single essential ingredient as in bipartite the maximally entangled state or EPR-state, but many different ones. In the multi-particle case, however, it is meaningful to introduce the concept of a minimal reversible entanglement generating set $(MREGS)$. An $MREGS$ $S$ is a set of pure states such that any other state can be generated from $S$ by means of reversible asymptotic LOCC. It must be minimal in the sense that no set
of smaller cardinality possesses the same property \cite{bprst,lpsw,gpv00}. Yet, it can be shown that merely to consider maximally entangled qubit pairs is not sufficient to construct an $MREGS$ \cite{lpsw}. To find general means for constructing $MREGS$ constitutes one of the challenging open problems of the field: as long as this question is generally unresolved, the development of a ``theory of multi-particle entanglement'' in the same way as in the bi-partite setting seems unfeasible.

\section{ Generation of entangled quantum systems : entanglement capacity }

Since entanglement cannot be increased (generated) via LOCC, the interaction between various parts of a quantum system is needed to generate and increase its entanglement. Thus a quantum system evolves to generate entanglement provided its parts interacts. Any Hamiltonian $\mathcal{H}_{AB}\neq \mathcal{H}_A+\mathcal{H}_B$ that is not a sum of local terms couples the systems $A$ and $B$. Together with local operations, the coupling can be used to generate entanglement \cite{dvclp01,bhls020,zzf00}, to transmit classical and quantum information \cite{bhls020,bgnp01,hvc020,bs020}, and more generally, to simulate the bipartite dynamics of some other Hamiltonian $ \acute{H}_{AB}$ and thus to perform arbitrary unitary gates on the composite space $\mathcal{H}_{AB}=\mathcal{H}_A\otimes \mathcal{H}_B$ \cite{dnbt02,bhls020,vc02}. Much experimental effort has been devoted to creating entangled states of quantum systems, including those in quantum optics, nuclear magnetic resonance, and condensed matter physics \cite{f00}. Determining the ability of a system to create entangled states provides a benchmark of the ``quantumness'' of the system. Furthermore,such states could ultimately be put to practical use in various quantum information processing tasks, such as superdense coding \cite{bw93}or quantum teleportation  \cite{bbcj95}.

The theory of optimal entanglement generation can be approached in different ways. For example, Ref. \cite{dvclp01} considers single-shot capacities. In the case of two-qubit interactions, and assuming that ancillary systems are not available, Ref. \cite{dvclp01} presents a closed form expression for the entanglement capacity and optimal protocols by which it can be achieved. In contrast, Ref. \cite{bhls020} considers the asymptotic entanglement capacity, allowing the use of ancillary systems, and shows that when ancillas are allowed, the single-shot and asymptotic capacities are in fact the same. However, such capacities can be difficult to calculate because the ancillary systems may be arbitrarily large.

In the past few years there has been a considerable increase in experimental activity aiming to create entangled quantum states. One reason is the potential applications of entanglement to quantum information processing. Creating entanglement has been possible in quantum optics for more than a decade; however, now many new communities, working in a variety of experimental areas (for example, $NMR$, condensed matter physics) are also joining the field \cite{f00}. In general, entanglement between two systems can be generated if they interact in a controlled way. However, in most experiments these interactions are weak which makes the production of entanglement a very difficult task. Thus, it would be very convenient to have a theory which would provide us with the best way of exploiting interactions to produce entanglement. we try to analyze the entanglement capabilities of Hamiltonians. In particular, we would like to answer questions such as the following: Given an interaction (Hamiltonian), what is the most efficient way of entangling particles? Can we make the process more efficient by supplementing the action of the Hamiltonian with some local unitary operations? Can we increase the entanglement more efficiently by using some ancillas? So far, much of the theoretical effort in quantum information theory has been devoted to the characterization and quantification of the entanglement of a given state. Very
recently, it has been realized that there is a parallel notion of the entanglement in the dynamics of a system \cite{ejpp}. In \cite{ejpp}, the authors consider the situation that one has a given unitary transformation and ask, for example, how much state entanglement is needed to produce it. Here we focus on a different issue: Given an interaction (i.e., a Hamiltonian) how can we make the most effective use of it \cite{zzf00}? What we propose here is to define and determine the entanglement capabilities of physical processes, in particular, of unitary evolutions \cite{ci01}. This is a very relevant problem not only from the theoretical point of view but also from the experimental one. Of course, this problem is even more difficult than the one of quantifying the entanglement of states. In any case, then we give the first steps in this direction by considering the case in which the physical process is acting on two qubits. From our results it turns out that (i) it is more efficient to produce entanglement if initially one already has some. (ii) The best initial entanglement is universal, i.e., independent of the physical process. (iii) One can improve the performance of a physical process by complementing it with fast local operations. (iv) One can also improve it (in certain cases) by using auxiliary systems. (v) All entangling Hamiltonians can simulate each other and are thus qualitatively equivalent; we also provide an upper bound on the time required for one Hamiltonian to simulate another. We consider two qubits interacting via a nonlocal Hamiltonian $H$. We want to determine the most efficient way in which we can use such an interaction to produce entanglement. We will characterize the entanglement of a state of
the qubits at a given time $t$, $\left|\psi(t)\right\rangle$, by some entanglement measure $E$. In order to quantify the entanglement production, we define the entanglement rate $\Gamma$ at a particular time $t$ of the interaction as follows:

\ben \label{ee3}
\Gamma\left(t\right)=\frac{dE\left(t\right)}{dt}.	
\een

This quantity depends on $\left|\psi\right\rangle$ not only through its entanglement $E$. The goal is then to find the conditions which must be satisfied in order to obtain a maximal entanglement rate. In particular, we will be interested in determining the following: (i)  For any initial entanglement $E$ of the two-qubit system, what is the state $\left|\psi(t)\right\rangle$, say $\left|\psi_E\right\rangle$, for which the interaction produces the maximal rate $\Gamma_E$? (ii) The maximal achievable entanglement rate $\Gamma_{max}$ ( note that the definitions of $\Gamma$ and $\Gamma_{max}$ are not restricted to qubits but are also valid for d-level systems, $d\geq 0$), $\Gamma_{max}\equiv max_E\Gamma_E$ and the state $\left|\psi_{max}\right\rangle$ for which $\Gamma=\Gamma_{max}$.

 In what follows we provide some essential background to appreciate the work presented in chapter 2. We review some definitions and known results. Let $\left|\psi\right\rangle$ be a state of the systems $A$ and $B$. This state can always be written using the Schmidt decomposition \cite{perb93},

\ben \label{ee4}
\left|\psi\right\rangle:=\sum_i \sqrt{\lambda_i }\left|\phi_i\right\rangle_A\otimes	\left|\eta_i\right\rangle_B
\een

where $\left\{\left|\phi_i\right\rangle\right\}$ and $\left\{\left|\eta_i\right\rangle\right\}$  are orthonormal sets of states, and with $\lambda_i>0$ $\sum_i \lambda_i =1$. The entanglement between $A$ and $B$ is defined as

\ben \label{ee5}
E\left(	\left|\psi\right\rangle\right):=-\sum_i \lambda_i \log_2 \lambda_i
\een

Reference \cite{dvclp01} considers maximizing the rate of increase of entanglement when a pure state is acted on by $e^{-iHt}$, the evolution according to a time-independent Hamiltonian $H$ (we set $\hbar=1 $ ). We refer to this maximal rate as the single-shot entanglement capacity. When no ancillas are used, this is given by

\ben \label{ee6}
\Gamma^{(1*)}_{H}:=\max_{\left|\psi\right\rangle\in \mathcal{H}_{AB}} \lim_{t\to 0}	 \frac{E\left(e^{-iHt}\left|\psi\right\rangle\right)- E \left(\left|\psi\right\rangle\right)}{t}
\een

Here the rate of increasing entanglement is optimized over all possible pure initial states of $\mathcal{H}_{AB}$ without ancillary systems. In fact, the single-shot capacity may be higher if ancillary systems $\acute{A}$ and $\acute{B}$, not acted on by $H$, are used. For this reason, we may consider the alternative single-shot entanglement capacity

\ben \label{ee7}
\Gamma^{(1)}_{H}:=\sup_{\left|\psi\right\rangle\in \mathcal{H}_{A\acute{A}B\acute{B}}} \lim_{t\to 0}	 \frac{E\left(e^{-iHt}\left|\psi\right\rangle\right)- E \left(\left|\psi\right\rangle\right)}{t}
\een

Note that in Eqs. ((\ref{ee3}) , (\ref{ee4})), the limit is the same from both sides even though it might be the case that $\Gamma^{(1*)}_{H}\neq \Gamma^{(1)}_{-H}$ in general (and similarly for $\Gamma^{(1)}_{H}$). For any two-qubit Hamiltonian $H$, Ref. \cite{dvclp01} shows that it is locally equivalent to a $canonical form$

\ben \label{ee8}
\sum_{i=x,y,z}\mu_i \sigma_i\otimes\sigma_i,\  \mu_x\geq\mu_y\geq\left|\mu_z\right|	
\een

In terms of this canonical form, the optimal single-shot entanglement capacity of any two-qubit interaction without ancillas is given by

\ben \label{ee9}
\Gamma^{(1*)}_{H}:=\alpha\left(\mu_x+\mu_y\right).
\een

\ben \label{ee10}
\alpha:= 2\max_x\sqrt{x\left(1-x \right)}	\log_2\left(\frac{x}{1-x}\right)\approx 1.9123.
\een

where the maximum is obtained at $x_0\approx 0.9168$. In addition, $\Gamma^{(1)}_{H}$ may be strictly larger than $\Gamma^{(1*)}_{H}$ when $\left|\mu_z\right|>0$
\cite{dvclp01}. Reference \cite{bhls020} considers the asymptotic entanglement capacity $\Gamma^{(1)}_{H}$ for an arbitrary Hamiltonian $H$.  $\Gamma^{(1)}_{H}$ is defined as the maximum average rate at which entanglement can be produced by using many interacting pairs of systems, in parallel or sequentially. These systems may be acted on by arbitrary collective local operations (attaching or discarding ancillary systems, unitary transformations, and measurements). Furthermore, classical communication between $A$ and $B$ and possibly mixed initial states are allowed. Reference \cite{bhls020} proves that the asymptotic entanglement capacity in this general setting turns out to be just the single-shot capacity in Ref. \cite{dvclp01}, $\Gamma_{H}=\Gamma^{(1)}_{H}$, for all $H$, so

\ben \label{ee11}
\Gamma_{H}:=\sup_{\left|\psi\right\rangle\in \mathcal{H}_{A\acute{A}B\acute{B}}} \lim_{t\to 0}	 \frac{E\left(e^{-iHt}\left|\psi\right\rangle\right)- E \left(\left|\psi\right\rangle\right)}{t}
\een

Note that the definition of the capacity involves a supremum over both all possible states and all possible interaction times, but in fact it can be expressed as a supremum over states and a limit as $t\to 0$, with the limit and the supremum taken in either order. Let $\left|\psi\right\rangle$ be the optimal input in Eq. ((\ref{ee4}) or (\ref{ee8})). When $\left|\psi\right\rangle$ is finite dimensional, the entanglement capacity can be achieved \cite{dvclp01,bhls020} by first inefficiently generating some $EPR$ pairs, and repeating the following three steps: (i) transform $nE\left(\left|\psi\right\rangle\right)$ $EPR$ pairs into $\left|\psi\right\rangle^{\otimes n}$ \cite{bbps96,lp99}, (ii) evolve each $\left|\psi\right\rangle$ according to $H$ for a short time $\delta t$, and  (iii) concentrate the entanglement into $n\left(E\left(\left|\psi\right\rangle\right)+\delta t\Gamma_H\right)$ $EPR$ pairs \cite{bbps96}.	It shows that $\Gamma^{(1*)}_{K}:=\Gamma^{(1)}_{K}$ for any two-qubit Hamiltonian with canonical form

\ben \label{ee12}
K:=\mu_x \sigma_x \otimes \sigma_x +\mu_x \sigma_y \otimes \sigma_y,\  \mu_x\geq\mu_y\geq 0	
\een

so that all three entanglement capacities are equal:

\ben \label{ee13}
\Gamma_{K}=\Gamma^{(1)}_{K}=\Gamma^{(1*)}_{K}	
\een

The optimal input is therefore a two-qubit state, and the optimal protocol applies. In particular, for these Hamiltonians, which include the Ising interaction $\sigma_z \otimes \sigma_z$ and the anisotropic Heisenberg interaction $\mu_x \sigma_x \otimes \sigma_x +\mu_x \sigma_y \otimes \sigma_y$, entanglement can be optimally generated from a two-qubit initial state $\left|\psi\right\rangle$ without ancillary systems $\acute{A}\acute{B}$. As mentioned above, this result is not generic, since ancillas increase the amount of entanglement generated by some two-qubit interactions, such as the isotropic Heisenberg interaction $\mu_x \sigma_x \otimes \sigma_x +\mu_x \sigma_y \otimes \sigma_y + \mu_z \sigma_z \otimes \sigma_z$ \cite{dvclp01}.

\section{ Entanglement in indistinguishable particle systems}

%************************fermions entanglement********************

%Joag manscript

Understanding and using entangled states of identical and indistinguishable particles \cite{z02,sikld01,wv03,esbl02,glts02,cp05,ccw07,gf01} generates many questions of fundamental nature.

The first of these problems is that of locality and nature of local operations \cite{z02}. for the case of distinguishable particles, by locality, we mean Einstein locality \cite{e48}, which is identically realized by different parts of a quantum system being space-like separated. However, the quantum particles making up a system are essentially indistinguishable as long as their wave functions overlap, that is, they are at short distances from one another \cite{sikld01} . Such a situation can arise, for example, in a quantum device based on quantum dot technology \cite{Ld98,bLd99,Pjtlylmhg05}. Here qubits are realized by the spins of the electrons in a system of quantum dots. The overlap between the electron wave functions in different dots can be varied by controlling parameters like gate voltages or magnetic fields, which change the tunneling amplitudes of the electrons from one dot to the other. For non-negligible overlaps, the entanglement between the qubits is then intimately connected to the electron entanglement, which is essentially that of indistinguishable fermions.

Second problem is to define separable (and hence entangled) states. For N-qubits, the state space is the tensor product of state-spaces of individual qubits. A N-qubits pure state is separable if it can be expressed as tensor product of individual qubit states \cite{z02,cp05}. For indistinguishable particles all physical states reside in the subspace of the tensor product state space, having appropriate symmetry (antisymmetric (symmetric) subspace for fermions (bosons)) \cite{sikld01,fc03}. This subspace cannot be expressed as the tensor product of the state spaces of  individual particles. Thus the usual definitions of separable states cannot be implemented here. a way out is to view N-particle state as separable if its Fock space representation contains a single term

\ben \label{ee14}
\left|n_1,n_2,...\right\rangle=\prod_{i} \frac{1}{\left(n_i!\right)^{\frac{1}{2}}}\left(a^{\dagger}_{i}\right)^{n_i}\left|0\right\rangle
\een

where $a_i(a^{\dagger}_{i})$ annihilates (creates) a particle with individual particle state $\left|\psi_i\right\rangle$. This is often expressed by saying that the state $\left|\psi\right\rangle$ has slater number 1. Note that the separable state of two indistinguishable fermions namely, $a^{\dagger}_{i}\left|0\right\rangle$ is given, in the first quantized version, as
$$ \frac{1}{\sqrt{2}}\left\{\left|10\right\rangle-\left|10\right\rangle\right\} $$

when viewed as a two qubit state, this is a maximally entangle state! However, as a two fermion state, this apparent entanglement comes about only as a consequence of the anti symmetrization requirement. such an apparent ``entanglement'' (i.e. the corresponding correlations ) cannot be used as a resource in a quantum information processing or quantum communication task. If this was possible, then the results of a local measurement on a fermion will be affected by the existence of identical fermions in the universe, which is not true.

Additional correlations in many-fermion systems arise if more than one terms occur in Eq. (\ref{ee14}) that is, if the Slater number for the state exceeds one. i.e., if there is no single-particle basis such that a given state of N indistinguishable fermions can be represented as an elementary Slater determinant (i.e., fully antisymmetric combination of N orthogonal single-particle states). These correlations are the analog of quantum entanglement in separated systems and are essential for quantum information processing in non separated systems. As an example consider a ``swap'' process exchanging the spin states of electrons on coupled quantum dots by gating the tunneling amplitude between them \cite{bLd99,slm01}. Before the gate is turned on, the two electrons in the neighboring quantum dots are in a state represented by a simple Slater determinant and can be regarded as distinguishable since they are separated by a large energy barrier. When the barrier is lowered, more complex correlations between the electrons due to the dynamics arise. Interestingly, as shown in Refs. \cite{bLd99,slm01}, during such a process the system must necessarily enter a highly correlated state that cannot be represented by a single Slater determinant. The final state of the gate operation, however, is, similarly as the initial one, essentially given by a single Slater determinant. Moreover, by adjusting the gating time appropriately one can also perform a ``square root of a swap'' which turns a single Slater determinant into a ``maximally'' correlated state in much the same way \cite{slm01}. At the end of such a process the electrons can again be viewed as effectively distinguishable, but are in a maximally entangled state in the usual sense of distinguishable separated particles. In this sense the highly correlated intermediate state can be viewed as a resource for the production of entangled states. We expect that similar scenarios apply to other schemes of quantum information processing that involve cold particles (bosons or fermions) interacting at microscopic distances at which the quantum statistics becomes essential. For instance, it should be of relevance for quantum computing models employing ultra cold atoms in optical lattices \cite{jbcgz99} or ultra cold atoms in arrays of optical micro traps \cite{bbde01}.

 A possible way to model and quantify entanglement in the system of $N$ identical and indistinguishable fermions is to map the corresponding Fock space to an isomorphic $N$-qubit space and measure and monitor the N-fermion entanglement in terms of that on the mapped $N$-qubit space. This way was suggested and used by Zanardi \cite{z02}. We use this approach to clarify the above problems and establish a quantitative measure for the entanglement in N-fermion systems. We emphasize that our measure can deal with multi-partite entanglement in N-fermion systems; an area almost untouched until now.

 Entanglement in fermionic systems has myriads of applications in the area of quantum devices and is also expected to play a fundamental role in many physical phenomena like quantum phase transitions, quantum Hall effect, and so on \cite{on02,lh08}, involving many-body quantum systems.

%********************END FERMIONS************

\section{Quantum discord}

%=== Joag manscript and our paper pulished in J. Phys. A ,Math, Theor ====

 Entanglement in a quantum state of a multipartite quantum system is a fundamental paradigm to isolate and understand quantum correlation implied by the state. For pure states and a large class of mixed states, entanglement corresponds to the non-local quantum correlations which break Bell inequalities. However, quantum correlations breaking Bell inequalities need not account for all quantum correlations in a composite quantum system in a given state. In order to account for the quantum correlation in a given state, we must find some means to divide the total correlation into a classical part and a purely quantum part. This is particularly important for mixed states, since their quantum correlations are many a time hidden by their classical correlations (CC). An answer to this requirement is given by quantum discord (QD), a measure of the quantumness of correlations introduced in Ref. \cite{oz01}. Quantum discord is built on the fact that two classically equivalent ways of defining the mutual information turn out to be inequivalent in the quantum domain. In addition to its conceptual role, some recent results \cite{kl98}, suggest that quantum discord and not entanglement may be responsible for the efficiency of a mixed state based quantum computer.

%===== luo paper (Quantum discord for two- qubit system) PRA 77, 042303(2008)======

The idea is to take advantage of the observation that in pursuing quantum analogs of classical notions, equivalent classical expressions often lead to different quantum analogs due to non commutativity of operators which represent quantum states and observables, and this difference can be exploited to characterize and quantify the ``quantumness'' of an object. In particular, Olliver and Zurek defined quantum discord, as the difference of two natural quantum extensions of the classical mutual information, and exhibited its applications in revealing quantum aspect of correlations in bipartite states including separable ones. The quantum discord is further used by Zurek in analyzing Maxwells demons \cite{zu03}. A closely related and important quantity has also been introduced by Henderson and Vedral from a different perspective \cite{hv01}. Other similar quantities with the same spirit have been extensively studied by Horodecki et al. \cite{hhhosss05}.

%======== Dakic and vedral paper Neacessary and sufficient condition for nonzero quantum discord PRL, 105, 190502 (2010) ======

Correlations between two random variables of classical systems $A$ and $B$ are in information theory quantified by the mutual information $I(A:B)=H(A)+H(B)-H(A,B)$. If $A$ and $B$ are classical systems, then $H(.)$ stands for the Shannon entropy  $H(p)=-\sum_i p_i \log_2 p_i$, where $\textbf{p}=(p_1,p_2,...)$, is the probability distribution vector, while $H(.,.)$  is the Shannon entropy of the joint probability distribution $p_{ij}$. For quantum systems $A$ and $B$, function $H(.)$  denotes the von Neumann entropy $S(\rho)=-Tr\rho\log_2 \rho$, where $\rho$ is the density matrix. In the classical case, we can use the Bayes rule and find an equivalent expression for the mutual information $I(A:B)=H(A)-H(A|B)$, where $H(A|B)$ is the Shannon entropy of A conditioned on the measurement outcome on $B$. For quantum systems, this quantity is different from the first expression for the mutual information and the difference defines the quantum discord. Consider a quantum composite system defined by the Hilbert space $\mathcal{H}_{AB}=\mathcal{H}_A\otimes\mathcal{H}_B$. Let dimensions of the local Hilbert spaces be $dim\mathcal{H}_A = d_A$ and $dim\mathcal{H}_B = d_B$, while $dim\mathcal{H}_{AB} = d_{AB}$. Given a state $\rho$ (density matrix) of a composite system, the total amount of correlations is quantified by quantum mutual information \cite{gpw05}:

\ben \label{ee15}
I\left(\rho\right)=S\left(\rho_A\right)+S\left(\rho_B\right)-S\left(\rho\right)	
\een
where $S\left(\rho\right)$ is the von Neumann entropy and $\rho_{A,B}=Tr_{B,A}\left(\rho\right)$ are reduced density matrices. A generalization of
the classical conditional entropy is $S\left(\rho_{B|A}\right)$, where $\rho_{B|A}$ is the state of $B$ given a measurement on $A$.By optimizing over
all possible measurements in A, we define an alternative
version of the mutual information

\ben \label{ee16}
\mathcal{Q}\left(\rho\right)=S\left(\rho_B\right)-\min_{ \left\{E_k\right\} }\sum_k p_k S\left(\rho_{B|A}\right)	
\een
where $\rho_{B|A}=Tr_A\left(E_k\otimes I_B \rho \right)/Tr\left(E_k\otimes I_B \rho\right)$ is the state of $B$ conditioned on outcome $k$ in $A$, and ${E_A}$ represents the set of positive operator valued measure elements. The discrepancy between the two measures of information defines the quantum discord \cite{oz01,hv01}:

\ben \label{ee17}
D_A\left(\rho\right)=I\left(\rho\right)-\mathcal{Q}\left(\rho\right)	
\een

The discord is always non-negative \cite{oz01} and reaches zero for the classically correlated states \cite{hv01}. Note that discord is not a symmetric quantity $D_A\left(\rho\right)\neq D_B\left(\rho\right)$  and $D_A\left(\rho\right)$ refers to the ``left'' discord, while $D_B\left(\rho\right)$ refers to the ``{\it right}'' discord. The state $\rho$ for which $D_A\left(\rho\right)= D_B\left(\rho\right)=0$ is completely classically correlated in the sense of \cite{op02,mpsvw10}. In this thesis, when we refer to the discord we mean the ``{\it left}'' discord $D_A\left(\rho\right)$. To give an example of a state with nonvanishing discord, consider the two-qubit separable state in which four nonorthogonal states of one qubit are correlated with four nonorthogonal states of the second qubit:

\ben \label{ee18}
\frac{1}{4}\left(\left|0\right\rangle\left\langle 0\right|\otimes\left|+\right\rangle\left\langle +\right|+\left|1\right\rangle\left\langle 1\right|\otimes\left|-\right\rangle\left\langle -\right|+\left|+\right\rangle\left\langle +\right|\otimes\left|1\right\rangle\left\langle 1\right|+\left|-\right\rangle\left\langle -\right|\otimes\left|0\right\rangle\left\langle 0\right|\right)	
\een

Unlike the state above, one can show that the state $\left(\rho\right)$ is of zero discord if and only if there exists a von Neumann measurement ${ \Pi_{k} =\left|\psi_k\right\rangle\left\langle\psi_k\right|}$ such that \cite{da10}

\ben \label{ee19}
\sum_{k}\left(\Pi_{k}\otimes I_B\right)\rho\left(\Pi_{k}\otimes I_B \right)	
\een

In other words, the zero-discord state is of the form $\sum_k p_k \left|\psi_k\right\rangle\left\langle\psi_k\right|\otimes\rho_k $, where ${\left|\psi_k\right\rangle}$ is some orthonormal basis set, $\rho_k$ are the quantum states in $B$, and $p_k$ are nonnegative numbers such that
$\sum_{k} p_k =1$.

{\it Easily implementable necessary and sufficient condition} \cite{dvb10}. Let us choose basis sets in local Hilbert-Schmidt spaces of Hermitian operators, ${A_n}$ and ${B_n}$ where $n=1,...,d^{2}_{A}$ and $m=1,...,d^{2}_{B}$. We decompose the state $\rho$ of the composite system into $\rho=\sum_{nm} r_{nm} A_n\otimes B_m$. The coefficients $r_{nm}$ define $d^{2}_{A}\times d^{2}_A$ real matrix $R$, which we call the correlation matrix. We can find its singular value decomposition (SVD), $URW^{T}=diag[c_1,c_2,...]$ where U and W are $d^{2}_{A}\times d^{2}_{A}$ and $d^{2}_{B}\times d^{2}_{B}$ orthogonal matrices, respectively, while $diag[c_1,c_2,...]$ is $d^{2}_{A}\times d^{2}_B$ diagonal matrix. SVD defines the new basis in local spaces $S_n=\sum_{\acute{n}} U_{n\acute{n}} A_{\acute{n}}$ and $F_m =\sum_{\acute{m}} W_{{m}\acute{m}} B_{\acute{m}}$. The state $\rho$ in the new basis is of the form $\rho=\sum^{L}_{n=1} c_n S_n \otimes F_n$, where $L=rank R$ is the rank of correlation matrix R (the number of nonzero eigenvalues $c_n$).
 The necessary and sufficient condition Eq. (\ref{ee19}) becomes $\sum^{L}_{n=1} c_n\left(\sum_{k}\Pi_{k}S_n\Pi_{k}\right)\otimes F_n=\sum^{L}_{n=1} c_n S_n\otimes F_n$ and it is equivalent to the set of conditions:

\ben \label{ee20}
\sum_{k}\Pi_{k} S_{n}\Pi_{k}=S_{n},\ \ n=1,...,L,
\een

or equivalently $[S_n,\Pi_{k}]=0$; for all k, n. This means that the set of operators ${S_n}$ has a common eigenbasis defined by the set of projectors ${\Pi_{k}}$. Therefore, the set  ${\Pi_{k}}$ exists if and only if

\ben \label{ee21}
\left[S_n,S_m\right]=0,\  \ n,m=1,...,L.
\een

In order to show zero discord we have to check at most $L(L-1)/2$ commutators, where $L=rank R \leq min\left\{ d^{2}_{A},d^{2}_{B}\right\}$. Now, recall that the state of zero discord is of the form $\rho =\sum^{d_A}_{k=1} p_{k}\Pi_{k}\otimes\rho_{k}$; therefore, it is a sum of at most $d_A$ product operators. This bounds the rank of the correlation tensor to $L\leq d_A$. Thus, the rank of the correlation tensor is the simple discord witness: If $L>d_A$, the state has a nonzero discord.
 A correlation matrix can be obtained directly by simple measurements usually involved in quantum state tomography. However, the detection of nonzero discord does not necessarily require measurement of all $(d_Ad_B)^2$ elements of the correlation matrix (full state tomography). It is sufficient
that the experimentalist measures that many elements of the correlation matrix until he finds $d_A+1$ linearly independent rows (or columns) of the correlation matrix.

 {\it Geometric measure of discord} \cite{dvb10}. Evaluation of quantum discord given by Eq. (\ref{ee17}) in general requires considerable numerical minimization. Different measures of quantum discord \cite{bt10} and their extensions to multipartite systems \cite{mpsvw10} have been proposed. However, analytical expression are known only for certain classes of states \cite{lu08}. Here we use the following geometric measure, proposed in Ref. \cite{dvb10}.

\ben \label{ee22}
D^{\left(2\right)}_{A}\left(\rho\right)=\min_{\chi\in\Omega_0}\left\|\rho-\chi\right\|^2,	
\een
where $\Omega_0$ denotes the set of zero-discord states and is the square norm in the $\left\|X-Y\right\|^2=Tr\left(X-Y\right)^2$ Hilbert-Schmidt space and $\chi$ denotes the classical state. We will show how to evaluate this quantity for an arbitrary two-qubit state.

 {\it Two-qubit case} \cite{dvb10}: Consider the case $\mathcal{H_A}=\mathcal{H_B}=\mathbb{C}$. We write a state $\rho$ in Bloch representation

\ben \label{ee23}
\rho=\frac{1}{4}\left( I \otimes I+\sum^{3}_{i=1} x_{i}\sigma_{i}\otimes I+\sum^{3}_{i=1} y_{i} I\otimes\sigma_{i} +
\sum^{3}_{i,j=1}T_{ij}\sigma_{i}\otimes\sigma_{j}\right)
\een
where $x_i=Tr\rho\left(\sigma_{i}\otimes I\right)$, $y_i=Tr\rho\left(I\otimes\sigma_{i}\right)$ are component of the local Bloch vectors, $T_{ij}=Tr\rho\left(\sigma_{i}\otimes\sigma_{j}\right)$ are component of the correlation tensor, and $\sigma_{i}$, $i\in \left[1,2,3\right]$, are the three Pauli matrices. To each state $\rho$ we associate the triple $\left\{\vec{x},\vec{y},T\right\}$. Now, we characterize the set $\Omega_0$. A zero discord state is of the form
$\chi=p_1\left|\psi_1\right\rangle\left\langle\psi_1\right|\otimes\rho_1+p_2\left|\psi_2\right\rangle\left\langle\psi_2\right|\otimes\rho_2$,
where $\left\{\left|\psi_1\right\rangle,\left|\psi_2\right\rangle\right\}$ is a single-qubit orthonormal basis, $\rho_{1,2}$ are $2\times2$ density matrices, and $p_{1,2}$ are non-negative numbers such that $p_1+p_2=1$.

 We define$t=p_1-p_2$ and three vectors

\ben \label{ee24}
\vec{e}=\left\langle \psi_1\left|\vec{\sigma}\right|\psi_1\right\rangle.
\een

\ben \label{ee25}
\vec{s}_{\pm}=Tr\left( p_1\rho_1 \pm p_2 \rho_2 \right)\vec{\sigma}.	
\een

It can easily be shown that $t\vec{e}$ and $\vec{s}_{+}$ represent the local Bloch vectors of the first and second qubit, respectively,while the vector $\vec{s}_{-}$ is directly related to the correlation tensor which is of the product form $T=\vec{e}\vec{s}^{T}_{-}$. Therefore, a state of zero discord $\chi$ has Bloch representation $\vec{\chi}=\left\{t\vec{e},\vec{s}_{+},T=\vec{e}\vec{s}^{T}_{-}\right\}$, where $\left\|\vec{e}\right\|=1$, $\left\|\vec{s}_{\pm}\right\|\leq 1$, and $t\in \left[-1,1\right]$. The distance between state $\rho$ and $\chi$ is given by

\ben \label{ee26}
\left\|\rho-\chi\right\|^2=\left\|\rho\right\|^2-2Tr\rho\chi+\left\|\chi\right\|^2
\een
$$=\frac{1}{4}\left(1+\left\|\vec{x}\right\|^2+\left\|\vec{y}\right\|^2+\left\|T\right\|^2\right)$$
$$-\frac{1}{2}\left(1+t\vec{x}\vec{e}+\vec{y}\vec{s_{+}}+\vec{e}T\vec{s}_{-}\right)$$
$$+\frac{1}{4}\left(1+t^2+\left\|\vec{s}_{+}\right\|^2+\left\|\vec{s}_{-}\right\|^2\right) $$
where $\left\|T\right\|^2=TrT^{T}T$. First, we optimize the distance over parameters $t$ and $\vec{s}_{pm}$. It is straightforward to see that its Hessian is a positive and nonsingular matrix. Therefore, the function has a unique global minimum. The minimum occurs when the derivative is zero:

\ben \label{ee27}
\frac{\left\|\rho-\chi\right\|^2}{\partial t}=\frac{1}{2}	\left(-\vec{x}\vec{e}+t\right)=0.
\een

\ben \label{ee28}
\frac{\left\|\rho-\chi\right\|^2}{\partial\vec{s}_{+}}=\frac{1}{2}	\left(-\vec{y}\vec{e}+\vec{s}_{+}\right)=0.
\een

\ben \label{ee29}
\frac{\left\|\rho-\chi\right\|^2}{\partial\vec{s}_{-}}=\frac{1}{2}	\left(-T^T\vec{e}+\vec{s}_{-}\right)=0.
\een
which gives the solution $t=\vec{x}\vec{e}$, $\vec{s}_{-}=\vec{y}$, $\vec{s}_{+}=T^T\vec{e}$. Since the solution lies within the range of parameter, $\left|\vec{x}\vec{e}\right|,\left\|\vec{y}\right\|, \left\|T^T\vec{e}\right\|\leq1$, it represents the global minimum. After substituting the solution we obtain

\ben \label{ee30}
\left\|\rho-\chi\right\|^2=\frac{1}{4}\left[\left\|\vec{x}\right\|^2+\left\|T\right\|^2-\vec{e}\left(\vec{x}\vec{x}^T+TT^T\right)\vec{e}\right],
\een
which attains the minimum when $\vec{e}$ is an eigenvector of matrix $K=\vec{x}\vec{x}^T+TT^T$ for the largest eigenvalue. Therefore, we have

\ben \label{ee31}
D^{\left(2\right)}_{A}\left(\rho\right)=\frac{1}{4}\left(\left\|\vec{x}\right\|^2+\left\|T\right\|^2-k_{max}\right)	
\een
where $k_{max}$ is the largest eigenvalue of matrix $K=\vec{x}\vec{x}^T+TT^T$.

\section{ Bloch Representation}

  Throughout this thesis we use the geometric approach to a density matrix via its Bloch representation. The determination of a state on the basis of the actual measurement (experimental data) is important both for experimentalists and theoreticians. In classical physics, it is trivial because there is a one-to-one correspondence between the state and the actual measurement. On the other hand, in quantum mechanics, where a density matrix is used to describe the state, it is generally nontrivial to connect them \cite{perb93,neum32,pgsh33,des76,w92}. the Bloch representation of the density matrix can be constructed experimently giving the required connection between the density matrix and experiments.

$N$-level quantum states are described by density operators, i.e. unit trace Hermitian positive semidefinite linear operators, which act on the Hilbert space $\mathcal{H} \simeq \mathbb{C}^N.$ The Hermitian operators acting on $\mathcal{H}$ constitute a Hilbert space themselves, the so-called Hilbert-Schmidt space denoted by $HS(\mathcal{H})$, with inner product $(\rho,\sigma)_{HS}=Tr(\rho^{\dagger}\sigma)$. Accordingly, the density operators can be expanded by any basis of this space. In particular, we can choose to expand $\rho$ in terms of the identity operator $I_N$ and the traceless Hermitian generators of $SU(N) \;\; \lambda_i \; (i=1,2,\cdots,N^2-1),$

\ben \label{ee32}
\rho=\frac{1}{N}(I_N+\sum^{N^2-1}_{i=1} r_i \lambda_i).
\een

The generators of $SU(N)$ satisfy the orthogonality relation

\ben \label{ee33}
(\lambda_i,\lambda_j)_{HS}= Tr(\lambda_i\lambda_j)=2\delta_{ij},	
\een
and they are characterized by the structure constants of the corresponding Lie algebra, $f_{ijk}$ and $g_{ijk}$, which are, respectively, completely antisymmetric and completely symmetric,

\ben \label{ee34}
\lambda_i\lambda_j = \frac{2}{N}\delta_{ij}I_N+if_{ijk}\lambda_k+g_{ijk}\lambda_k.	
\een

The generators can be easily constructed from any orthonormal basis $\{|j\rangle\}^{N-1}_{j=0}$ in $\mathcal{H}$ \cite{he81}. The (orthogonal) generators are given by

\ben \label{ee35}
\{\lambda_i\}^{N^2-1}_{i=1}=\{u_{jk},v_{jk},w_l\},	
\een
when $i=1,\cdots,N-1$

\ben \label{ee36}
\lambda_i=w_l=\sqrt{\frac{2}{l(l+1)}}\sum^l_{j=1}(|j\rangle\langle j|-l|l+1\rangle\langle l+1|), \; 1 \le l \le N-1,	 
\een
while for $i=N,\cdots,(N+2)(N-1)/2$

$$\lambda_i=u_{jk}=|j\rangle \langle k|+|k \rangle \langle j|,$$
and for $i=N(N+1)/2,\cdots,N^2-1$
 $$\lambda_i= v_{jk}=-i(|j\rangle \langle k|- |k\rangle \langle j|),$$
 $1 \le j \le k\le N.$\\
The orthogonality relation Eq. (\ref{ee33}) implies that the coefficients in Eq. (\ref{ee32}) are given by
$$r_i=\frac{N}{2}Tr(\rho\lambda_i).$$
Notice that the coefficient of $I_N$ is fixed due to the unit trace condition. The vector $ \textbf{r}=(r_1r_2\cdots r_{N^2-1})^t \in \mathbb{R}^{N^2-1}$, which completely characterizes the density operator, is called Bloch vector or coherence vector. The representation Eq. (\ref{ee32}) was introduced by Bloch \cite{bloc46} in the $N=2$ case and generalized to arbitrary dimensions in \cite{he81}. Any density matrix in two-level systems turns out to be characterized uniquely by a three-dimensional real vector where the length satisfies

\ben \label{ee37}
|\mathbf{\lambda}| \equiv \sqrt{\lambda_i \lambda_i}\le 1.
\een

 Therefore, if we define the Bloch-vector space $B(\mathbb{R}^3)$ as a ball with radius 1:
$$B(\mathbb{R}^3)=\{\mathbf{\lambda}=(\lambda_1,\lambda_2,\lambda_3) \in \mathbb{R}^3 : |\mathbf{\lambda}| \le 1\},$$
its element gives an equivalent description of the density matrix with the following bijection (one-to-one and onto) map from $B(\mathbb{R}^3)$ to the set of density matrices. $$\mathbf{\lambda} \longrightarrow \rho=\frac{1}{2}I_2+\frac{1}{2}\lambda_i \sigma_i$$
$B(\mathbb{R}^3)$ is called the Bloch ball, its surface the Bloch sphere and its element the Bloch vector. The equality in Eq. (\ref{ee37}) (i.e., $|\mathbf{\lambda}|=1$), the surface of the ball (the Bloch sphere) which constitutes the set of extreme points of Bloch ball, corresponds to the set of pure states, the points interior to the Bloch ball correspond to mixed states. It has an interesting appeal from the experimentalist point of view, since in this way it becomes clear how the density operator can be constructed from the expectation values of the operators $\lambda_i$,

\ben \label{ee38}
\langle \lambda_i \rangle = Tr(\rho \lambda_i)=\frac{2}{N} r_i.	
\een

As we have seen, every density operator admits a representation as in Eq. ({\ref{ee32}); however, the converse is not true. A matrix of the form Eq. (\ref{ee32}) is of unit trace and Hermitian, but it might not be positive semidefinite, so to guarantee this property further restrictions must be added to the coherence vector. The set of all the Bloch vectors that constitute a density operator is known as the Bloch-vector space $B(\mathbb{R}^{N^2-1}).$ from above discussion it is known that in the case $N=2$ this space equals the unit ball in $\mathbb{R}^3$ and pure states are represented by vectors on the unit sphere. The problem of determining $B(\mathbb{R}^{N^2-1})$ when $N \ge 3$ is still open and a subject of current research \cite{kk05}. However, many of its properties are known. For instance, for pure states $(\rho^2=\rho)$ it must hold

\ben \label{ee39}
||\mathbf{r}||_2=\sqrt{\frac{N(N-1)}{2}},\;\;r_ir_jg_{ijk}=(N-2)r_k,	
\een

where $||.||_2$ is the Euclidean norm on $\mathbb{R}^{N^2-1}$. In the case of mixed states, the conditions that the coherence vector must satisfy in order to represent a density operator have been recently provided in \cite{kim03,bk03}. Regretfully, their mathematical expression is rather cumbersome. It is also known \cite{har78,koss03} that $B(\mathbb{R}^{N^2-1})$ is a subset of the ball $D_R(\mathbb{R}^{N^2-1})$ of radius $R=\sqrt{\frac{N(N-1)}{2}}$, which is the minimum ball containing it, and that the ball $D_r(\mathbb{R}^{N^2-1})$ of radius $r=\sqrt{\frac{N}{2(N-1)}}$ is included in $B(\mathbb{R}^{N^2-1})$. that is,

\ben \label{ee40}
D_r(\mathbb{R}^{N^2-1}) \subset B(\mathbb{R}^{N^2-1}) \subset D_R(\mathbb{R}^{N^2-1}).	
\een
In the case of bipartite quantum systems of dimensions $M\times N\;(\mathcal{H} \simeq \mathbb{C}^M\otimes \mathbb{C}^N)$ composed of subsystems $A$ and $B$, we can analogously represent the density operators as

\ben \label{ee41}
\rho=\frac{1}{MN}(I_M\otimes I_N + \sum_i r_i \lambda_i\otimes I_N +\sum_j s_j I_M \otimes  \tilde{\lambda_j} + \sum_{ij} \lambda_i \otimes \tilde{\lambda_j}),	
\een

where $\lambda_i \; (\tilde{\lambda_j})$ are the generators of $SU(M) \; (SU(N))$. Notice that $\mathbf{r} \in \mathbb{R}^{M^2-1}$ and $\mathbf{s}\in \mathbb{R}^{N^2-1}$ are the coherence vectors of the subsystems, so that they can be determined locally,

\ben \label{ee42}
\rho_A=Tr_B\rho=\frac{1}{M}(I_M+\sum_i r_i \lambda_i), \;\;\rho_B=Tr_A\rho=\frac{1}{N}(I_N+\sum_i s_i \tilde{\lambda_i}).	
\een

The coefficients $t_{ij}$, responsible for the possible correlations, form the real matrix $T \in \mathbb{R}{(M^2-1)\times(N^2-1)}$, and, as before, they can be easily obtained by $t_{ij}=\frac{MN}{4}Tr(\rho\lambda_i\otimes  \tilde{\lambda_j})=\frac{MN}{4}\langle \lambda_i\otimes  \tilde{\lambda_j}\rangle.$

\section{Chapterwise Summary}

This thesis is concerned with: (a) Production of genuine multipartite entanglement quantum systems ($> 2$ qubits) as well as bipartite entangled quantum systems (2 qubits and qutrits). (b) Quantification of both bipartite entanglement and genuine multipartite entanglement for indistinguishable spin fermions. (c) Thermal quantum discord and classical correlations in a two qubit XX model (Heisenberg chain) in a non-uniform external magnetic field.
(d) Finding a geometric measure of quantum discord for an arbitrary state of a bipartite quantum system. The chapters are arranged as follows :

\textbf{Chapter 2} : In this chapter,we develop a geometric approach to quantify the capability of creating entanglement for a general physical interaction acting on two qubits, two qutrits and three qubits. We use the entanglement measure proposed by A. S. Hassan and P. S. Joag for $N$-qubit pure states (Phys. Rev. A \textbf{77}, 062334 (2008)). This geometric method has the distinct advantage that it gives the experimentally implementable criteria to ensure the optimal entanglement production rate without requiring a detailed knowledge of the state of the two qubit system. For the production of entanglement in practice, we need criteria for optimal entanglement production which can be checked {\it in situ} without any need to know the state, as experimentally finding out the state of a quantum system is generally a formidable task. Further, we use our method to quantify the entanglement capacity in higher level and multipartite systems. We quantify the entanglement capacity for two qutrits and find the maximal entanglement generation rate and the corresponding state for the general isotropic interaction between qutrits, using the entanglement measure of $N$-qudit pure states proposed by  A. S. Hassan and P. S. Joag (Phys. Rev. A \textbf{80}, 042302 (2009)). Next we quantify the genuine three qubit entanglement capacity for a general interaction between qubits. We obtain the maximum entanglement generation rate and the corresponding three qubit state for a general isotropic interaction between qubits. The state maximizing the entanglement generation rate is of the GHZ class. To the best of our knowledge, the entanglement capacities for two qutrit and three qubit systems have not been reported earlier. \\

\textbf{Chapter 3} : This chapter reports our work on multipartite entanglement in a system consisting of indistinguishable fermions. Specifically, we have proposed a geometric entanglement measure for $N$ spin-$\frac{1}{2}$ fermions distributed over $2L$ modes (single particle states). The measure is defined on the $2L$ qubit space isomorphic to the Fock space for $2L$ single particle states. This entanglement measure is defined for a given partition of $2L$ modes containing $m\geq 2$ subsets. Thus this measure applies to $m\leq 2L$ partite fermionic system where $L$ is any finite number, giving the number of sites. The Hilbert spaces associated with these subsets may have different dimensions.  Further, we have defined the local quantum operations with respect to a given partition of modes. This definition is generic and unifies different ways of dividing a fermionic system into subsystems. We have shown, using a representative case, that the geometric measure is invariant under local unitaries corresponding to a given partition. We explicitly demonstrate the use of the measure to calculate multipartite entanglement in some correlated electron systems. To the best of our knowledge, there is no usable entanglement measure of $m>3$ partite fermionic systems in the literature, so that this is the first measure of multipartite entanglement for fermionic systems going beyond the bipartite and tripartite cases.

\textbf{Chapter 4} : In this chapter,we investigate how thermal quantum discord $(QD)$ and classical correlations $(CC)$ of  a two qubit one-dimensional XX Heisenberg chain in thermal equilibrium depend on temperature of the bath as well as on nonuniform external magnetic fields applied to two qubits and varied separately. We show that the behavior of $QD$ differs in many unexpected ways from thermal entanglement $(EOF)$. For the nonuniform case, $(B_1= - B_2)$ we find that $QD$ and $CC$ are equal for all values of $(B_1=-B_2)$ and for different temperatures. We show that, in this case, the thermal states of the system belong to a class of mixed states and satisfy certain conditions under which $QD$ and $CC$ are equal. The specification of this class and the corresponding conditions are completely general and apply to any quantum system in a state in this class and satisfying these conditions. We further find that the relative contributions of $QD$ and $CC$ can be controlled easily by changing the relative magnitudes of $B_1$ and $B_2$. Finally, we connect our results with the monogamy relations between the EOF, classical correlations and the quantum discord of two qubits and the environment. \\

\textbf{Chapter 5} : Quantum discord, as introduced by Olliver and Zurek [Phys. Rev. Lett. \textbf{88}, 017901 (2001)], is a measure of
the discrepancy between quantum versions of two classically equivalent expressions for mutual information.
Dakic, Vedral, and Brukner [Phys. Rev. Lett. 105, 190502 (2010)] introduced a geometric measure of quantum discord and
derived an explicit formula for any two-qubit state. Luo and Fu [Phys. Rev. A \textbf{82}, 034302 (2010)] introduced another (equivalent) form for geometric measure of quantum discord. We find an exact formula for the geometric measure of quantum discord for an arbitrary state of a $m\times n$ bipartite quantum system, using the form for geometric measure of quantum discord given by Luo and Wu.\\

\textbf{Chapter 6} : In this chapter, we summarize the work presented in this thesis and give the possible ways in which this work may be developed further.

\chapter{Entanglement capacity of nonlocal Hamiltonians: A geometric approach}

\begin{figure}[!ht]
\begin{center}
\includegraphics[width=5cm,height=.75cm]{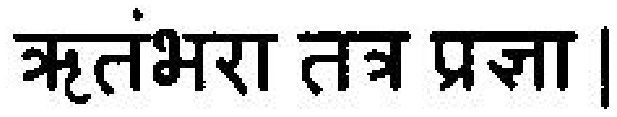}

Yoga Sutras of Patanjali\\
\scriptsize\textsc{ Creativity of such a person gets one with the universal creation, transcending all limitations of an individual. }
\end{center}
\end{figure}

  In this chapter, We develop a geometric approach to quantify the capability of creating entanglement for a general physical interaction acting on two qubits. We use the geometric entanglement measure for N-qubit pure states \cite{hj08}. This geometric method has the distinct advantage that it gives the experimentally implementable criteria to ensure the optimal entanglement production rate without requiring a detailed knowledge of the state of the two qubit system. For the production of entanglement in practice, we need criteria for optimal entanglement production, which can be checked in situ without any need to know the state, as experimentally finding out the state of a quantum system is generally a formidable task. Further, we use our method to quantify the entanglement capacity in higher level and multipartite systems. We quantify the entanglement capacity for two qutrits and find the maximal entanglement generation rate and the corresponding state for the general isotropic interaction between qutrits, using the entanglement measure of N-qudit pure states \cite{hj09}. Next we quantify the genuine three qubit entanglement capacity for a general interaction between qubits. We obtain the maximum entanglement generation rate and the corresponding three qubit state for a general isotropic interaction between qubits. The state maximizing the entanglement generation rate is of the Greenberger-Horne-Zeilinger class. To the best of our knowledge, the entanglement capacities
for two qutrit and three qubit systems have not been reported earlier.

We have already seen in chap.1 that a quantum system evolves to generate entanglement provided its parts interact. For such an interaction, the Hamiltonian of the total system is not just a sum of the Hamiltonians pertaining to each part (local Hamiltonians). Thus, for a bipartite system $AB,$ $H_{AB}\neq H_{A}+H_{B}$ but has a term which couples the two parts $A$ and $B.$ Together with local operations, this coupling can be used to generate entanglement \cite{dvclp01,bhls03,zzf00}, to transmit classical and quantum information \cite{bhls03,bgnp01,hvc02,bs03} and more generally, to simulate the dynamics of some other Hamiltonian (say $H^{\prime}_{AB}$) and thus to perform arbitrary unitary gates on the composite space $\mathcal{H}_{AB}=\mathcal{H}_{A}\otimes \mathcal{H}_{B}$ \cite{dnbt02,bclllpv02,vc02}.

A lot of experimental work is devoted to creating entangled states of quantum systems, including those in quantum optics, nuclear magnetic resonance and condensed matter physics \cite{f00}. Determining the ability of a system to create entangled states provides a benchmark of the ``quantumness'' of the system. Furthermore, such states can ultimately be put to some information processing task like superdense coding \cite{bw92}, or quantum teleportation \cite{bbcjpw93}.

The theory of optimal entanglement generation can be approached in different ways. Ref. \cite{dvclp01} considers {\it single shot} capacities. For two qubit interaction, without any ancilla qubits, Ref.\cite{dvclp01} presents a closed form expression for the entangling capability and optimal protocols by which it can be achieved. In contrast, Ref.\cite{bhls03} considers the {\it asymptotic} entanglement capacity, allowing the use of ancillary systems and shows that when ancillas are allowed, the single shot and asymptotic capacities are in fact the same. However, such capacities are difficult to calculate because the ancillary systems may be arbitrarily large. In this chapter we exclusively deal with the single shot entanglement capacity. Throughout this chapter, we take $\hbar=1.$

The chapter is organized as follows. In section 2.1 we deal with entanglement capacity for two qubit states while in section 2.2 we deal with this problem involving two qutrits. In section 2.3 we address the problem of the entanglement capacity involving the genuine tripartite entanglement for three qubits. The discussion of the results for two qubit, two qutrit and three qubit cases is included separately in sections 2.1, 2.2 and 2.3 respectively.

\section{The two qubit case}

We develop a geometric approach to calculate the entanglement capacity of any two qubit system ( for the case of pure state) interacting via a Hamiltonian which is locally equivalent to

\ben \label{en1}
H_{I}= \mu_1\si^{A}_1\otimes \si^{B}_1+ \mu_2\si^{A}_2\otimes \si^{B}_2+ \mu_3\si^{A}_3\otimes \si^{B}_3.
\een

Here $\mu_1\geq\mu_2\geq\mu_3.$ In this section we omit qubit and system identifiers $A,B$ and $AB.$

 We define the single shot entanglement capacity by

\ben \label{en2}
\Gamma^{max}=\max_{\substack{|\psi\ran\in\mathcal{H}_{I}}}
\lim_{t\to 0}\frac{E\Big(e^{-iHt}|\psi\ran\Big)-E (|\psi\ran)}{t}.
\een

The Hamiltonian $H_{I}$ in Eq. (\ref{en2}) is given by Eq. (\ref{en1}).
$E(|\psi\ran)$ in Eq. (\ref{en2})  stands for the two qubit pure state entanglement measure given by us and is shown to have all the essential (as well as many desirable, e.g., superadditivity and continuity) properties expected of a good entanglement measure \cite{hj08}. For a $N$-qubit pure state $|\psi\ran ,$
$$E(|\psi\ran)=||\mathcal{T}^{(N)}||-1$$ where $||\mathcal{T}^{(N)}||$ is the Hilbert-Schmidt (Euclidean) norm of the $N$ way array $\mathcal{T}^{(N)}$ occurring in the Bloch representation of $|\psi\ran\lan\psi|$ \cite{hj08,hjq08}.

The scenario we address, is as follows \cite{dvclp01}.  The idea is to supplement the interaction Hamiltonian $H_{I}$ with appropriate local unitary operations in such a way that the state of the qubits at any time $t$ is precisely $|\psi_{E(t)}\ran$, for which the increase of entanglement is optimal. In order to construct such a procedure, we consider the evolution given by $H_{I}$ to proceed in very small time steps $\delta t$. Let us also assume that the qubits are initially disentangled. Using local operations,  we can always prepare the state $|\psi_0\ran$ that is, the product state which most efficiently becomes entangled under the action of $H_{I}$. After a time step $\delta t$, the state will change and its entanglement will increase to $\delta E$. Then, we use (fast) local unitary operations to transform the new state of the qubits into the state $|\psi_{\delta E}\ran$ for which $\Gamma$ is optimal. Note that this is always possible, since for qubits all states with the same value of $E$, say $\delta E$, are connected by local unitary transformations. By proceeding in the same way after every time step, and taking the continuous time limit $\delta t\rightarrow 0$, we obtain that the state of the qubits at time $t$ is always the optimal one, $|\psi_{E(t)}\ran$. Obviously, in an experimental realization, this procedure requires that we can apply the appropriate local transformations in times which are short compared to the typical time scale $\tau_{H_{I}}$ associated with $H_{I}$, $\tau_{H}=(e_{max}-e_{min})^{-1}$, where $e_{max}$ and $e_{min}$ are the maximum and minimum eigenvalues of $H_{I}$. Note that Eq. (\ref{en2}) defines the entanglement capacity as the maximum achievable entanglement rate for a given system with given interactions. We are also interested in finding the state $|\psi_{max}\ran$ for which the entanglement rate is maximum, (denoted by $\Gamma^{max}$ in Eq. (\ref{en2})).

We consider two qubits interacting via the Hamiltonian $H_{I}$ in Eq. (\ref{en1}), which represents general interaction between two qubits \cite{dvclp01}. First we find the entanglement rate $\Gamma$ given by

\ben \label{en3}
\Gamma = \lim_{t\to 0}\left[\frac{E\Big(e^{-iHt}|\psi\ran\Big)-E (|\psi\ran)}{t}\right] \equiv \frac{dE}{dt}.
\een

Here $|\psi\ran$ is given by a general two qubit state in the Bloch representation \cite{hj08,hjq08},

\ben \label{en4}
\rho=|\psi\ran\lan\psi|=\frac{1}{4}\left(I\otimes I + \sum_{k}r_{k} \si_{k}\otimes I + \sum_{l}s_{l} I\otimes \si_{l}
+ \sum_{k,l}\tau_{kl}\si_{k}\otimes \si_{l}\right),
\een
where $\si_{k,l},\;k,l=1,2,3$ are the Pauli operators.
We denote by $\mathcal{T}=[\tau_{ij}]$ the correlation matrix occurring in the last term of Eq. (\ref{en4}). $\tau_{ij}$ are defined by

\ben \label{en5}
\tau_{ij}=Tr(\si_i\otimes\si_j \rho)=\lan\psi|\si_i\otimes\si_j|\psi\ran.
\een 	

$r_{k}$ and $s_{l}$ are the components of the Bloch vectors \cite{hjq08} of the reduced density operators $\rho_{A}$ and $\rho_{B}$ respectively, given by

\ben\label{en6a}
r_{k}=Tr(\si_{k}\rho_{A})=\lan\psi|\si_k \otimes I |\psi\ran,\  \ \ \left(a\right)
\een

\label{en6b}
$$s_{l}=Tr(\si_{l}\rho_{B})=\lan\psi| I \otimes \si_{l} |\psi\ran. \  \ \ \left(b\right)$$

We define the entanglement of the state $|\psi\ran$ as \cite{hj08}

\ben \label{en7}
E(|\psi\ran)=||\mathcal{T}||-1,
\een
where $||\mathcal{T}||=\sqrt{\sum_{ij=1}^{3}\tau_{ij}^2}$ is the Euclidean norm of $\mathcal{T}.$ For two qubits, this measure is related to concurrence \cite{hj08} and hence to the Von Neumann entropy of the reduced density matrix.

After finding $\Gamma$ we maximize it, using a simple geometric argument. It is heartening to see that the scenario described above emerges naturally out of this geometric method.

The entanglement rate $\Gamma$ is given by (See Eq. (\ref{en3})) $$\Gamma= \frac{dE}{dt}=\frac{d||\mathcal{T}||}{dt}=
\frac{1}{||\mathcal{T}||} \sum_{ij}\tau_{ij}\dot{\tau}_{ij},$$ with $\tau_{ij}$ given by Eq. (\ref{en5}). We evaluate $\dot{\tau}_{ij}$ as follows. $$\dot{\tau}_{ij}=\frac{d\tau_{ij}}{dt}=\frac{d}{dt}(Tr(\si_i\otimes\si_j \rho))=Tr(\si_i\otimes\si_j\fr{d\rho}{dt}).$$ We now use the equation of motion ,
$$i\frac{d\rho}{dt}=[H_{I},\rho],$$ where the Hamiltonian $H_{I}$ is defined via Eq. (\ref{en1}), to get \cite{mw95},
 $$\fr{d\t_{ij}}{dt}=-iTr(\si_i\otimes\si_j [H_{I},\rho])=iTr(H_{I}[\si_i\otimes\si_j,\rho]).$$  Substituting $\rho$ from Eq. (\ref{en4}) and using the commutation relations \cite{mw95} $$[\si_i\otimes\si_j,\si_k\otimes\si_l]=\frac{1}{2}[\si_i,\si_k]\otimes\{\si_j,\si_l\}
+\frac{1}{2}\{\si_i,\si_k\}\otimes [\si_j,\si_l],$$ we get, using $[\si_i,\si_j]=2i\vep_{ijk}\si_k ,$ $\{\si_i,\si_j\}=
2\delta_{ij}$ and the expression of $H_{I}$ in Eq. (\ref{en1}), after a bit of algebra, $$\frac{d\tau_{ij}}{dt}=-2\left[\sum_{k,n}
r_k\vep_{ikn}\mu_n\del_{nj}+ \sum_{l,n} s_l\vep_{jln}\mu_n\del_{ni}\right].$$ This gives $$\sum_{ij}\tau_{ij}\dot{\tau}_{ij}=-2\left[\sum_{i,k,n}\tau_{in}r_k\vep_{ikn}\mu_n+ \sum_{j,k,n}\tau_{nj}s_l\vep_{jln}\mu_n\right].$$ Thus we get, for the entanglement rate $\G ,$

\ben \label{en8}
\G = \frac{2}{||\mathcal{T}||}\sum_{n} \left[(\vec{r}\times\vec{\tau}_{:n})_{n} + (\vec{s}\times\vec{\tau}_{n:})_{n} \right]\mu_n .
\een

Here $\vec{\tau}_{:n}$ and $\vec{\tau}_{n:}$ are, respectively, the $n$ th column and row vectors of the correlation matrix $\mathcal{T}=[\tau_{ij}] .$

The entanglement generation rate $\G$ expressed in Eq. (\ref{en8}) is obtained via the temporal evolution of the initial state by the interaction Hamiltonian $H_{I}.$
This expression for $\G$ does not depend on any local unitary transformation applied to a qubit. Following the general scenario described above, (see the third paragraph of this section), we now lock on to an instant of time and apply the local unitary transformations to qubits, in order to find the conditions for optimal $\G$ and the corresponding two qubit state $|\psi_{E}\ran.$ The experimental meaning of this sentence is described as a part of the scenario above. In the geometrical approach we have adopted, local unitary transformations amount to rotations of vectors in Eq. (\ref{en8}), which are the vectors in the Bloch space of individual qubits. We expect the entanglement to remain unultered by the local unitaries, which turns out to be the case. The entanglement measure in Eq. (\ref{en7}) is not affected by local unitaries, as proved in \cite{hj08}.

Obviously, $\G$ will be maximum if the components of the vector products occurring in Eq. (\ref{en8}) are replaced by the magnitudes of these vector products and the factors in these products are mutually perpendicular. Geometrically, this means that the vector products themselves are in the directions of the components occurring in Eq. (\ref{en8}) with the other two orthogonal components zero. For example, $(\vec{r}\times\vec{\tau}_{:1})$ is along its first component, i.e. along $x$ axis, with its $y$ and $z$ components zero. Thus, in order to maximize the first term in Eq. (\ref{en8}), namely,
$$\sum_{n} (\vec{r}\times\vec{\tau}_{:n})_{n}\mu_n=(\vec{r}\times\vec{\tau}_{:1})_{1}\mu_1+(\vec{r}\times\vec{\tau}_{:2})_{2}\mu_2+(\vec{r}\times\vec{\tau}_{:3})_{3}\mu_3,$$ we must have vectors $(\vec{r}\times\vec{\tau}_{:1}),$
$(\vec{r}\times\vec{\tau}_{:2})$ and $(\vec{r}\times\vec{\tau}_{:3})$ along $x,y,z$ axes respectively. This can be done only when one of the vector products is zero. Since $\mu_1\geq\mu_2\geq\mu_3,$
we choose $(\vec{r}\times\vec{\tau}_{:3})=0.$ Given the vector $(\vec{r}\times\vec{\tau}_{:1})$ along the $x$ axis and the vector $(\vec{r}\times\vec{\tau}_{:2})$ along the $y$ axis, we can choose $\vec{r}$ to be along the $z$ axis and vectors $\vec{\tau}_{:1}$ and $\vec{\tau}_{:2}$ along the $y$ and $x$ axes respectively. In exactly the same way, maximization of the second term in Eq. (\ref{en8}), $\sum_{n}(\vec{s}\times\vec{\tau}_{n:})_{n}\mu_n,$ makes the vector $\vec{s}$ along the $z$ axis and vectors $\vec{\tau}_{1:}$ and $\vec{\tau}_{2:}$ along the $y$ and $x$ axes respectively. Writing explicitly the components of the vector products in the expression for $\G$ (Eq. (\ref{en8})) and putting $r_{1,2}=0=s_{1,2}$ we get,
$$\G=\frac{2}{||\mathcal{T}||}((-r_3\tau_{21}-s_3\tau_{12})\mu_1+(r_3\tau_{12}+s_3\tau_{21})\mu_2).$$ Since we are dealing with the two qubit pure states we have $||\vec{r}||=||\vec{s}||$ \cite{v08}, so that $r_3=\pm s_3.$ Choosing

\ben \label{en9}
r_3=- s_3
\een
we get, $$\G = \frac{2}{||\mathcal{T}||}r_3(\tau_{12}-\tau_{21}) (\mu_1+\mu_2).$$
  The expression $(\tau_{12}-\tau_{21})$ becomes maximum when

\ben \label{en10}
\tau_{12}=-\tau_{21}.
\een

Finally, we note that this maximization procedure does not change $||\mathcal{T}||=\sqrt{\sum_{n}||\tau_{:n}||^2}$ and hence the entanglement value given by Eq. (\ref{en7}). Further, choosing the cross products along their components appearing in Eq. (\ref{en8}) corresponds to the rotations in Bloch space, generating local unitaries on the system. Therefore, the maximum of $\G$ over the states with same entanglement, that is, $\G_{E},$ is given by

\ben \label{en11}
\G_{E}=\frac{4}{||\mathcal{T}||}r_3\tau_{12}(\mu_1+\mu_2).
\een

To get the state $|\p_{E}\ran$ corresponding to $\G_{E},$ we seek the state satisfying conditions Eq. (\ref{en9}) and Eq. (\ref{en10}). We start with the general state $|\p\ran =\sum_{i,j=0}^{1}c_{ij}|ij\ran$ and calculate $\tau_{12}$ and $\tau_{21}.$ In order to satisfy Eq. (\ref{en10}), the state $|\p\ran$ should be

\ben\label{en12}
|\p_{E}\ran = |c_{01}| |01\ran + i|c_{10}| |10\ran\;;\;|c_{01}|^2+|c_{10}|^2=1,
\een
which is the same as $|\p_{E}\ran$ obtained in Ref \cite{dvclp01} if we identify $|c_{01}|=\sqrt{p}.$ Further, we can write $\G_{E}$ (Eq. (\ref{en11})) as the product of two factors $$\G_{E}=f(p)h_{max}$$ with
\ben \label{en13}
h_{max}=(\mu_1+\mu_2)$$ and $$f(p)=\frac{4 r_3\tau_{12}}{||\mathcal{T}||}.
\een
To get $f(p)$ as a function of $p,$ we calculate $r_3,$ $\tau_{12}$ and $||\mathcal{T}||$ using the state $|\p_{E}\ran$ (Eq. (\ref{en12})) so that
$$f(p)=\frac{4 r_3\tau_{12}}{||\mathcal{T}||}=\frac{8(1-2p)\sqrt{p(1-p)}}{\sqrt{1+8p(1-p)}}.$$
Fig. (2.1a) depicts this $f(p)$ verses $p,$ while Fig. (2.1b) plots the analogous $f(p)$ obtained using Von Neumann entropy as the entanglement measure. Note that $f(p)$ and hence $\G_{E}$
vanishes for the maximally entangled state ($p=\frac{1}{2}$) about which it is antisymmetric $f(\frac{1}{2}+x)=-f(\frac{1}{2}-x).$ We see that, as $p$ increases from $0$ to $\fr{1}{2}$, $\G_E>0$ makes the entanglement increase, until is maximal at $p=\fr{1}{2}$, after which $\G<0$, making entanglement decrease to zero as $p$ approaches $1$.
\
\\
\
\\
\
\\
\
\\
\
\\
\
\\
\
\\
\
\\
\begin{figure}[!ht]
\begin{center}
\includegraphics[width=11cm,height=8.3cm]{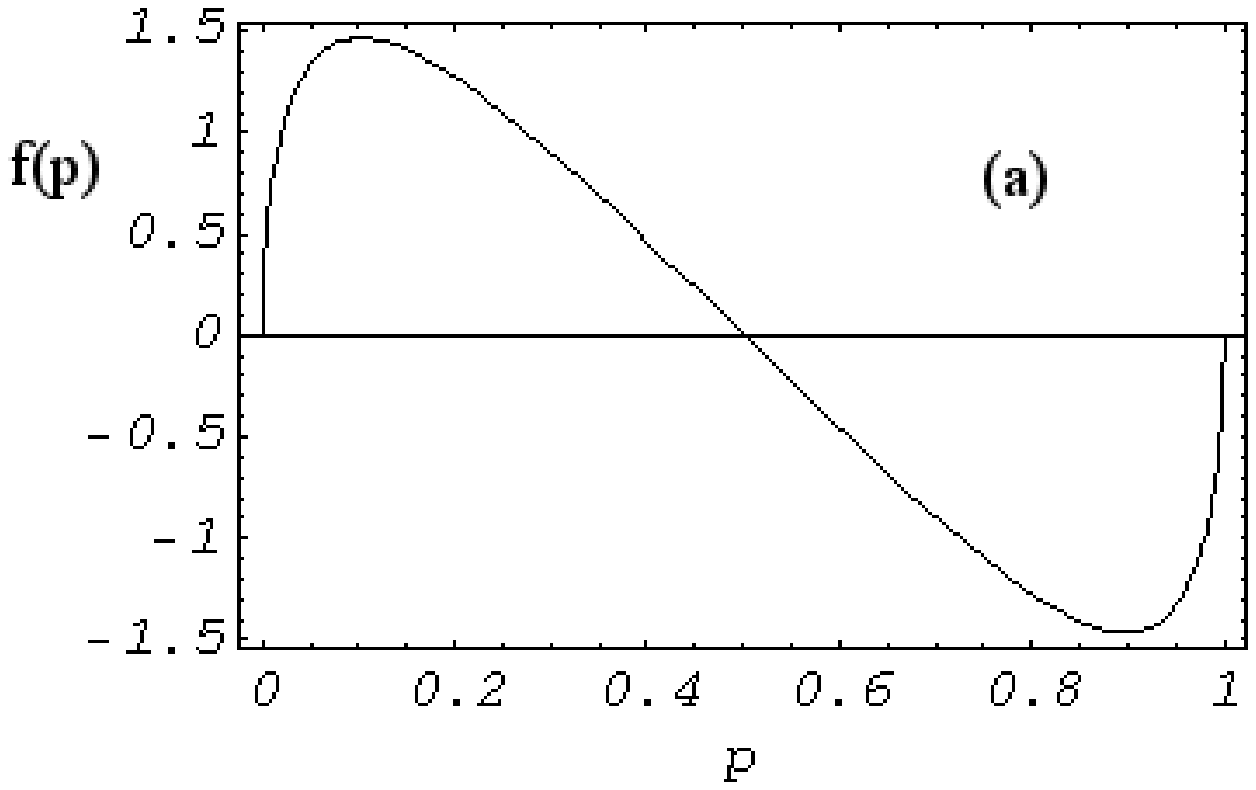}

Fig. (2.1a)
\end{center}
\end{figure}

\begin{figure}[!ht]
\begin{center}
\includegraphics[width=10.5cm,height=8.3cm]{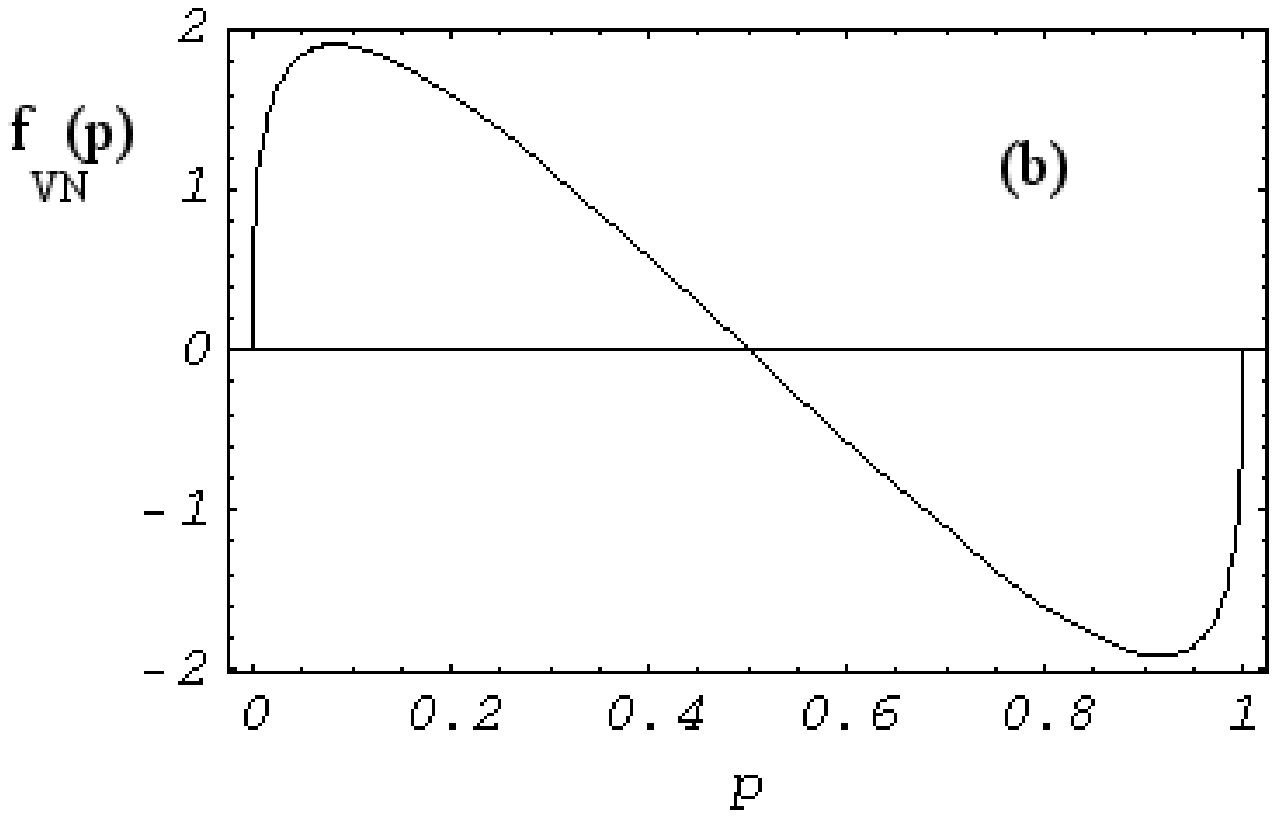}

Fig. (2.1b)

Fig. (2.1): (a) $f(p)$ vs. $p$ for entanglement measure in Eq. (\ref{en7}) and (b) $f_{VN}(p)$ vs. $p$ for Von Neumann entropy of the reduced density matrix (see text).
\end{center}
\end{figure}

Thus we see that, $\G_{E}$ is the product of the function which depends only on the state, (via $p$) and the factor $h_{max}$ which depends only on the interaction strengths $\mu_1$ and $\mu_2$, that is, on the interaction Hamiltonian. Note that $h_{max}$ is independent of the entanglement measure. The form of $f(p)$ for the entanglement measure in Eq. (\ref{en7}) and that for the Von Neumann entropy, (Fig. (2.1b)) also turns out to be the same. To get  $\Gamma^{max}$ we have to find $p_0$ at which $f(p)$ is maximum. To do this, we invoke the relation between $f_{VN}(p),$ which is the analog of $f(p)$ in Eq. (\ref{en13}) obtained via Von Neumann entropy of the reduced density operator of the two qubit pure state \cite{dvclp01} and $f(p)$ in Eq. (\ref{en13}) obtained via $E(|\psi\ran)$ in Eq. (\ref{en7}). This is
\setcounter{equation}{13}
\begin{equation}
f_{VN}(p)=f(p)\left(\frac{dE_{VN}}{dp}\Big/\frac{dE}{dp}\right), \label{en14}
\end{equation}
where $E_{VN}$ is the Von Neumann entropy of the reduced density operator and $E$ is given by Eq. (\ref{en7}). Maximizing the RHS of Eq. (\ref{en14}) we get
$p_0\approx 0.0832217$ and $\G^{max}\approx 1.9123$. The state $|\p_{max}\ran$ corresponding to $\Gamma^{max}$ is the state $|\p_{E}\ran$ with $p=p_0.$

From the definitions of $\tau_{ij}$ and $r_k$(Eq. (\ref{en5}) and (\ref{en6a})), these quantities are averages of the Pauli operators in a state $|\psi\ran$, which can be obtained using experimentally measured values of the corresponding operators on two qubits. Therefore, {\it the geometric method presented here has the advantage that the conditions for $\G_{E},$ Eqs. (\ref{ee9}) and (\ref{ee10}), can be tested experimentally giving us an experimental way to check out whether the system has reached the state $|\p_{E}\ran$.} The value of $\G_{E}$ can also be experimentally estimated via Eq. (\ref{en11}), for given $\mu_1+\mu_2.$  Further, the function $f(p)$ can be estimated experimentally via Eq. (\ref{en13}) as the system evolves, under the given Hamiltonian, toward $|\p_{max}\ran,$ or under the local unitaries toward $|\p_{E}\ran$. {\it These experimental estimations can be carried out without a detailed a priori knowledge of the quantum state at any time during the evolution of the two qubit system.} These facts can be of great advantage in a practical implementation of any scheme to entangle two qubits interacting via some Hamiltonian or quantum gates \cite{ejpp00}. For the production of entanglement in practice, we need criteria for optimal entanglement production which can be checked {\it in situ} without any need to know the state, as experimentally finding out the state of a quantum system is generally a formidable task. We note that, in order to achieve such an experimental determination of optimal entanglement production rate using the model in ref \cite{dvclp01}, we have to experimentally obtain the values of the Schmidt coefficients of the evolving two qubit state, which requires the experimental determination of the two qubit state itself. This requires more experimental effort and resources ($d+1$ different joint measurements, $d=$ dimension of the joint Hilbert space \cite{p93}) as compared to measuring the quantities in Eqs. (\ref{ee9}) and (\ref{ee10}), which are simply the average values of the Pauli operators in the state.  \\

\section{The two qutrit case}

  The entanglement measure in Eq. (\ref{en7}) can be generalized to the $N$ qudit pure states which satisfies all the essential (and many desirable, e.g., superadditivity and continuity) properties expected of a good entanglement measure \cite{hj09}. Therefore, we can use it to obtain the entanglement generation rates for the multipartite $d$ level systems. Here we find the entanglement generation rate for two qutrits (labeled $A$ and $B$) interacting via the Hamiltonian
\begin{equation}
H_{I}= \sum_{p=1}^{8}\mu_p\la^{A}_p\otimes \la^{B}_p, \label{en29}
\end{equation}
where $\mu_p$s are the interaction strengths satisfying $\mu_k\geq\mu_l$ for $k<l, k,l=1,\ldots, 8.$ Here $\la_p,\;\;p=1,\ldots,8$ are the generators of the $SU(3)$ group satisfying $Tr(\la_i\la_j)=2\del_{ij}$ and are characterized by the structure constants of the corresponding Lie algebra, $f_{ijk}\;\textrm{and}\;g_{ijk},$ which are, respectively, completely antisymmetric and completely symmetric.
\begin{equation}
\la_i\la_j=\fr{2}{3}\del_{ij}I_3+if_{ijk}\la_k+g_{ijk}\la_k. \label{en30}
\end{equation}
Other useful relations are
\ben
4if_{jkl}=Tr([\la_i,\la_k]\la_l) \label{en31}
\een
\ben
4g_{ikp}=Tr(\{\la_i,\la_k\}\la_p). \label{en32}
\een

We give here the  generators of $SU(3)$ in the $|1\ran,\;|2\ran,\;|3\ran$ basis \cite{mw95} to be used below.\\
$\lambda_1=|1\ran\lan 2|+|2\ran \lan 1|$\\
$\lambda_2=-i(|1\ran\lan 2|-|2\ran \lan 1|)$\\
$\lambda_3=|1\ran\lan 1|-|2\ran \lan 2|$\\
$\lambda_4=|1\ran\lan 3|+|3\ran \lan 1|$\\
$\lambda_5=-i(|1\ran\lan 3|-|3\ran \lan 1|)$\\
$\lambda_6=|2\ran\lan 3|+|3\ran \lan 2|$\\
$\lambda_7=-i(|2\ran\lan 3|-|3\ran \lan 2|)$\\
$\lambda_8=\fr{1}{\sqrt{3}}(|1\ran\lan 1|+|2\ran \lan 2|-2 |3\ran\lan3|).$\\

The action of these generators on the basis states $\{|1\ran,|2\ran,|3\ran\}$ is given by the following.

$\lambda_1 |1\ran=|2\ran,\; \lambda_1 |2\ran=|1\ran,\; \lambda_1 |3\ran=0$\\
$\lambda_2 |1\ran=i|2\ran,\; \lambda_2 |2\ran=-i|1\ran,\; \lambda_2 |3\ran=0$\\
$\lambda_3 |1\ran=|1\ran,\; \lambda_3 |2\ran=-|2\ran,\; \lambda_3 |3\ran=0$\\
$\lambda_4 |1\ran=|3\ran,\; \lambda_4 |2\ran=0,\; \lambda_4 |3\ran=|1\ran$\\
$\lambda_5 |1\ran=i|3\ran,\; \lambda_1 |2\ran=0,\; \lambda_1 |3\ran=-i|1\ran$\\
$\lambda_6 |1\ran=0,\; \lambda_6 |2\ran=|3\ran,\; \lambda_6 |3\ran=|2\ran$\\
$\lambda_7 |1\ran=0,\; \lambda_7 |2\ran=i|3\ran,\; \lambda_7 |3\ran=-i|2\ran$\\
$\lambda_8 |1\ran=\fr{1}{\sqrt{3}}|1\ran,\; \lambda_8 |2\ran=\fr{1}{\sqrt{3}}|2\ran,\; \lambda_8 |3\ran=-\fr{2}{\sqrt{3}}|3\ran .$\\
We use these equations to get the vectors $\vec{\Lambda}^{A}$ and $\vec{\Lambda}^{B}$ in $\mathbb{R}^8$ whose components are the averages $\Lambda_{i}^{A}=\lan\psi|\la_{i}\otimes I|\psi\ran=Tr(\la_{i}\otimes I\rho),\;i=1,\ldots,8$ and $\Lambda_{i}^{B}=\lan\psi|I\otimes\la_{i}|\psi\ran=Tr(I\otimes\la_{i}\rho),\;i=1,\ldots,8$ respectively,
where $|\psi\ran,$ ($\rho=|\psi\ran\lan\psi|$) is a two qutrit pure state (see Eq. (\ref{en36}) and the discussion following it).

The pure state entanglement for two qutrits is given by \cite{hj09},
\begin{equation}
E(|\psi\ran)=||\mathcal{T}||-3  \nn
\end{equation}
where $||\mathcal{T}||=\sqrt{\sum_{ij=1}^{8}\tau_{ij}^2}$ is the Euclidean norm of $\mathcal{T}.$ The general two qutrit pure state $\rho$ has the following Bloch representation.

\benr
\rho &=& |\psi\ran\lan\psi|    \nn \\
&=&\frac{1}{9}\big(I^{A}\otimes I^{B} +\fr{3}{2}\big( \sum_{k}\lan \la_{k}^{A}\ran \la_{k}^{A}\otimes I^{B} + \sum_{l}\lan \la_{l}^{B}\ran I^{A}\otimes \la_{l}^{B}\big) \nn \\
 &+& \fr{9}{4}\sum_{k,l}\tau_{kl}\la_{k}^{A}\otimes \la_{l}^{B}\big),  \nn
\eenr

Here $\lan\la^{A,B}\ran=Tr(\rho_{A,B}\la^{A,B})$ with $\la^{A,B}$ and $\rho_{A,B}$ (the reduced density operator) apply to the qutrit A and B respectively, while $\tau_{kl}=(9/4)Tr(\la_k^{A}\otimes\la_l^{B}\rho).$ The definitions of $\dot{\tau}_{kl}$ and $\G$ are
\begin{equation}
\G=\frac{1}{||\mathcal{T}||} \sum_{ij}\tau_{ij}\dot{\tau}_{ij},  \nn
\end{equation}
\begin{equation}
\dot{\tau}_{ij}=iTr(H[\la_i\otimes\la_j, \rho]),  \nn
\end{equation}
where we have used the Heisenberg equation of motion as in the two qubit case. Using Eq. (\ref{en30}, (\ref{en31}), (\ref{en32}) and the elements of the tensors $f_{ijk}$ and $g_{ijk}$ in \cite{mw95}, we get, after some algebra, the following expression for $\G$
\ben
\G=-3\left(\fr{1}{||\mathcal{T}||}\right)\sum_{k,p,l=1}^{8}\mu_p f_{klp}\left(\tau_{kp}\la_l^{A}+\tau_{pk}\la_l^{B}\right).
\label{en35}
\een
Expanding the sum in Eq. (\ref{en35}) and rearranging, we get,
\ben
\G= -3\left(\fr{1}{||\mathcal{T}||}\right)\sum_{S}\al(S)\sum_{p\in S}\mu_p[(\vec{\tau}_{:p}\times\vec{\la}^{A})_p+
(\vec{\tau}_{p:}\times\vec{\la}^{B})_p], \label{en36}
\een
where $S$ runs over the triplets
$$(1,4,7),(2,1,6),(3,1,5),(3,2,4),(2,5,7),(3,7,6),(5,4,6),(3,6,8),(2,5,8)$$ and $\al(S)$ has
values $1,1/2,1/2,1/2,1/2,1/2,1/2,\sq{3}/2,\sq{3}/2$ respectively for these triplets. $\vec{\tau_{:p}}$ and $\vec{\tau_{p:}}$
are the vectors in $\mathbb{R}^3$ with $p\in S$ where $S$ is one of the above triplets and the index $:$ varies over a given $S$ for fixed $p.$ $\vec{\la}^{A,B}$  are vectors in $\mathbb{R}^3$ respectively comprising the components of $\vec{\Lambda}^{A,B}$ indexed by one of the triplets $S.$ There are in all $54$ terms in Eq. (\ref{en36}). Unfortunately, all these terms are coupled and a simple geometrical procedure to maximize $\G,$ as in the two qubit case, seems very difficult. However, it is straightforward to maximize $\G$ numerically over the coefficients $c_{ij}, \;\;i,j=0,1,2$, by expressing all the terms in the expression for $\G$ (Eq. (\ref{en36})) as averages in the general two qutrit state $|\psi\ran=\sum_{ij}c_{ij}|ij\ran,\;\;i,j=0,1,2.$ We can carry out the numerical maximization for the general Hamiltonian in Eq. (\ref{en29}), where the strengths of interaction $\mu_k$ have different values. In that case, $\G$ does not have the simple structure analogous to $\G=f(p)h_{max}$ as in the two qubit case. Therefore, we assume isotropic interactions so that all interaction strengths are equal to a common value $\mu .$ In this case, $\G$ has a simple form
\ben
\G=h_{max}(c_{ij};\;i,j=0,1,2)\mu.\nn
\een
Therefore, we maximize $\G$ assuming the isotropic interactions. The result is
\ben
\G^{max}\approx 3.90495\mu  \nn
\een
 and the corresponding (normalized) state is given by $$c_{00}=-0.28317+i0.148948;\;c_{01}=-0.433055+i0.382479$$
$$c_{02}=-0.117778+i0.274948;\;c_{10}=0.0625717-i0.144584$$ $$c_{11}=0.102783-i0.0787094;\;c_{12}=-0.340939-i0.324717 $$ $$c_{20}=0.25066-i0.167261;\;c_{21}=0.0344755-i0.244282$$
$$c_{22}=0.227159-i0.088347.$$
After converting this state to the Schmidt canonical form we get the state giving the maximal entanglement generating rate for two qutrits, under a general isotropic interaction, as
\ben
|\psi_{max}\ran=0.884297|00\ran+0.448838|11\ran+0.128697|22\ran. \nn
\een
We find that $E(|\psi_{max}\ran)=0.677882.$ This shows that, in order to increase the entanglement of a two qutrit system in an optimal way, it is better to start with an initially entangled state rather than a product state, at least when all the interaction strengths in $H_{I}$ (Eq. (\ref{en29}))
are equal. We also note that the optimal entanglement $E(|\psi_{max}\ran)$ is independent of $H_{I},$ provided, again, that all interaction strengths in $H_{I}$ (Eq. (\ref{en29})) are equal. \\

\section{ The three qubit case}

We now deal with the problem of entanglement generation capacity for three qubits. We emphasize that this is the capacity to generate genuine three qubit entanglement and not the bipartite entanglement between any two parts of the three qubit system. In this case also, the entanglement measure given by Eq. (\ref{en7}) can be used as this entanglement measure applies to $N$-qubit pure states and has all the essential (and many desirable, eg superadditivity and continuity) properties expected of a good entanglement measure \cite{hj08}. For the three qubit case, $\tau$ in Eq. (\ref{en7}) is the three qubit correlation tensor appearing in the Bloch representation of the state. Here $\tau$ is a three way array while for two qubits $\tau$ was a matrix. The Bloch representation of a general three qubit pure state is
{\setlength\arraycolsep{2pt}
\begin{eqnarray}
\rho & = & |\psi\ran\lan\psi|=\frac{1}{8}\big(I\otimes I\otimes I + \sum_{l}r_l  \si_l\otimes I\otimes I   {}  \nn \\
& & {} + \sum_{n}s_n  I\otimes \si_n\otimes I+ \sum_{m}q_m  I\otimes I\otimes \si_m  {}  \nn \\
& & {} +\sum_{ln}t_{ln}^{(AB)}\si_l\otimes \si_n \otimes I + \sum_{lm}t_{lm}^{(AC)}\si_l\otimes I\otimes \si_m   {} \nn \\
 & & {}   +\sum_{nm}t_{nm}^{(BC)}I\otimes \si_n\otimes \si_m  +\sum_{lmn}\tau_{lmn}\si_l\otimes\si_n\otimes \si_m \big).  \label{en15}
\end{eqnarray}}
Here $\tau=[\tau_{ijk}]$ is a three way array while $t^{(\cdot\cdot)}=[t^{(\cdot\cdot)}_{ij}]$ are matrices. The definitions of various symbols in $\rho$ are as follows.
\begin{eqnarray}
r_l & = & Tr(\si_{l}^{A}\rho_{A})  =  Tr(\si_l^{A}\otimes I\otimes I \;\rho)\nonumber \\
s_n & = & Tr(\si_{n}^{B}\rho_{B})  =  Tr( I\otimes \si_n^{B}\otimes I \;\rho)\nonumber \\
q_m & = & Tr(\si_{m}^{C}\rho_{C})  = Tr(I\otimes I\otimes \si_m^{C}   \;\rho),  \nn
\end{eqnarray}
\begin{eqnarray}
t_{ln}^{AB} & = & Tr(\si_l^{A}\otimes\si_{n}^{B}\rho_{AB}) = Tr(\si_l^{A}\otimes\si_{n}^{B}\otimes I^{C}\rho) \nn \\
t_{lm}^{AC} & = & Tr(\si_l^{A}\otimes\si_{m}^{C}\rho_{AC}) = Tr(\si_l^{A}\otimes I^{B} \otimes\si_{m}^{C}\rho) \nn \\
t_{nm}^{BC} & = & Tr(\si_n^{B}\otimes\si_{m}^{C}\rho_{BC}) = Tr(I^{A}\otimes\si_n^{B}  \otimes\si_{m}^{C}\rho),  \nn
\end{eqnarray}
\begin{equation}
\tau_{lmn}=Tr(\si_l^{A}\otimes\si_n^{B}\otimes\si_m^{C}\rho), \nn
\end{equation}
where $\rho_{A,B,C}$ and $\rho_{AB,AC,BC}$ are the appropriate reduced density operators. We consider the general interaction between qubits which can be reduced by the singular value decomposition to the Hamiltonian
\begin{equation}
H_{I}=H_{AB}+H_{AC}+H_{BC}   \label{en19}
\end{equation}
where
\begin{eqnarray}
H_{AB} & = & \sum_{s=1}^{3}\mu_s^{AB}\si_s^{A}\otimes\si_{s}^{B}\otimes I^{C} \nn \\
H_{BC} & = & \sum_{s=1}^{3}\mu_s^{BC}I^{A}\otimes\si_s^{B}  \otimes\si_{s}^{C} \nn \\
H_{AC} & = & \sum_{s=1}^{3}\mu_s^{AC}\si_s^{A}\otimes I^{B} \otimes\si_{s}^{C}.  \label{en20}
\end{eqnarray}
It is helpful to imagine that the three spins are at the vortices of a triangle. If they are arranged on a line, we expect on physical grounds that one of the terms can be neglected in comparison with the other two, as it gives the next nearest neighbor interaction. For all subsystems we have $\mu_1\geq\mu_2\geq\mu_3.$
Using the definition of the entanglement generation rate $\G$ in Eq. (\ref{en3}) and the definition of the entanglement measure in Eq. (\ref{en7}) we get,
\begin{equation}
\G=\frac{1}{||\mathcal{T}||}\sum_{i,j,k=1}^{3}\tau_{ijk}\dot{\tau}_{ijk}. \nn
\end{equation}
Using the Heisenberg equation of motion,
\begin{equation}
i\frac{d\rho}{dt}=[H_{I},\rho], \nn
\end{equation}
where the Hamiltonian $H_{I}$ is defined via Eq.s (\ref{en19}), (\ref{en20}), we get
\begin{equation}
\dot{\tau}_{ijk}=iTr(H_{I}[\si_i\otimes\si_j\otimes\si_k, \rho]). \nn
\end{equation}
We now use the commutator identity
%{\setlength\arraycolsep{2pt}
%begin{eqnarray}
%\begin{displaymath}
%[A\otimes B\otimes C, D\otimes E\otimes F]=
%\begin{array}{ll}
%\frac{1}{4}([A,D]\otimes[B,E]\otimes[C,F]+[A,D]\otimes\{B,E\}\otimes\{C,F\}+ \\
%+\{A,D\}\otimes[B,E]\otimes\{C,F\}
%+\{A,D\}\otimes\{B,E\}\otimes[C,F]) \\  \label{en24}
%\end{array}
%\end{displaymath}
%\end{eqnarray}}
%\begin{eqnarray}
$$[A\otimes B\otimes C, D\otimes E\otimes F]= \frac{1}{4}([A,D]\otimes[B,E]\otimes[C,F]+[A,D]\otimes\{B,E\}\otimes\{C,F\}$$
 $$+\{A,D\}\otimes[B,E]\otimes\{C,F\}+\{A,D\}\otimes\{B,E\}\otimes[C,F])$$
%\end{eqnarray}
and the definitions of $\rho$ and $H_{I}$ in Eq.s (\ref{en15}) and (\ref{en19}) respectively to get,
{\setlength\arraycolsep{2pt}
\begin{eqnarray}
\dot{\tau}_{ijk}=-2\Big[\mu_k^{AC}\sum_{j^{\prime}=1}^{3}t_{j^{\prime}j}^{AB}\varepsilon_{ij^{\prime}k}+\mu_k^{BC}\sum_{k^{\prime}=1}^{3}t_{ik^{\prime}}^{AB}\varepsilon_{jk^{\prime}k}+ {} \nn \\
+\mu_j^{BC}\sum_{l^{\prime}=1}^{3}t_{il^{\prime}}^{AC}\varepsilon_{kl^{\prime}j}+
\mu_i^{AC}\sum_{l^{\prime}=1}^{3}t_{jl^{\prime}}^{BC}\varepsilon_{kl^{\prime}i}+ {} \nn \\
+\mu_i^{AB}\sum_{k^{\prime}=1}^{3}t_{k^{\prime}k}^{BC}\varepsilon_{jk^{\prime}i}
+\mu_j^{AB}\sum_{j^{\prime}=1}^{3}t_{j^{\prime}k}^{AC}\varepsilon_{ij^{\prime}j}\Big],  \label{en25}
\end{eqnarray}}
where $\varepsilon$ s are the Levi-Civita symbols. Substitution of Eq. (\ref{en25}) in the expression for $\G$ gives,
{\setlength\arraycolsep{2pt}
\begin{eqnarray}
\G=\frac{-2}{||\mathcal{T}||}\Big[\sum_{k,s=1}^{3}\left[(\vec{\tau}_{:sk}\times\vec{t}_{:k}^{AC})_s + (\vec{\tau}_{s:k}\times\vec{t}_{:k}^{BC})_s\right]\mu_s^{AB} + {} \nn \\
+\sum_{i,s=1}^{3}\left[(\vec{\tau}_{i:s}\times\vec{t}_{i:}^{AB})_s + (\vec{\tau}_{is:}\times\vec{t}_{i:}^{AC})_s\right]\mu_s^{BC} + {} \nn \\   +\sum_{j,s=1}^{3}\left[(\vec{\tau}_{:js}\times\vec{t}_{:j}^{AB})_s +
(\vec{\tau}_{sj:}\times\vec{t}_{j:}^{BC})_s\right]\mu_s^{AC}\Big],  \label{en26}
\end{eqnarray}}
where $\vec{\tau}_{:sk}=[\tau_{1sk},\tau_{2sk},\tau_{3sk}]^{T},$ for example, is a vector in $\mathbb{R}^3$ for fixed $s$ and $k.$ Similarly,
$\vec{t}_{:k}^{\cdot\cdot}$ and $\vec{t}_{j:}^{\cdot\cdot}$ are the $k$th column and the $j$th row vectors of the matrix $t^{(\cdot\cdot)}.$
The expression for the entanglement generation rate $\G$ for three qubits (Eq. (\ref{en26})) has 54 coupled terms, each term being a component of the cross product of two vectors. A geometric argument to maximize $\G ,$ as in the two qubit case, seems to be very difficult. However, it is quite straightforward to maximize $\G$ numerically, by writing the elements of the three way array $\mathcal{T}$ and the matrices $t^{(\cdot\cdot)}$ as the appropriate averages in the general three qubit state

\begin{equation}
|\psi\ran = \sum_{i=0}^{7}c_i|i\ran\;\;\;;\;\;\;\sum_i|c_i|^2=1  \label{en27}
\end{equation}
where $i$ labeling the product basis ket $|i\ran$ is the binary representation of the index $i.$
We can numerically optimize $\G$ for the general Hamiltonian in Eq. (\ref{en19}). However, for the general case, where the interaction is anisotropic, that is, the strengths of interaction $\mu_k^{(\cdot\cdot)}$ have different values, $\G$ does not have the simple structure $\G=f(p)h_{max}$ as in the two qubit case. Therefore, we assume isotropic interactions so that all interaction strengths are equal to a common value $\mu .$ In this case, after evaluating all the terms in Eq. (\ref{en26}) in the state $|\psi\ran$ given by Eq. (\ref{en27}), $\G$ can be written as
\begin{equation}
\G= h(c_0,\ldots,c_7)\mu. \nn
\end{equation}
After the numerical optimization of $\G$ as a function of $c_i, \; i=0,1,\ldots,7,$ we get,
\begin{equation}
\G^{max}= 5.72523 \mu.  \nn
\end{equation}
The (normalized) state corresponding to this $\G^{max}$ is given by
%{\setlength\arraycolsep{2pt}
%\begin{eqnarray}
$$|\psi_{max}\ran= (0.033768-i0.168758|000)\ran+(0.574022-i0.0709471)|001\ran+$$ $$+(0.0218412-i0.111565)|010\ran+
 (0.672021-i0.0754116)|011\ran+$$ $$+(-0.0603488+i0.172566)|100\ran
+(-0.0051137-i0.183831)|101\ran+ $$
$$+(0.0556843+i0.151888)|110\ran+(0.0700719-i0.259423)|111\ran .$$
%end{eqnarray}}
This state has the following Acin canonical form, expressed by the two fold degenerate sets of entanglement parameters \cite{aajt01}.
%{\setlength\arraycolsep{2pt}
%\begin{eqnarray}
$$|\psi^+\ran=0.610291|000\ran+0.67402\exp(i2.51395)|100\ran+0.394893|101\ran +$$
$$+0.110357|110\ran+0.0715772|111\ran ,$$
%\end{eqnarray}}
or,
%{\setlength\arraycolsep{2pt}
%\begin{eqnarray}
$$|\psi^-\ran=0.329873|000\ran+0.546087\exp(i0.402558)|100\ran+0.730583|101\ran+$$
$$+0.20417|110\ran+0.132424|111\ran .$$
%\end{eqnarray}}
We see that the state with the maximal entanglement generation rate $\G^{max}$ belongs to the GHZ class. Further, we find that $E(|\psi_{max}\ran)=0.258918.$ This means that, given the isotropic interaction, it is beneficial to start with an entangled three qubit state for optimal entanglement generation. Also, we note that the optimal entanglement is independent of $H_{I},$ provided the corresponding interaction is isotropic.

Thus we see that, for three qubits, the geometric method based on the entanglement measure given by Eq. (\ref{en7}) can be numerically implemented to get the state with maximal entanglement generation rate. This program can be carried out for the general interaction Hamiltonian in Eq. (\ref{en19}) although we have restricted to the isotropic interactions. This procedure can be suitably carried out in a laboratory using quantum circuits. Every quantum circuit acts unitarily on a quantum state and we can always find a Hamiltonian corresponding to such a circuit \cite{dnbt02,aajt01}. On the other hand, given a (interaction) Hamiltonian  for a three qubit system, we may construct a circuit implementing the corresponding evolution using universal quantum gates. We note that, for the isotropic interaction, the maximal entanglement generation rate $\G^{max}$ is proportional to the interaction strength $\mu$ and the corresponding state is independent of $\mu,$ as in the two qubit case. When the interactions are anisotropic, the scenario for two qubits does not apply to the three qubit case, as $\G$ does not factor into the product of a state dependent function and an expression involving only the interaction strengths. Even when the interactions are isotropic, we do not know the explicit form of such a state dependent function. In other words, we do not know whether it is possible to separately account for the contribution due to the state and that due to the interactions. Thus a general procedure for the maximization of the entanglement generation rate for the higher dimensional and multipartite systems still seems to be an open question. These observations ensue from the fact that the terms in the expression for $\G$ could not be decoupled. This difficulty seems to be generic, as it may be a consequence of the difficulties in the geometric interpretation of the Bloch space for the multipartite and higher dimensional systems \cite{kc01,kmsm97}.  All the  remarks in this paragraph apply to the two qutrit case as well. \\

\chapter{Multipartite entanglement in fermionic systems via a geometric measure}

\begin{center}
\scriptsize\textsc{Before I came here I was confused about this subject. \\ Having listened to your lecture I am still confused. But on a higher level.\\  { Enrico Fermi}}
\end{center}

\section{Introduction}

  In this chapter, We study multipartite entanglement in a system consisting of indistinguishable fermions. Specifically, we have proposed a geometric entanglement measure for $N$ spin-$\fr{1}{2}$ fermions distributed over $2L$ modes (single particle states). The measure is defined on the $2L$ qubit space isomorphic to the Fock space for $2L$ single particle states. This entanglement measure is defined for a given partition of $2L$ modes containing $m\geq 2$ subsets. Thus this measure applies to $m\leq 2L$ partite fermionic system where $L$ is any finite number, giving the number of sites. The Hilbert spaces associated with these subsets may have different dimensions.  Further, we have defined the local quantum operations with respect to a given partition of modes. This definition is generic and unifies different ways of dividing a fermionic system into subsystems. We have shown, using a representative case, that the geometric measure is invariant under local unitaries corresponding to a given partition. We explicitly demonstrate the use of the measure to calculate multipartite entanglement in some correlated electron systems.

We study multipartite entanglement in a system consisting of identical particles. We use the idea due to Zanardi \cite{z02} whereby the Fock space of a system of fermions is mapped to the isomorphic qubit or `mode' space. We then discuss entanglement in this `mode' space via a geometric measure. The idea is to use the Bloch representation of the state of the $m$-partite quantum system \cite{kk05}. The measure is defined by the Euclidean norm of the $m$-partite correlation tensor in the Bloch representation. This correlation tensor contains all information of genuine $m$-partite entanglement (see section \ref{sec:basic}). Such a measure was proposed earlier by P. S. Joag and A. S. M. Hassan for $N$ qubit and $N$ qudit pure states\cite{hj08,hj09} and shown to satisfy most of the properties expected of a good measure. The Bloch representation of a quantum state has a natural geometric interpretation, which is why we call this measure a geometric measure \cite{kk05}. Other geometric measures are based on the distance of the given state from the set of separable states in the Hilbert space \cite{vprk97,wg03}. An important question in the context of quantum entanglement is that of locality. For indistinguishable particles distributed over `modes' (which are taken to be single particle states of particles constituting the system), local operations have meaning only in the context of partitions over modes. A quantum operation confined to a single subset in a given partition is then a local operation. We therefore define entanglement in such a system with respect to partitions and require it to be invariant under local unitaries defined with respect to a given partition. We explicitly demonstrate the use of the measure to calculate entanglement in some correlated electron systems.

 The outline of the chapter is as follows: In Sec.\ref{sec:basic},  we begin by briefly reviewing some details about the isomorphism between the Fock space of a system of indistinguishable particles and the `mode' space \cite{z02}. This will help us in defining various quantities and also set up notation necessary for the subsequent analysis. In  Sec.\ref{sec:g-measure}, we define and construct the geometric measure. Various properties of the measure are also discussed with reference to a specific example in Sec.\ref{sec:fourmode}. In Sec.\ref{sec:appl}, we use the measure to study entanglement in the Hubbard dimer and trimer.

\section{\label{sec:basic}Mapping between Fock space and qudit space}

We deal with $N$ spin-$\fr{1}{2}$ fermions on a $L$ site lattice. The total number of available (localized) single particle states are then $2L$ in number. The fermionic Fock space in the occupation number representation has basis states of the form $|n_1n_2\ldots n_{2L}\ran\;\;(n_i=0,1\;;\;i=1,\ldots ,2L)$. We further assume that the total number of particles is conserved. This means that we only deal with subspaces of the Fock space corresponding to a fixed eigenvalue for the total number operator. We shall refer to this number super-selection rule as N-SSR. For a $N$-fermion system, we call such a subspace of the Fock space `$N$-sector' and denote it by $F_N$. The $N$-sector of a $2L$ mode system is the subspace $F_N$ of dimension $\binom{2L}{N}$ of the Fock space with the dimension of the Fock space for $2L$ single particle states $(N=0,1,2,\ldots ,2L)$ being
\ben \label{e1}
\sum_{N=0}^{2L}\binom{2L}{N}\;=\;2^{2L}
\een
Since a $2L$ qubit Hilbert space $(\mathbb{C}^{2})^{\otimes 2L}$ has exactly this dimension, it is possible to construct an isomorphism between the Fock space and the $2L$ qubit Hilbert space $(\mathbb{C}^{2})^{\otimes 2L}$ \cite{h87}. The particular isomorphism we implement is
\ben  \label{e2}
|n_1n_2\ldots n_{2L}\ran\ra |n_1\ran\otimes |n_2\ran\otimes \cdots \otimes |n_{2L}\ran \;;\;n_i=0,1\;;\;i=1,2,\ldots ,2L
\een
where, in qubit space we associate $|0\ran\lr |\up\ran$ and $|1\ran\lr |\dn\ran .$ Note that the Slater rank of the Fock basis states $|n_1n_2\ldots n_{2L}\ran $  is $1$ so that these are separable states. Thus the above isomorphism maps separable basis states in Fock space to the separable basis states in qubit space. Further, the subspace structure of the Fock space namely,
\ben  \label{e3}
F_{2L}=\bigoplus _{N=0}^{2L}F_{N}
\een
is carried over to the qubit space under mapping (Eq. (\r{e2})) because each subspace of the Fock space with conserved fermion number $N$ is mapped to a subspace of the $2L$ qubit space spanned by the basis vectors with $N$ ones and $2L-N$ zeros. We can write
\ben  \label{e4}
H_{2L}=(\mathbb{C}^{2})^{\otimes 2L}=\bigoplus _{N=0}^{2L}H_{2L}(N)
\een
where $H_{2L}(N)$ is the image of $F_{N}$ in Eq. (\r{e3}) under the map given by Eq. (\r{e2}).

Next crucial step is to transfer the action of the creation and annihilation operators on Fock space to the qubit space, under the isomorphism given by Eq. (\r{e2}) \cite{cp05}. We need the creation and annihilation operators $a$ and $a^{\dg}$ acting on a single qubit state,
\benr \label{e5}
a|0\ran  =  0,\;\;a|1\ran  =  |0\ran \;\;\ && \nonumber \\
a^{\dg}|0\ran  =  |1\ran,\;\;a^{\dg}|1\ran  =  0 \nonumber  \\
\eenr
such that,
\benr  \label{e6}
a_i & \ra & I\otimes I\otimes\cdots\otimes \underbrace{a}_{i th\; qubit}\otimes \cdots\otimes I  \;\; \nonumber   \\
a_{i}^{\dg} & \ra & I\otimes\cdots\otimes\underbrace{a^{\dg}}_{i th\; qubit}\otimes \cdots\otimes I  \nonumber   \\
\eenr
Here $a_i$ ($a_i^{\dg}$) is the annihilation (creation) operator acting on Fock space $F_{2L},$ annihilating (creating) a fermion in $i$th mode. $I$ is the identity on single qubit space. The tensor product satisfying the correspondence in Eq. (\r{e6}) must be consistent with the anti-commutation property of the Fock space creation and annihilation operators,
\ben  \label{e7}
\{a_i,a_j^{\dg}\}=\del_{ij}\;\;\{a_i,a_j\}=0=\{a_i^{\dg},a_j^{\dg}\}
\een
This requirement leads to the following action of the tensor product operators on the $2L$ qubit states
\benr \label{e8}
(I\otimes I\otimes\cdots\otimes \underbrace{a(a^{\dg})}_{ith\;place}\otimes \cdots\otimes I)(|n_1\ran\otimes \cdots \otimes|n_i\ran \otimes \cdots \otimes|n_{2L}\ran)     \nonumber  \\
=(-1)^{\sum_{j=i+1}^{2L}n_j} (|n_1\ran\otimes  \cdots \otimes\underbrace{a(a^{\dg})|n_i\ran}_{ith\; qubit} \otimes\cdots\otimes|n_{2L}\ran)   \nonumber
 \\
\eenr
Here $n_i\in\{0,1\}\;;\;i\in\{1,2,\ldots ,2L\}$ and $\sum_{j=i+1}^{2L}n_j$ is evaluated$\mod{2}.$ Using Eq. (\r{e8}), it is straightforward to see that

\
\\

$$\{I\otimes I\otimes\cdots\otimes \underbrace{a(a^{\dg})}_{ith\;place}\otimes \cdots\otimes I\;,\; I\otimes I\otimes\cdots\otimes \underbrace{a(a^{\dg})}_{jth\;place}\otimes \cdots\otimes I\}$$
\ben  \label{e9}
(|n_1\ran\otimes \cdots \otimes|n_i\ran \otimes \cdots \otimes|n_{j}\ran\otimes \cdots \otimes|n_{2L}\ran)=0
\een
and
\benr  \label{e10}
\{I\otimes I\otimes\cdots\otimes \underbrace{a}_{ith\;place}\otimes \cdots\otimes I\;,\; I\otimes I\otimes\cdots\otimes \underbrace{a^{\dg}}_{jth\;place}\otimes \cdots\otimes I\}&&  \nonumber   \\
(|n_1\ran\otimes \cdots \otimes|n_i\ran \otimes \cdots \otimes|n_{j}\ran\otimes \cdots \otimes
|n_{2L}\ran) =  \underbrace{(I\otimes \cdots \otimes I)}_{2L\;factors}\del_{ij}   \nonumber   \\
\eenr
We note that the phase factors appearing in Eq. (\r{e8}) are the consequence of the conservation of the parity operator \cite{bcw07}
\ben \label{e11}
\hat{P}=\Pi_{i=1}^{2L}(1-2a_i^{\dg}a_i).
\een

 Henceforth, in this chapter, by `fermions' we mean spin-$\fr{1}{2}$ fermions. Further, we call a single particle state a mode. Thus two spin-$\fr{1}{2}$ fermions on two sites is a four mode system. In general, $N$ spin-$\fr{1}{2}$ fermions on $L$ sites is equivalent to $N$ fermions on $K=2L$ modes. For example, two spin-$\fr{1}{2}$ fermions on a two site lattice, $A,B$ say, generate four single particle states or modes $|A\up\ran,|A\dn\ran,|B\up\ran,|B\dn\ran$. In this work, we deal with entanglement between subsets forming a partition of a $2L$-mode fermionic system. We define the entanglement measure for any such partition of a $2L$-mode system without any restriction on the number and the size of the subsets forming the partition. These subsystems may involve different degrees of freedom, for example, we can deal with the entanglement between spins and sites or entanglement between two spins on the same site (intrasite entanglement). Or if each of the subsets partitioning the $2L$ modes comprises modes with common site label we have the entanglement between sites or the `site entanglement'. Thus all physically realizable subsystems of $N$ fermions over $2L$ single particle states can be addressed by dividing the $2L$ modes into suitable partitions, for example the `particle entanglement' defined in \cite{wv03}.

 We now define the local and non-local operations on the $2L$ mode fermionic system \cite{z02,gf01,wv03,zhs08,p10}. We do this by using the corresponding operations on the isomorphic qubit space $H_{2L}.$ We note that, due to isomorphism between $F_{2L}$ (Eq. (\r{e2})) and $H_{2L}$ (Eq. (\r{e4})), partitioning $2L$ modes is equivalent to the corresponding partitioning of the $2L$ qubit system into subsystems. Locality is defined with respect to the partition of $2L$ qubits (or, the corresponding partition of $2L$ modes) between whose subsets we are seeking entanglement. The operations on the state space of a single subset in a partition of $2L$ qubits is taken to be local. The entanglement measure defined with respect to a partition must be invariant under a unitary operation which is local with respect to that partition. We will illustrate this point later, using the geometric entanglement measure defined below. Henceforth `mode' and `qubit' are taken to be synonymous and we shall use the expression `modes' instead of `qubits'. In other words, the spaces $F_{2L}$ and $H_{2L}$ are taken to be the same.

\section{\label{sec:g-measure}Geometric measure for entanglement.}
\subsection{\label{sec:def}Definition}

   We define a geometric measure for the partitions of the $2L$ mode $N$ fermion systems in pure states. Although the definition of the measure is quite general, the fermion number super-selection rule restricts the pure states to the appropriate subspace corresponding to $N$ fermions, namely $H_{2L}(N)$ (Eq. (\r{e4})). Thus a state $|0110\ran \in H_{4}(2)$ may be partitioned as $|01\ran\otimes|10\ran$ or as $|011\ran\otimes|0\ran$ etc where the definition of the tensor product is consistent with Eq. (\r{e8}). We use the Bloch representation of $N-$partite states (drawn from $H_{2L}(N)$) to get this measure \cite{hj08,hj09}.

 First we assume that a partition equally divides $2L$ modes into subsets, i.e. all subsets in the partition contain equal number of modes, say $n.$ This corresponds to the case of $H_{2L}$ divided into subspaces of dimension $d=2^n ,\;n$ being some divisor of the number of modes $2L.$ To get the entanglement measure we expand the state $\rho=|\p\ran\lan\p|$ of the system, supported in the appropriate $H_{2L}(N),$ in its Bloch representation.

  In order to give the Bloch representation of a density operator acting on the Hilbert space $\mathbb{C}^{d} \otimes \mathbb{C}^{d} \otimes \cdots \otimes \mathbb{C}^{d}$ of an $m=2L/n$-qudit quantum system, we introduce following notation \cite{hj09}. We use $k$, $k_i \; (i=1,2,\ldots)$ to denote a qudit chosen from $m$ qudits, so that $k$, $k_i \; (i=1,2,\ldots)$ take values in the set  $\mathcal{N}=\{1,2,\ldots,m\}$. The variables $\alpha_k \;\mbox{or} \; \alpha_{k_i}$ for a given $k$ or $k_i$ span the set of generators of $SU(d)$ group  for the $k$th or $k_i$th qudit, namely the set $\{\la_1,\la_2,\cdots,\la_{{d}^2-1}\}$ for the $k_i$th qudit. For two qudits $k_1$ and $k_2$ we define

   $$\lambda^{(k_1)}_{\alpha_{k_1}}=(I_{d}\otimes I_{d}\otimes \dots \otimes \lambda_{\alpha_{k_1}}\otimes I_{d}\otimes \dots \otimes I_{d})   $$
   $$\lambda^{(k_2)}_{\alpha_{k_2}}=(I_{d}\otimes I_{d}\otimes \dots \otimes \lambda_{\alpha_{k_2}}\otimes I_{d}\otimes \dots \otimes I_{d})  $$
   $$\lambda^{(k_1)}_{\alpha_{k_1}} \lambda^{(k_2)}_{\alpha_{k_2}}=(I_{d}\otimes I_{d}\otimes \dots \otimes \lambda_{\alpha_{k_1}}\otimes I_{d}\otimes \dots \otimes \lambda_{\alpha_{k_2}}\otimes I_{d}\otimes I_{d})   $$
where  $\lambda_{\alpha_{k_1}}$ and $\lambda_{\alpha_{k_2}}$ occur at the $k_1$th and $k_2$th places (corresponding to $k_1$th and $k_2$th qudits respectively) in the tensor product and are the $\alpha_{k_1}$th and  $\alpha_{k_2}$th generators of $SU(d),\; \alpha_{k_{1,2}}=1,2,\dots,d^2-1\;$. Then we can write

$$\rho=\fr{1}{d^N} \{ I_{d}^{\otimes^m}+ \sum_{k \in \mathcal{N}}\sum_{\alpha_{k}}s_{\alpha_{k}}\lambda^{(k)}_{\alpha_{k}} +\sum_{\{k_1,k_2\}}\sum_{\alpha_{k_1}\alpha_{k_2}}t_{\alpha_{k_1}\alpha_{k_2}}\lambda^{(k_1)}_{\alpha_{k_1}} \lambda^{(k_2)}_{\alpha_{k_2}}+\cdots +$$
$$\sum_{\{k_1,k_2,\cdots,k_M\}}\sum_{\alpha_{k_1}\alpha_{k_2}\cdots \alpha_{k_M}}t_{\alpha_{k_1}\alpha_{k_2}\cdots \alpha_{k_M}}\lambda^{(k_1)}_{\alpha_{k_1}} \lambda^{(k_2)}_{\alpha_{k_2}}\cdots \lambda^{(k_M)}_{\alpha_{k_M}}+ \cdots $$

\ben \label{eq:erho}
+\sum_{\alpha_{1}\alpha_{2}\cdots \alpha_{N}}t_{\alpha_{1}\alpha_{2}\cdots \alpha_{N}}\lambda^{(1)}_{\alpha_{1}} \lambda^{(2)}_{\alpha_{2}}\cdots \lambda^{(m)}_{\alpha_{N}}\}
\een
where $\textbf{s}^{(k)}$ is a Bloch vector corresponding to $k$th qudit, $\textbf{s}^{(k)} =[s_{\alpha_{k}}]_{\alpha_{k}=1}^{d^2-1} $ which is a tensor of order one defined by
 $$s_{\alpha_{k}}=\fr{d}{2} Tr[\rho \lambda^{(k)}_{\alpha_{k}}]= \fr{d}{2} Tr[\rho_k \lambda_{\alpha_{k}}],$$ where $\rho_k$ is the reduced density matrix \cite{nc00} for the $k$th qudit. Here $$\{k_1,k_2,\ldots,k_M\},\; 2 \le M \le m,$$ is a subset of $\mathcal{N}$ and can be chosen in $\binom{m}{M}$  ways, contributing $\binom{m}{M}$ terms in the sum $\sum_{\{k_1,k_2,\cdots,k_M\}}$ in Eq. (\r{eq:erho}), each containing a tensor of order $M$. The total number of terms in the Bloch representation of $\rho$ is $2^m$. We denote the tensors occurring in the sum $\sum_{\{k_1,k_2,\cdots,k_M\}},\; (2 \le M \le m)$ by $\mathcal{T}^{\{k_1,k_2,\cdots,k_M\}}=[t_{\alpha_{k_1}\alpha_{k_2}\cdots \alpha_{k_M}}]$ which  are defined by

 $$t_{\alpha_{k_1}\alpha_{k_2}\dots\alpha_{k_M}}=\fr{d^M}{2^M} Tr[\rho \lambda^{(k_1)}_{\alpha_{k_1}} \lambda^{(k_2)}_{\alpha_{k_2}}\cdots \lambda^{(k_M)}_{\alpha_{k_M}}]$$

$$ =\fr{d^M}{2^M} Tr[\rho_{k_1k_2\dots k_M} (\lambda_{\alpha_{k_1}}\otimes\lambda_{\alpha_{k_2}}\otimes\dots \otimes\lambda_{\alpha_{k_M}})]   $$
where $\rho_{k_1k_2\dots k_M}$ is the reduced density matrix for the subsystem $\{k_1, k_2, \dots \\ ,k_M\}$. Each of the  $\binom{m}{M}$ tensors of order $M$, occurring in the Bloch representation of $\rho$, contains all information about entanglement of the corresponding set of $M$ subsystems. All information on the entanglement contained in $\rho$ is coded in the tensors occurring in the Bloch representation of $\rho$. The tensor in last term in Eq. (\r{eq:erho}), we call it $\mathcal{T}^{(m)}$, contains all the  information of genuine $m$-partite entanglement. This follows from the observation that all other terms in the Bloch representation of $\rho$ (Eq. (\r{eq:erho})) correspond to subsystems comprising  $M<m$ qudits and the density operator contains all possible information about the state of the system. \\

The operators $\la_{\al_{k}}, \al_{k}= 1,2,\ldots,d^2-1$ are given by \cite{mw95}

\ben
\mf{\hat{\la}}=\{\hat{u}_{12},\hat{u}_{13}, \hat{u}_{23},\ldots , \hat{v}_{12},\hat{v}_{13}, \hat{v}_{23},\ldots ,\hat{w}_1,\hat{w}_1,\ldots ,\hat{w}_{d-1}\}
\een
with
\ben
\hat{u}_{jk}=\hat{P}_{jk}+\hat{P}_{kj}   \nonumber  \\
\een
\ben
\hat{v}_{jk}=-i(\hat{P}_{jk}-\hat{P}_{kj})   \nonumber  \\
\een
\benr
w_{l}=\sqrt{\fr{2}{l(l+1)}}(\hat{P}_{11}+\cdots+\hat{P}_{ll}-l\hat{P}_{l+1,l+1})&&     \nonumber  \\
1\leq j < k\leq d \;\; ; \;\; 1\leq l\leq d-1    \nonumber    \\
\eenr
where
\ben
\hat{P}_{kl}=|k\ran\lan l|\;\;\;(k,l=1,2,\ldots ,d).      \nonumber  \\
\een
Note that each of the generators of the $SU(d)$ group $\la_i\;,\;i=1,2,\ldots ,d^2-1$ acts on a single qudit space and hence is local \ref{sec:basic}, apart from the phase factor contributed by their action, as given by Eq. (\r{e8}). We assume these phase factors to be absorbed in the coefficients in the expansion of the density operator $\rho.$

Let a $2L$ mode $N$ fermion system be partitioned by $m=2L/n$ subsets, each containing n modes. Then for this partition, we define the entanglement measure for a state $|\p\ran\in H_{2L}(N)$ by \cite{hj09}
\ben  \label{e12}
E=||\t||-||\t||_{sep}
\een
where
 \ben   \label{e13}
 ||\t||=\sqrt{\sum_{\al_1\cdots\al_m =1}^{d^2-1}t_{\al_1\cdots\al_m}^{2}}
 \een
and $||\t||_{sep}$ is $||\t||$ for separable (product) $m$ qudit state

\ben  \label{e14}
||\t||_{sep}=\left(\fr{d(d-1)}{2}\right)^{m/2}
\een

\subsection{\label{sec:unequal} Entanglement in partitions with unequal subsets}

  We can also generalize the definition of the entanglement to the case where the corresponding qubit subsystems have unequal dimensions. We discuss the simplest case of bi-partite entanglement with partitions having unequal dimensions, say $d_1$ and $d_2$. In this case, the definition of the geometric entanglement measure generalizes to
\ben \label{37}
E=||\t||-||\t||_{sep}
\een
 where
 \ben   \label{e38}
 ||\t||=\sqrt{\sum_{i=1}^{d_1^2-1}\sum_{j=1}^{d_2^2-1}t_{ij}^{2}}
 \een
 where
 \ben \label{e39}
 t_{ij}=\left(\fr{d_1}{2}\right)\left(\fr{d_2}{2}\right)\lan\p|\h{\la}_i\ot\h{\la}_j |\p\ran=\left(\fr{d_1}{2}\right)\left(\fr{d_2}{2}\right)K_{ij}.
 \een
 Here $\h{\la}_i \;(i=1,\ldots ,d_1^2-1)$ and $\h{\la}_j \; (j=1,\ldots ,d_2^2-1)$ are the generators of $SU(d_1)$ and $SU(d_2)$ respectively. $||\t_{sep}||$ is given by
\ben \label{e40}
||\t_{sep}||^2 = ||\mf{s}^{(1)}||.||\mf{s}^{(2)}|| = \left(\fr{d_1(d_1-1)}{2}\right)\left(\fr{d_2(d_2-1)}{2}\right).
\een
Here $||\mf{s}^{(1)}||$ and $||\mf{s}^{(2)}||$ are the norms of the Bloch vectors of the reduced density operators for each subsystem.

It is straightforward to extend these definitions to partitions containing more than two subsets.

\section{\label{sec:fourmode}Entanglement in a four mode system}

With the entanglement measure defined as above, we give an example wherein we can compute the entanglement for different partitions and  illustrate our comments on local and non-local operations. We also compute the upper bounds on the entanglement.

\subsection{\label{sec:l-nlocal}Local and Non-local operations}
 Consider a four mode system and the normalized state $|\p\ran\in H_4(2)$ defined as
\ben  \label{e15}
|\p\ran=\fr{1}{\sqrt{6}}\{ i\al |1100\ran + |1001\ran + |0110\ran + |0011\ran + \b |0101\ran + |1010\ran \}\;\;\al^2 + \b^2=2
\een
where $\al,$ $\b$ are real. Note that $|\p\ran$ can be treated as a member of the Fock space $F_4 (2)$ with the kets appearing in it being its basis states. Consider the evolution of the system in state $|\p\ran \in F_4(2)$ via the Hamiltonian
\ben  \label{e16}
H = f(a_1^{\dg}a_4+a_4^{\dg}a_1)+q\h{n}_1\h{n}_2 + \G \h{n}_1 + \g \h{n}_3 + \eta(a_1^{\dg}a_2+a_2^{\dg}a_1)
\een
acting on $F_4 .$ Here $f$ term is the interaction between two modes on different sites (intersite interaction), $\eta$ term is the interaction between two modes on the same site (intrasite interaction). $\G$ and $\g$ correspond to single mode on site $A$ and $B$ respectively. q term involves number operators $\h{n}_i = a_i^{\dg}a_i\;;\;i=1,2$ for first two modes, on $A$ site. We have included all the different kinds of typical interactions encountered in condensed matter systems, respecting number super-selection rule. After an infinitesimal unitary evolution via this Hamiltonian, the state $|\p\ran$ evolves to
\ben   \label{f1}
|\p^{\pr}\ran=|\p\ran-i\ep H|\p\ran
\een
By employing the mapping of annihilation and creation operators in Eq. (\r{e6}) and Eq. (\r{e8}) and that of Fock space basis states in Eq. (\r{e2}), we get, for $|\p^{\pr}\ran$

$$|\p^{\pr}\ran = \fr{1}{\sqrt{6}}\{(i\al+i\ep f+\al q\ep)|1100\ran +(1-i\G\ep -i\ep\eta\b)|1001\ran $$
$$+(1-i\g\ep -i\ep\eta)|0110\ran + (1-i\ep f -i\ep\g)|0011\ran  $$
\begin{equation}{\label{e17}}
+(\b-\ep f\b-i\ep\eta)|0101\ran + (1+i\ep f-i\ep\G-i\ep\g - i\ep\eta)|1010\ran \}
\end{equation}

Now we find the entanglement for different partitions of this four mode system, using the geometric entanglement measure, Eq. (\r{e12}). We first partition four modes into four subsets, each containing one mode. This case gives genuine entanglement between four modes, which is more general than only the bipartite entanglement considered in the literature. For this case $d=2,$ so that $||\t||_{sep}=1$ and we get, for the genuine four mode entanglement,
 \ben  \label{e18}
 E=||\t||-1
 \een
 where
 \ben   \label{e19}
 ||\t||=\sqrt{\sum_{i,j,k, l=1}^{3}t_{ijkl}^2}
 \een
 with
 \ben  \label{e20}
 t_{ijkl}=Tr[\rho\;\si_i\ot\si_j\ot\si_k\ot\si_l]=\lan\p|\si_i\ot\si_j\ot\si_k\ot\si_l|\p\ran .
 \een
 where $\{\si_i\}\;i=0,1,2,3$ are the generators of the $SU(2)$ group (Pauli operators). The resulting entanglement in $|\p^{\pr}\ran$ is
\begin{flushleft}
$E_g(|\p^{\pr}\ran)=$
\end{flushleft}
\begin{center}
$\fr{1}{6}\left(-6+\sq{88+64\al^2+32\b+64\b^2+10\al^2\b^2+\b^4}\right)-$
\end{center}
 \ben \label{e21}
\fr{4\left(4f\al-2q\al(1+\b)+f\al\b(\al^2-\b^2)+4\al\eta(1+\b)\right)\ep}{\left(-6+\sq{88+64\al^2+32\b+64\b^2+10\al^2\b^2+\b^4}\right)}+O[\ep^2]
 \een

where the first term gives the entanglement $E(|\p\ran)$ for the state $\psi$ as defined in Eq. (\r{e15}). For this partition, the operations on a single mode are the only local operations, while all others are non-local. Therefore, the terms $\G \h{n}_1$ and $\g \h{n}_3$ are the only local interactions. Therefore, we expect that the four mode genuine entanglement should not depend on $\G$ or $\g$ to the first order in $\ep,$ which is the case, as seen from Eq. (\r{e17}).

   Next, we consider the partition consisting of two subsets, each containing two modes on each site, $\{A\up ,A\dn\}$ and $\{B\up ,B\dn\}$ (site partition). Thus we have two subsystems with $d=4$ corresponding to a $SU(4)\ot SU(4)$ qudit system. Further, $n=2$ giving $m=(2L/n)=2$ so that the geometric entanglement is
 \ben \label{e22}
 E_s(|\p^{\pr}\ran)=||\t||-6
 \een
 where
 \ben
 ||\t||=4\sq{\sum_{j,k=1}^{15}K_{jk}^{2}}  \nonumber  \\
 \een
 with
 \ben
 K_{jk}=\lan\p|\h{\la}_j\ot\h{\la}_k|\p\ran  \nonumber  \\
 \een
where $\h{\la}_j \;;\;j=1,\ldots ,15$ are the generators of $SU(4).$ The entanglement of $|\p^{\pr}\ran$ in Eq. (\r{e17}) is then given by
\begin{flushleft}
$E_s(|\p^{\pr}\ran)=\fr{1}{3}\left(-18+\sq{208+136\al^2+9\al^4-32\b+104\b^2+34\al^2\b^2+9\b^4}\right)$
\end{flushleft}
 \ben \label{e23}
-\fr{16(-f\al+f\al\b(\al^2-\b^2-2))\ep}{3(\sq{208+136\al^2+9\al^4-32\b+104\b^2+34\al^2\b^2+9\b^4})}+O[\ep^2]
 \een

According to the `site partition', in addition to the operations on single modes, the operations on the pair of modes having the same site label are also local. Therefore, the resulting entanglement cannot change under the intrasite operations in the Hamiltonian, namely the $q$ term, the $\eta$ term and as before, $\G$ and $\g$ terms. Thus, to the first order in $\ep ,$ the entanglement is expected to depend only on the non-local part of the Hamiltonian, that is, on the $f$ parameter. From Eq. (\r{e23}) we see that this is the case. The intersite entanglement quantified using Von-Neumann entropy has been reported earlier\cite{z02}.

Thus we see that the geometric measure has the capability to quantify the genuine multi-mode fermionic entanglement as against the mainly bipartite entanglement reported in the literature. Also, the geometric entanglement measure, for the given partition of modes, shows the correct behavior under local and non-local unitary operations.

\subsection{\label{sec:bound}Upper bound for the geometric measure in a four mode system.}

We discuss here some upper bounds that one can find for the entanglement for the four mode system and compare with existing results
obtained from other measures. We find that the general state which leads to a maximum intersite entanglement computed via the geometric measure also leads to a maximum for the von-Neumann entropy.

We treat the entanglement for the given partition to be the function of the coefficients of the general state $|\p\ran\in H_4(2),$ namely,
\ben \label{e24}
|\p\ran=\sum_{k=3,5,6,9,10,12}c_k |k\ran
\een
where each ket is labeled by the (four bit) binary representation of $k.$ We then maximize $E(|\p\ran)$ with respect to the coefficients $c_k $.

 For the site partition (see above) the entanglement (Eq. (\r{e22}) and two equations following it) is a function of the coefficients $c_k$ in state
$|\p\ran\in H_4(2)$ given in Eq. (\r{e24}). We find that the the maximum value of the entanglement is given by
\ben
E_{max}=1.74593
\een
The Schmidt canonical form (with respect to the $H_{4}(2)$ basis) of $|\p_{max}\ran$ corresponding to this $E_{max}$ is
\ben \label{e27}
 |\p_{max}\ran=\fr{1}{2}\left(\sum_{k=3,5,10,12}|k\ran\right)
 \een
 or,
 \ben  \label{e28}
 |\p_{max}\ran=\fr{1}{2}\left(\sum_{k=3,6,9,12}|k\ran\right)
 \een
 where each ket is labeled by the four bit binary representation of $k.$ Also, both these forms lead to the von-Neumann entropy $=2$ which is the maximum  possible\cite{gmw02}.

For the four mode entanglement (Eq. (\r{e18})) we find that entanglement is maximum for the state given by
\begin{eqnarray}
&&c_3=0.06701-0.33751i;\;\; c_5=0.16442-0.51281i    \nn   \\
&&c_6=-0.1006+0.2854i;\;\; c_9=0.29967-0.04206i     \nn   \\
&&c_{10}=-0.53522+0.05955i;\;\;c_{12}=-0.34410-0.00118i   \nn   \\
\end{eqnarray}
and the maximum value is found to be
\ben
E_{max}=2
\een
However, we do not have any entanglement measure to compare with the geometric measure. We also note that, for four modes, no canonical form like the Schmidt or Acin canonical form (for two and three modes respectively) is available.

\section{\label{sec:appl}Applications}
We now consider some correlated lattice models and discuss multi-mode entanglement in these models using the geometric measure.

\subsection{\label{sec:Hdimer}Hubbard dimer.}

 The Hubbard dimer model is a simple model for a number of physical systems, including the electrons in a $H_2$ molecule, double quantum dots, etc\cite{z02,wv03,hdimer}. The Hamiltonian can be written as
\ben \label{e29}
H\;=\;-t\sum_{\si=\up,\dn}\left(c^{\dg}_{A\si}c_{B\si}+c^{\dg}_{B\si}c_{A\si}\right)+U\sum_{j=A,B}\h{n}_{j\up}\h{n}_{j\dn}
\een
where $A,$ $B$ are the site labels and $\up$ and $\dn$ are spin labels. $t$ is the hopping coefficient measuring hopping between two sites while conserving spin and $U$ quantifies Coulomb interaction between fermions on the same site. By varying $\left(\fr{U}{4t}\right)$ we can vary the relative contributions of hopping and Coulomb mechanisms.

The ground state of the system at zero temperature can be easily obtained as \cite{z02,wv03}
\ben \label{e30}
|\p_0\ran=N\h{G}_0 |vac\ran
\een
where $N=\lan \p_0|\p_0\ran^{-1/2}$ is the normalization factor and
\ben \label{e31}
\h{G}_0=c^{\dg}_{A\up}c^{\dg}_{A\dn}+c^{\dg}_{B\up}c^{\dg}_{B\dn}+\al\left(\fr{U}{4t}\right)\left(c^{\dg}_{A\up}c^{\dg}_{B\dn}-c^{\dg}_{A\dn}c^{\dg}_{B\up}\right)
\een
with $\al(x)=x+\sq{1+x^2} .$ By mapping to $H_{2L}$ via Eq. (\r{e2}) we get,
\ben \label{e32}
|vac\ran\ra|0\ran\ot|0\ran\ot|0\ran\ot|0\ran=|0000\ran
\een
while mapping between the operators, (Eq. (\r{e6}) and Eq. (\r{e8})) gives
\benr \label{e33}
c^{\dg}_{A\up}\ra a^{\dg}\ot I_2\ot I_2\ot I_2\;\;\;c^{\dg}_{A\dn}\ra I_2\ot a^{\dg}\ot I_2\ot I_2 &&   \nonumber    \\
c^{\dg}_{B\up}\ra  I_2\ot I_2\ot a^{\dg}\ot I_2\;\;\;c^{\dg}_{B\dn}\ra I_2\ot I_2\ot I_2\ot a^{\dg}      \nonumber    \\
\eenr
The normalized ground state can be expressed in the qubit space as
\ben  \label{e34}
|\p_0\ran=\fr{-1}{\sq{2(1+\al^2)}}\left\{|1100\ran+|0011\ran+\al |1001\ran-\al |0110\ran\right\}
\een

 The four-partite entanglement in any state $|\psi\ran$ can be calculated by using Eqs. ((\r{e18}), (\r{e19}), (\r{e20})) as,
\ben \label{e35}
 E_g=||\t||-1
 \een
 where
 \ben   \label{e-fourp}
 ||\t||=\sq{\sum_{i,j,k,l=1}^{3}t_{ijkl}^{2}}
 \een
 with
 \ben  \label{e-fourp1}
 t_{ijkl}=\lan\p|\si_i\ot \si_j\ot \si_k\ot \si_l|\p\ran .
 \een
The ground state entanglement can be then calculated to be
\ben  \label{e-fourp2}
E_g=\fr{3}{(1+\al^2)}\sq{1+\fr{2}{9}\al^2+\al^4}-1
\een
We plot the four-partite entanglement as a function of $U$ and $t$ (Fig. (2.1a)) and as a function of $\al$ (Fig. (2.1b)). The entanglement is seen to  monotonically increase as a function of $\alpha$, saturating at large values of $\alpha$ to the maximum value $2$. The saturation to the maximum value can be obtained either for very large values of $U$ or very small values of $t$. We can interpret this result in the following way: since the total particle number is fixed to be $2$, the four mode entanglement essentially measures the correlations between the spins. The entanglement increases as a function of $\alpha$ because the spin correlations increase with $\alpha$.

\begin{figure}[!ht]
\begin{center}

\includegraphics[width=7cm,height=5cm]{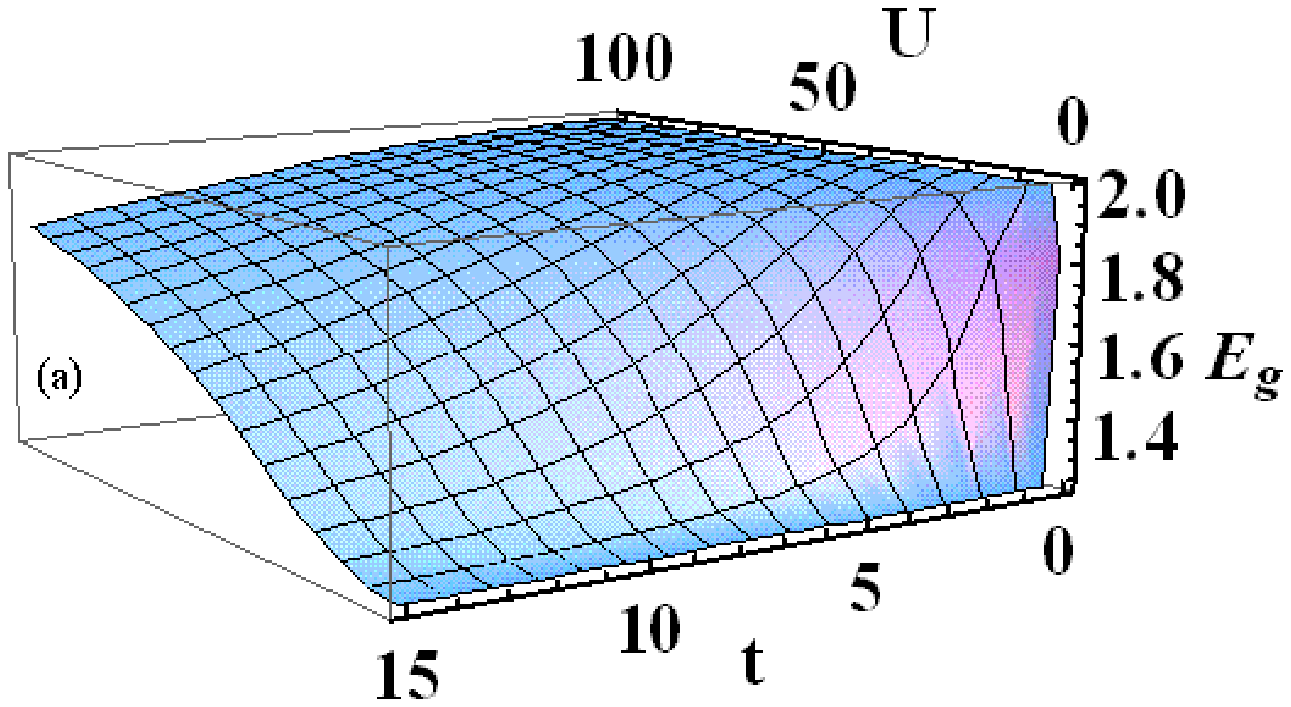}
\includegraphics[width=7cm,height=5cm]{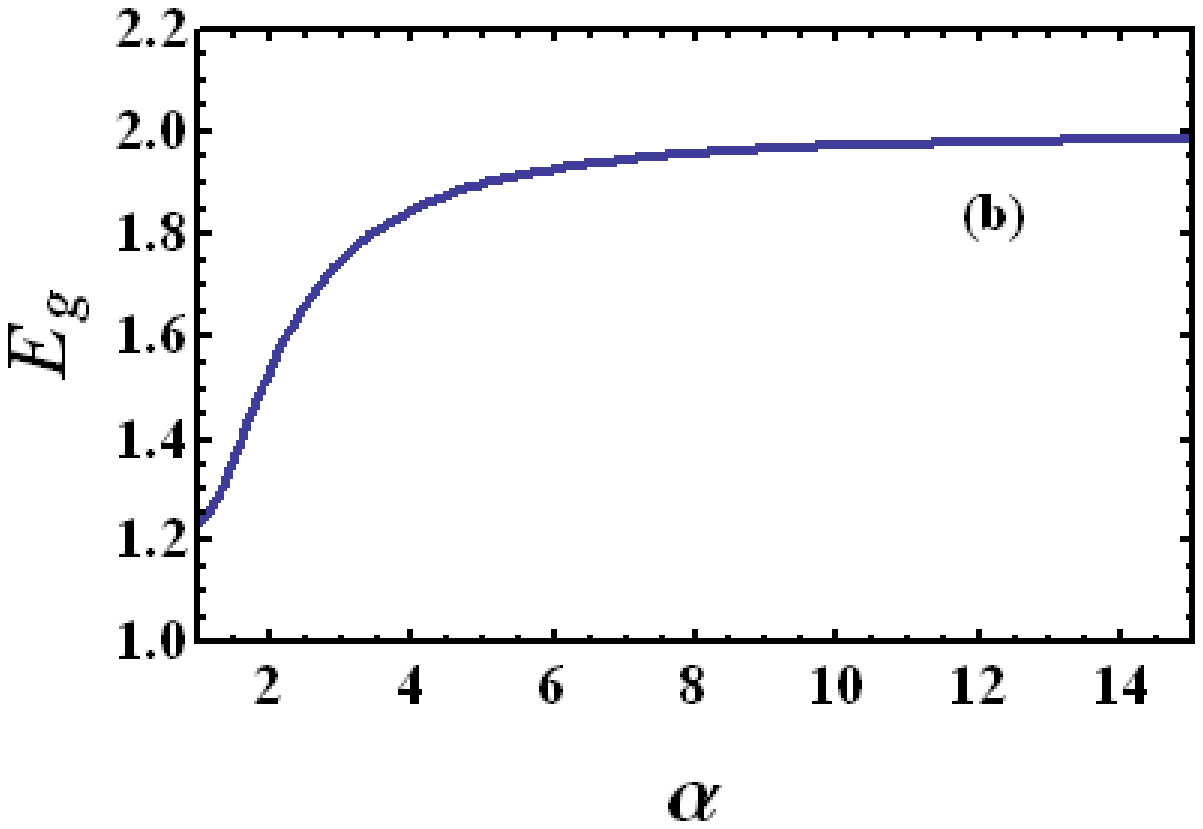}

 Fig. (3.1):  Four-partite entanglement for the Hubbard dimer (at half filling) as a function of (a) $U$ and $t$ (both in energy units) and (b) as a function of $\al(\frac{U}{4t}$ ) where $\al(x)=x+\sq{1+x^2}.$ ($\al\ge1$).

\end{center}
\end{figure}

We can also calculate the bipartite entanglement between sites $A$ and $B$ using the geometric measure Eqs. ((\r{e12}), (\r{e13}), (\r{e14})) considering the partitions to be $\{\{A\up \; A\dn\} ; \{B\up \; B\dn\}\} $
 \ben  \label{e48}
 E_s =||\t||-||\t||_{sep}
 \een
 where
 \ben  \label{e49}
 ||\t||=\sq{\sum_{i,j= 1}^{15} t_{ij}^2}; \,\,\,\,\,  t_{ij}=\left(\fr{d}{2}\right)^2 \lan\p|\h{\la}_i\ot \h{\la}_j |\p\ran
 \een
and $||\t||_{sep}=\left(\fr{d(d-1)}{2}\right)^{m/2}$.
Here $\h{\la}$s are the generators of $SU(4),$ there are $m=2$ partitions (same as the number of sites) and each partition has dimension $d=4.$ This leads to an intersite entanglement of the form
\ben \label{e36}
E_{s}=\fr{2}{(1+\al^2)}\sq{13\al^4+34\al^2+13}  - 6
\een
The bi-partite entanglement between sites $A$ and $B$ was calculated earlier using the von-Neumann entropy \cite{z02}
\ben \label{e55}
E_{VN}=\fr{1}{(1+\al^2)}\left\{\log_2\left[2(1+\al^2)\right]-\al^2\log_2\left[\fr{\al^2}{2(1+\al^2)}\right]\right\}   \\
\een
We plot the intersite entanglement (the von-Neumann entropy is also plotted for comparison) as a function of $\al$ in Fig. (3.2). It is seen that both measures show qualitatively similar behavior, i.e. a monotonically decreasing entanglement as a function of $\alpha$ saturating at very large values of $\al$. The entanglement between the sites $A$ and $B$ decreases as a function of $\alpha$ because  with increasing on-site repulsion $U$, the four dimensional local state space at each site gets reduced to a two dimensional local state space\cite{z02} due to a suppression of charge fluctuations or in other words, as $\al \to \iny$ the $SU(4)\ot SU(4)$ partition goes over to a $SU(2)\ot SU(2)$ partition. We have explicitly checked that the entanglement obtained in the $\al \to \iny$ limit matches with that obtained for the $SU(2)\ot SU(2)$ partition.\\
\begin{figure}[!ht]
%\begin{center}

\includegraphics[width=7cm,height=5cm]{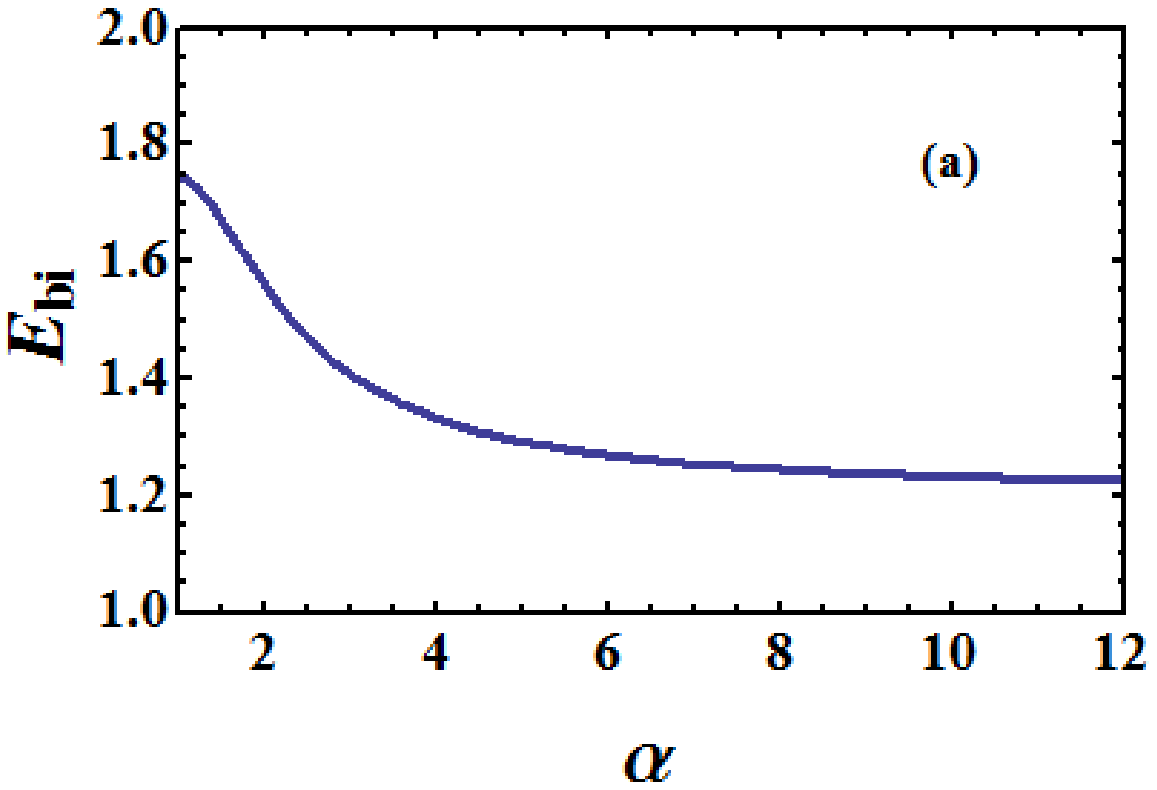}
\includegraphics[width=7cm,height=5cm]{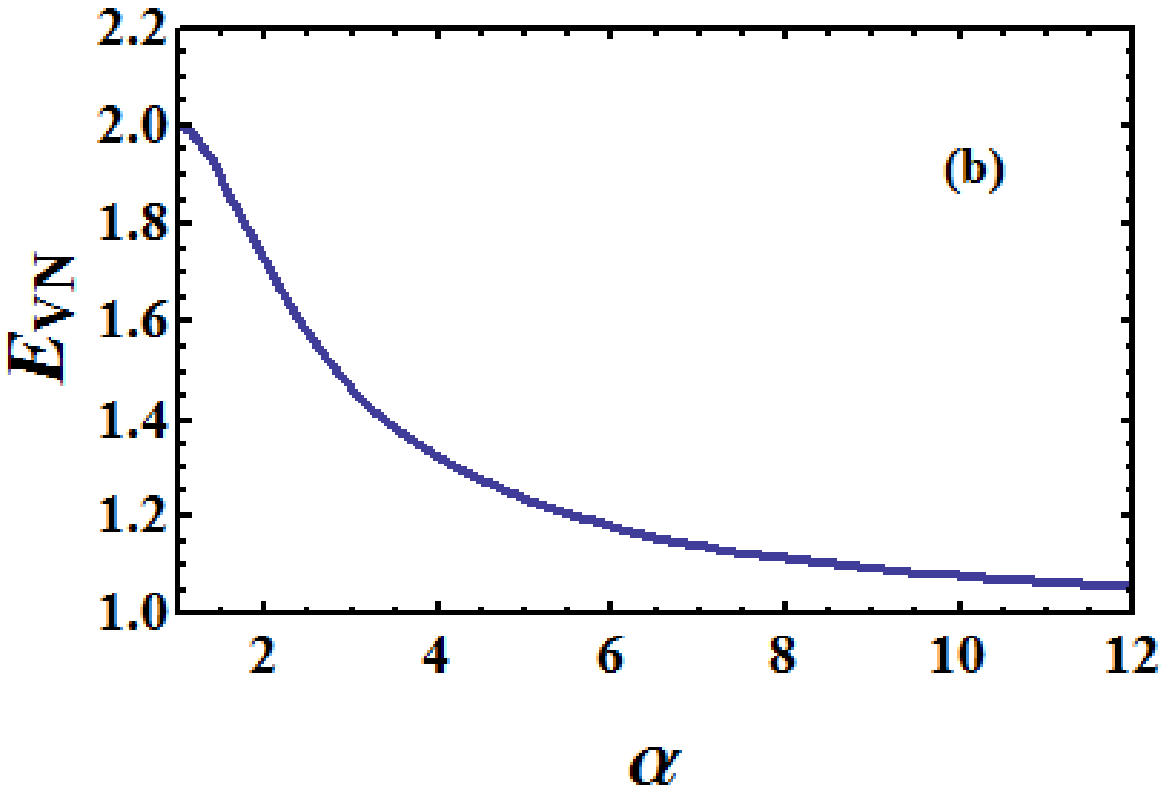}

Fig. (3.2) : (a) The bi-partite entanglement between sites A and B calculated with the geometric measure as a function of $\al$ ($\al\ge1$) for the Hubbard dimer at half filling. (b) The corresponding von-Neumann entropy as a function of $\al$ ($\al\ge1$).

%\end{center}

\end{figure}

We can also discuss bi-partite entanglement with unequal partitioning. Consider four modes partitioned into two subsets containing one and three modes respectively. This would correspond for example to one observer controlling a register which measures the occupancy of the spin up at site $A$ and the other observer controlling a register which measures the occupancy of the spin down at site $A$ as well as the occupancy at site $B$. The bi-partite entanglement in this can be obtained as discussed in Sec.(\ref{sec:unequal}). In the present case, the two partitions have dimension as $d_1=2$ and $d_2=8$ respectively. This gives, for the entanglement,
\ben  \label{e41}
E=4\sq{\sum_{j=1}^{3}\sum_{k=1}^{63}K_{jk}^2}-\sq{28}
\een
or, using $|\psi\ran$ in Eq. (\r{e34}),
\ben \label{e42}
E=1.6367
\een
 Interestingly, the entanglement is independent of $\al$. It turns out that this is the maximum value possible for the entanglement (Eq. (\r{e41}))( we checked this by maximizing the entanglement given by Eq. (\r{e41}) as a function of the coefficients in the general state Eq. (\r{e24}). We have also checked that von-Neumann entropy in this case is also independent of $\al$ and has the maximum possible value, i.e. $2$.

\subsection{ Three electrons on three sites.}

We next consider the Hubbard trimer, i.e. electrons on three sites with the sites $A,B,C$  with periodic boundary conditions. Ignoring the chemical potential , the Hamiltonian is
\ben\label{e43}
H(\beta)=-t\left[\sum_{j=A,B,C}(c_{j\up}^{\dg}c_{j+1\up}+c_{j\dn}^{\dg}c_{j+1\dn}+h.c.)-\beta \sum_{j=A,B,C}\h{n}_{j\up}\h{n}_{j\dn}\right]
\een
 where $t>0$ is the hopping parameter and $\beta =\frac{U}{t}$. We begin our analysis as earlier by mapping from the fermionic to the six qubit space corresponding to $(A\up \; A\dn \; B\up \; B\dn \; C\up \; C\dn )$. In the qubit space, the basis respecting number super-selection rule is given by
\ben  \label{e54}
\{|k\ran\}\;\; k=7,11,13,14,19,21,22,25,26,28,35,37,38,41,42,44,49,50,52,56
\een
where the twenty basis states are labeled by the six bit binary representation of the $k$ values. We numerically diagonalize the Hamiltonian $H(\beta)$ in Eq. (\r{e43}) in the basis given by Eq. (\r{e54}) at different values of $\beta$ as $\beta$ increases from zero. The total spin quantum number $S$ and the $z$ component of the total spin $S_Z$ commute with the Hamiltonian and can therefore be used as good quantum numbers to characterize the states. The ground state has total $S$ value $=1/2$. The triangular geometry of the three site model (with periodic boundary conditions) also leads to an additional symmetry under reflection about one of the medians of the triangle. This leads to a two fold degeneracy for the ground state (for a fixed $S_Z$ value). Since these symmetries are preserved even in the presence of the interaction $U$, the ground state remains two-fold degenerate for all $\beta$ values.

The entanglement in any state can be calculated in a similar manner as shown previously (Sec.\r{sec:Hdimer}).
We now show the results of our calculations for the entanglement in one of the ground states.

The six-mode entanglement as a function of the interaction parameter $\beta$ is shown in Fig (3.3a) while the tripartite entanglement (between the sites $A, B, C$) is shown in Fig. (3.3b) as a function of $\beta$.  The tripartite entanglement between the sites $A,B,C$ is seen to decrease with increasing $\beta$ - we interpret this result  in a similar way as that for the dimer as due to the fact that the local state space at each site decreases with increasing $\beta$.  We find that the six-mode entanglement increases with $\beta$, saturating at large values, however, the behavior is not monotonic. Such non- monotonic behavior as a function of $\beta$ is also shown by the bipartite entanglement between sites $A$ and $ BC$ calculated using the geometric measure as well as the von-Neumann entropy (Figs. (3.3c) and (3.3d)).

  We can understand the non-monotonic behavior in the following way: the bi-partite entanglement measures the entanglement between the the sites $A$ and the sites $BC$. With small increase in $\beta$ from zero, there is an increase in the spin fluctuations between $B$ and $C$ which shows up as an initial increase in the entanglement between $A$ and $BC$. However, with further increase in $\beta$, the charge fluctuations get completely suppressed leading to an asymptotic behavior similar to that for the dimer. One can understand the non-monotonic behavior of the six-mode entanglement in a similar way. For small $\beta$, there is a larger correlation between two sites which leads to a decrease in the overall entanglement, however with further increase in $\beta$, the contribution to the entanglement is solely due to spin fluctuations which increases with $\beta$ leading to the observed increase in the entanglement as well.
\begin{figure}[!ht]
\begin{center}

\includegraphics[width=7cm,height=5cm]{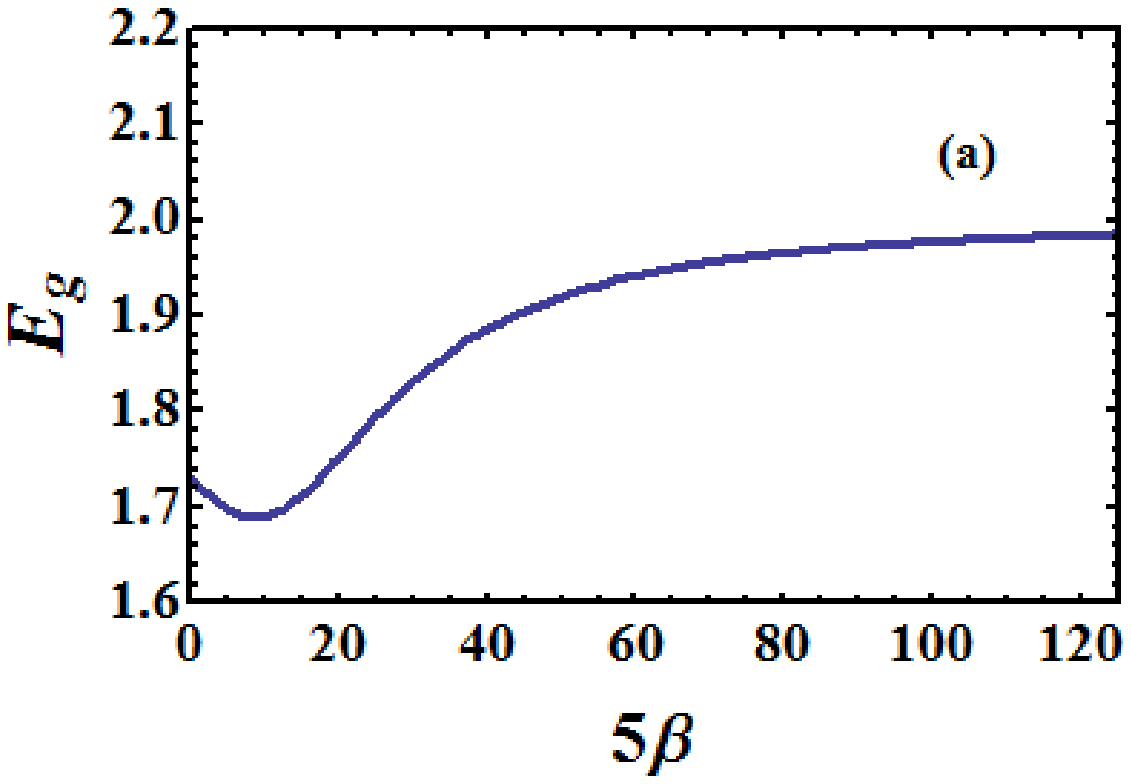}
\includegraphics[width=7cm,height=5cm]{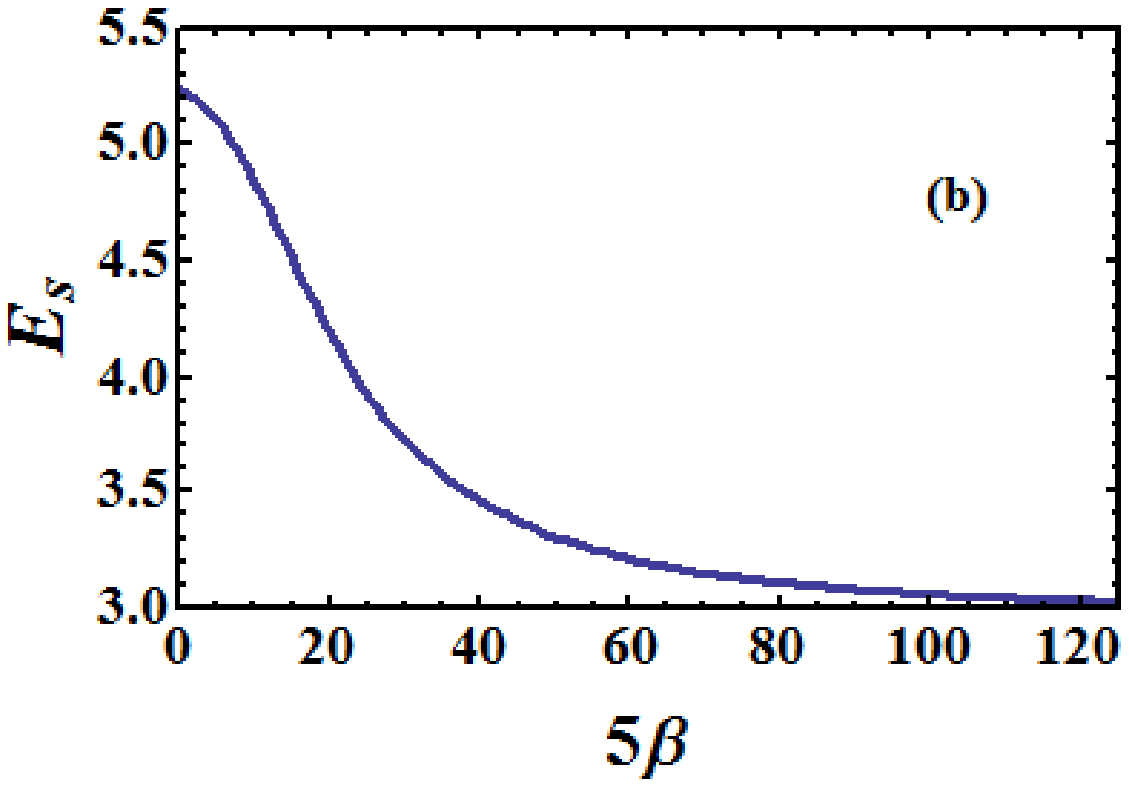}
\includegraphics[width=7cm,height=5cm]{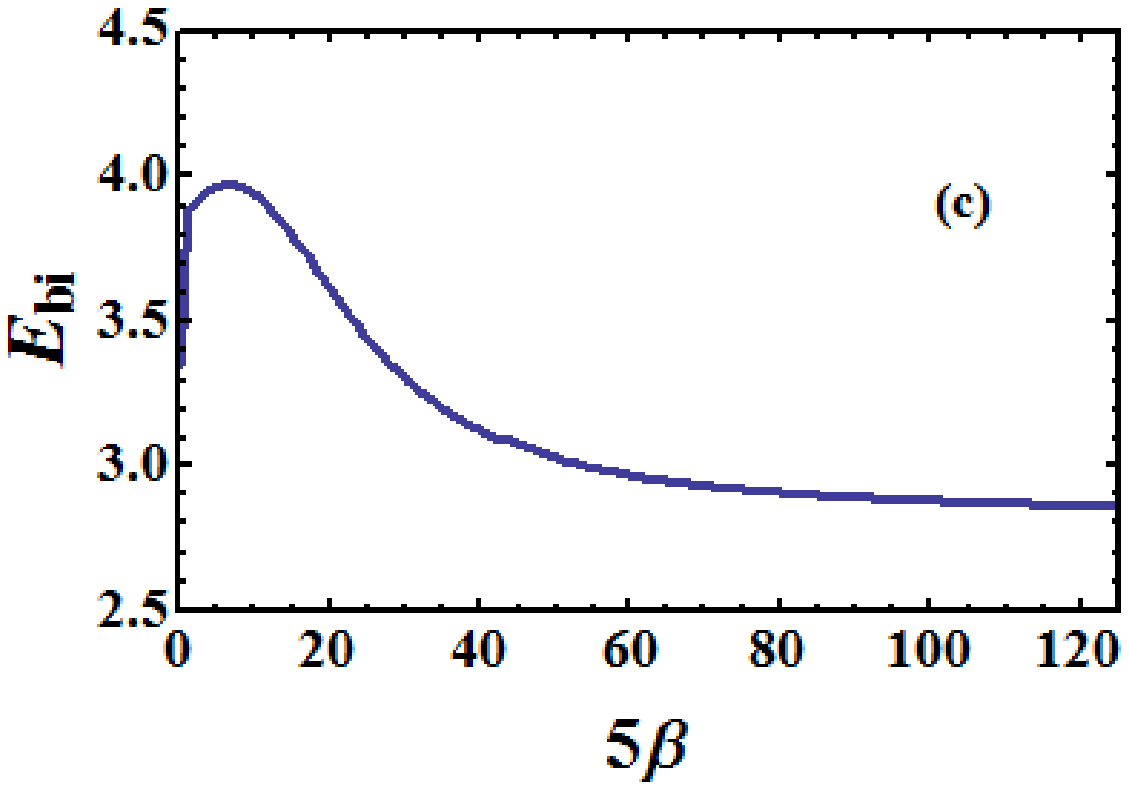}
\includegraphics[width=7cm,height=5cm]{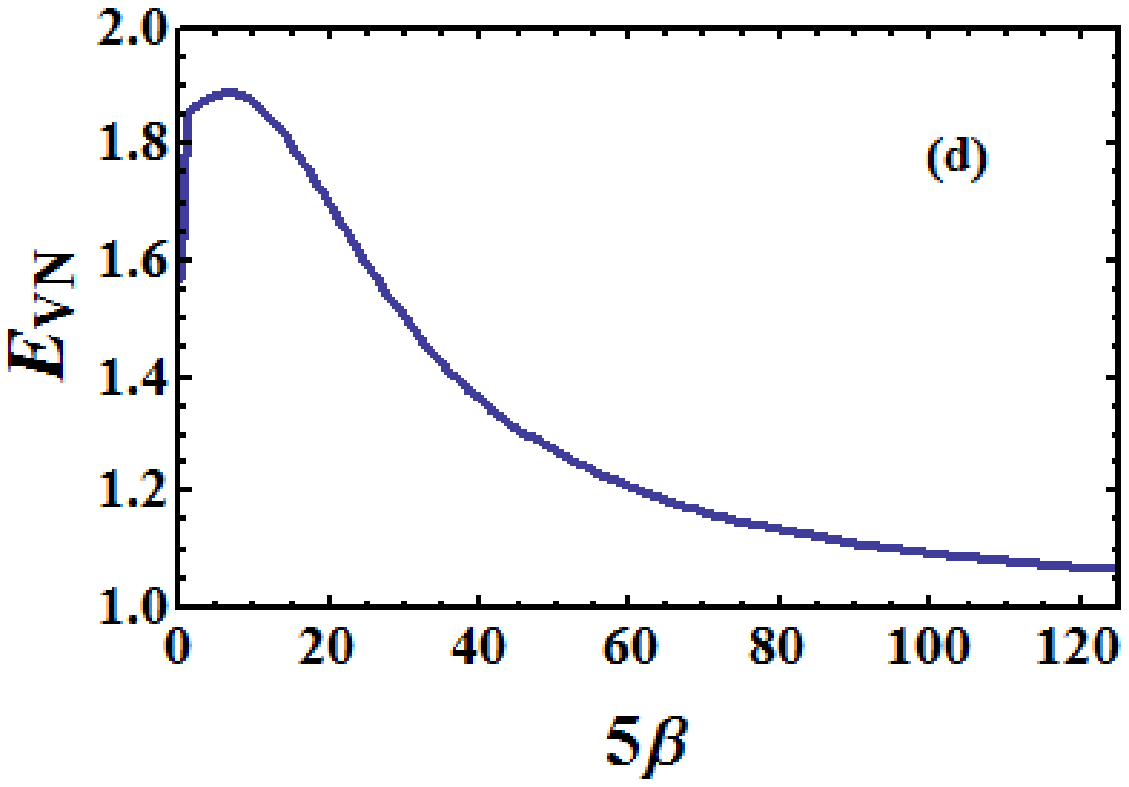}
\end{center}
Fig. (3.3): (a) Six-partite entanglement for the Hubbard trimer as a function of $\beta$ at half filling, (b) Tripartite (site) entanglement as a function of $\beta$ for the Hubbard trimer at half filling, (c) The bipartite entanglement between site A and sites BC (see text) as a function of $\beta$ for the Hubbard trimer at half filling and (d) The corresponding von-Neumann entropy as a function of $\beta$ for the Hubbard trimer at half filling.
\end{figure}\\
We have also calculated the upper bounds for the entanglements in various partitions in the same way as in Sec.\ref{sec:fourmode}. These are $E_g^{max}=4.42218,$ $E_{bi}^{max}=4.15105,$ $E_s^{max}=6.08767.$\\
\
\\
\
\\
\
\\
\
\\
\
\\

\section{\label{sec:summ}Summary and comments}

  We have proposed a multipartite entanglement measure for $N$ fermions distributed over $2L$ modes (single particle states). The measure is defined on the $2L$ qubit space isomorphic to the Fock space for $2L$ single particle states. The entanglement measure is defined for a given partition of $2L$ modes containing $m\geq 2$ subsets, using the Euclidean norm of the $m$-partite correlation tensor in the Bloch representation of the corresponding multi-mode state, viewed as a $m$-partite state (see sec. IV and V). The Hilbert spaces associated with these subsets may have different dimensions. The quantum operations confined to a subset of a given partition are local operations. This way of defining entanglement and local operations gives us the flexibility to deal with entanglement and its dynamics in various physical situations governed by different Hamiltonians.  We note that the concept of locality for indistinguishable fermions is distinct from that for distinguishable particles. In the latter case, locality applies to spatially separated subsystems. However, spatially separated fermions become distinguishable. We have shown, using a representative case, that the geometric measure is invariant under local unitaries corresponding to a given partition. As an application, we have also considered some correlated electron systems and demonstrated the use of the multipartite measure in these systems. In particular, we have calculated the multipartite entanglement in the Hubbard dimer and trimer (at half filling). We find that the bipartite entanglement between sites as computed with the geometric measure has a qualitatively similar behavior as a function of the interaction as that of the conventional von-Neumann entropy. We have also calculated the four(six)- partite and the two (three) site entanglement for the Hubbard dimer(trimer). We find that the multi-partite entanglement gives complementary information to that of the site entanglement in both the cases.

  Although the entanglement measures given in this chapter have been mainly applied to the study of the multipartite entanglement structure in Hubbard dimers and trimers, these measures are completely general and can be  applied to other fermionic systems. We have also shown \cite{hj08,hj09} that, viewed as a measure on qubit space, this measure has most of the properties required of a good entanglement measure, including monotonicity. To the best of our knowledge, this is the first measure of multipartite entanglement in fermionic systems going beyond the bipartite and even the tripartite case. Further, in this chapter we have restricted to applications involving ground states of $2L$ mode systems with $2L\leq 6$. It is straightforward to extend the calculations for large number of modes ($2L>6$) except for the length of the computation. We have shown earlier that \cite{hj08,hj09} for antisymmetric states, the computational complexity of computing the measure goes polynomially with the number of parts of the system (here, the number of partitions of $2L$ modes). It would be interesting to extend these calculations to larger system sizes where new and interesting results might be expected.

%@@@@@@@@@@@@@@@@@@@@@@@@@@@@@@@@@@@@@@@@@@@@@@@@@@@@@@@@@@@@@@@@@@@@@@@@@@@@@@@@@@@@@@@@@@@@@@@@@@@@@@@@@@@@@@@@@@@@@@@@@@@@@@@@@@@@@@@@@@@@@@@2

\chapter{Thermal quantum  and classical correlations in two qubit XX model in a nonuniform external magnetic field}

\begin{figure}[!ht]
\begin{center}
\includegraphics[width=8cm,height=1.5cm]{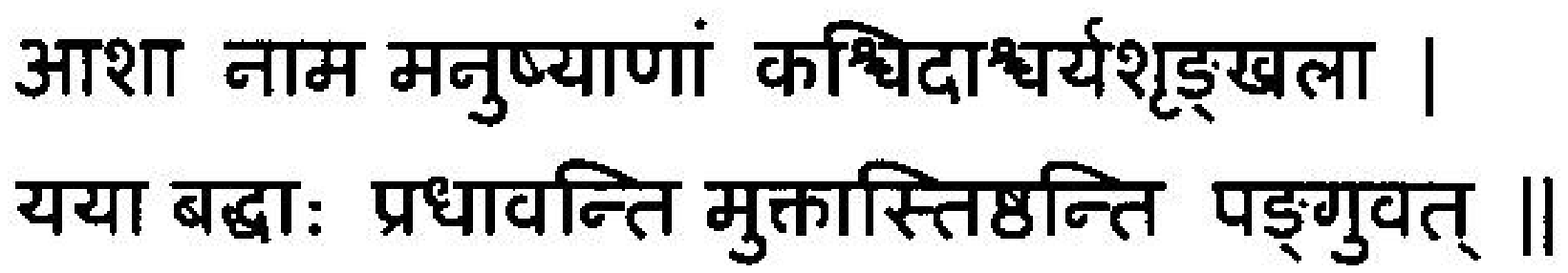}

A Sanskrit saying (Subhashit)\\
\scriptsize\textsc{People are bound by an amazing chain called hope. Those bound by it keep running. Others stay put as if disabled. }
\end{center}
\end{figure}

\section{Introduction}

  We have seen in chapter I that composite quantum systems can be in a class of states, called entangled states, in which the correlations between the constituents of the system cannot be achieved in a classical world \cite{nc00,w89}. Although all pure entangled states possess such nonlocal quantum correlations, there are mixed entangled states which do not, in the sense of violating Bell inequalities \cite{bel64}. The entanglement in quantum states and the resulting nonlocal quantum correlations form an area of intense research, due to their huge technological promise, especially in the areas of quantum communication and cryptography \cite{e91}. However, quantum correlations breaking Bell inequalities need not account for all quantum correlations in a composite quantum system in a given state. In order to account for the quantum correlation in a given state, we must find some means to divide the total correlation into a classical part and a purely quantum part. This is particularly important for mixed states, since their quantum correlations are many a time hidden by their classical correlations $(CC)$. An answer to this requirement is given by quantum discord $(QD)$, a measure of the quantumness of correlations introduced in Ref. \cite{oz01}. Quantum discord is built on the fact that two classically equivalent ways of defining the mutual information turn out to be inequivalent in the quantum domain. In addition to its conceptual role, some recent results \cite{kl98}, suggest that quantum discord and not entanglement may be responsible for the efficiency of a mixed state based quantum computer. We believe that $QD$ will turn out to be a very useful tool to analyze mixed state quantum correlations and their consequences, as  mixed state entanglement is very difficult and eluding to deal with \cite{hjq08}. To  realize this hope we need a viable relation between mixed state entanglement and quantum discord \cite{mpsvw10}. The pointers towards such a relation may be obtained by studying these properties for various quantum systems.

  Motivated by these considerations, we present , in this chapter, the results of our investigation of the amount of $QD$ and $CC$ in a two qubit Heisenberg XX chain at finite temperature subjected to nonuniform external magnetic fields $B_1$ and $B_2$ acting separately on each qubit. We study two distinct cases namely $B_1=-B_2$ (nonuniform field) and $B_1=B_2$ (uniform field). In each case, we obtain the dependence of $QD, CC$ and entanglement $(EOF)$ in the system on the external magnetic field and temperature. Such a model is realized, for example, by a pair of qubits (spin 1/2) within a solid at finite temperature experiencing a spatially varying magnetic field. Such Heisenberg models can describe fairly well the magnetic properties of real solids \cite{h99} and are well adapted to the study of the interplay of disorder and entanglement as well as of entanglement and quantum phase transitions \cite{s00,s09}. The variation of entanglement \cite{scc03} and $QD$ \cite{wr10,mgcss10} in a two qubit Heisnberg XX chain with external magnetic field is already reported.

 In order to quantify entanglement in a thermally mixed two qubit state, we use concurrence given by \cite{hw97}

\ben\label{eqno1}
C=\max\{\lambda_1-\lambda_2-\lambda_3-\lambda_4,0\}
\een
where $\lambda_i (i=1,2,3,4)$ are the square roots of the eigenvalues
of the operator $\rho \tilde{\rho}$ in descending order

\ben\label{eqno2}
\tilde{\rho}=(\sigma_1^y\otimes \sigma_2^y)\rho^{\ast}(\sigma_1^y\otimes \sigma_2^y),
\een
with $ \lambda_1\geq \lambda_2 \geq \lambda_3 \geq \lambda_4,$ and $\rho$ is the density matrix of the pair qubits; $\sigma_1^y$ and $\sigma_2^y$ are the normal Pauli operators. The entanglement is related to the concurrence by $$EN=h\left(\frac{1+\sqrt{1-C^2}}{2}\right),$$ where $h(x)=-x\log_2 x-(1-x)\log_2 (1-x).$ Henceforth, in this chapter, we denote the entanglement of formation ($EOF$) by $EN.$ The concurrence $C=0$ corresponds to an unentangled state and $C=1$ corresponds to a maximally entangled state. \\

\section{The thermalized Heisenberg system.}
The model Hamiltonian we study is given by
\ben\label{eqno3}
H=J(S_1^xS_2^x+S_1^yS_2^y)+B_1S_1^z+B_2S_2^z,
\een
where $S^{\alpha}\equiv\sigma^{\alpha}/2,~~~(\alpha=x,y,z)$ are the spin $1/2$ operators, $\sigma^{\alpha}$ are the Pauli operators and $J$ is the strength of Heisenberg interaction. $B_1$ and $B_2$ are external magnetic fields. As stated in the introduction, by changing $B_1$ and $B_2$ separately, we want to study the effects of magnetic field on the thermal $QD, CC$ and $EN$ in a very general way. The eigenvalues and eigenvectors of H are

$$H |00\rangle = -(B_1+B_2)|00\rangle,$$
$$H |11\rangle= (B_1+B_2)|11\rangle,$$
\ben\label{eqno4}
H |\psi^{\pm}\rangle=\pm D |\psi^{\pm}\rangle,
\een
 where $D^2= (B_1-B_2)^2+J^2$
 and $|\psi^{\pm}\rangle=\frac{1}{N_{\pm}}[|01\rangle+\frac{(B_1-B_2)\pm D}{J} |10\rangle].$
 We denote the eigenvalues corresponding to $|00\rangle, |11\rangle,|\psi^{\pm}\rangle$ by $E_{00},E_{11},E_{\pm}$ respectively. In the standard basis, $\{ |00\rangle, |01\rangle, |10\rangle, |11\rangle\},$ the density matrix $\rho(T)$ is given by
\ben\label{eqno5}
\rho(T)=\frac{1}{Z} \left[ \ba{rrrr} u_1 & 0 & 0 & 0 \\ 0 & w_1 & v & 0 \\ 0 & v & w_2 & 0 \\ 0 & 0 & 0 & u_2 \ea \right],
\een
where
$$u_1=e^{(B_1+B_2)/{kT}},$$
$$u_2=e^{-(B_1+B_2)/{kT}},$$
$$w_1=\cosh(\frac{D}{kT})+\frac{(B_1-B_2)}{D}\sinh(\frac{D}{kT}),$$
$$w_2=\cosh(\frac{D}{kT})-\frac{(B_1-B_2)}{D}\sinh(\frac{D}{kT}),$$

\ben\label{eqno6}
v=-\frac{J\sinh(\frac{D}{kT})}{D},
\een
and $Z=Tr[\exp(\frac{-H}{kT})]$ is the partition function. In the following we select $|J|$ as the energy unit and set $k=1.$\\

\section{Quantum Discord.}
 We suummarize here the concept of quantum discord, which already discussed in chapter 1. In classical information theory (CIT) the total correlation between two systems (two sets of random variables) A and B described by a joint distribution probability $p(A,B)$ is given by the mutual information (MI),
\ben\label{eqno7}
I(A, B) = H(A) + H(B) - H(A, B),
\een
with the Shannon entropy $H(p) = -\sum_j p_j log_2 p_j$. Here $p_j$ represents the probability of an event $j$ associated to
systems $A, B,$ or to the joint system $AB$. Using Bayes's rule we may write MI as
\ben\label{eqno8}
I(A, B) = H(A) - H(A|B),
\een
where $H(A|B)$ is the classical conditional entropy. In CIT these two expressions are equivalent but in the quantum domain this is no longer true \cite{oz01,lbaw08}. The first quantum
extension of MI, denoted by $I (\rho)$, is obtained directly replacing the Shannon entropy in Eq. (\ref{eqno7}) with the von Neumann entropy, $S (\rho) = -Tr (\rho log_2 \rho)$, with $\rho$, a density matrix, replacing probability distributions. To obtain a quantum version of Eq. (\ref{eqno8}) it is necessary to generalize the classical conditional entropy. This is done recognizing $H(A|B)$ as a measure of our ignorance about system $A$ after we make a set of measurements on $B$. When $B$ is a quantum system the choice of measurements determines the amount of information we can extract from it. We restrict ourselves to von Neumann measurements on $B$ described by a complete set of orthogonal projectors, $\Pi_j$, corresponding to outcomes $j$.

After a measurement, the quantum state $\rho$ changes to $\rho_j = [(I\otimes \Pi_j)\rho (I\otimes \Pi_j)]/p_j$, with $I$ the identity operator
for system $A$ and $p_j = Tr[(I\otimes \Pi_j)\rho (I\otimes \Pi_j)]$. Thus, one defines the quantum analog of the conditional entropy as $S (\rho|\{\Pi_j\}) = \sum_j p_j S (\rho_j)$ and, consequently, the second quantum extension of the classical MI as \cite{oz01}

 $\mathcal{J}(\rho|\{\Pi_j\}) = S(\rho^A)- S(\rho|\{\Pi_j\})$. The value of $\mathcal{J}(\rho|\{\Pi_j\})$ depends on the choice of $\{\Pi_j\}$.

 Henderson and Vedral \cite{oz01} have shown that the maximum of $\mathcal{J}(\rho|\{\Pi_j\})$ with respect to $\{\Pi_j\}$ can be interpreted as a measure of classical correlations. Therefore, the difference between the total correlations $I (\rho)$ and the classical correlations
$\mathcal{Q}(\rho) = sup_{\{\Pi_j\}} \mathcal{J}(\rho|\{\Pi_j\}) $ is defined as

\ben\label{eqno9}
D(\rho) = I (\rho) - Q(\rho),
\een
giving, finally, a measure of quantum correlations \cite{oz01,wr10} called quantum discord $(QD)$. For pure states $QD$ reduces to entropy of entanglement \cite{bbps96}, highlighting that in this case all correlations come from entanglement.
However, it is possible to find separable (not-entangled) mixed states with nonzero $QD$ \cite{oz01,lbaw08}, meaning that entanglement
does not cause all nonclassical correlations contained in a composite quantum system. Also, $QD$ can be operationally seen as the difference of work that can be extracted from a heat bath using a bipartite system acting either globally or only locally \cite{zu03}.

\section{Results and discussion.}

\subsection{Case I: $B_1=-B_2,$ and $(J>0)$.}
In this case, $|\psi^-\rangle$ is the ground state with eigenvalue $E_-=-\sqrt{4B_1^2+J^2}.$ Other eigenvalues are $\{0, 0, \sqrt{4B_1^2+J^2}\}$ for eigenvectors $\{|00\rangle, |11\rangle, |\psi^+\rangle\}$ respectively.
In this case the variation of $QD, CC$ and  $EN$  with $B_1$ at different  temperatures (T=0.2, 0.9, 1.5) is depicted in Figs. (4.1), (4.2) and (4.3), respectively. We observe that $QD$ and $CC$ in the thermal state coincide for all values of $B_1$ for different temperatures (T=0.2, 0.9, 1.5).

In order to understand this observation, we take a close look at the thermal state.  At temperature T, the thermal state is given by

$$\rho=\frac{1}{Z}\big{[}|00\rangle\langle00|+|11\rangle\langle11|+e^{\sqrt{4B_1^2+J^2}/{T}}$$

\ben\label{eqno10}
|\psi^-\rangle\langle\psi^-|+e^{-\sqrt{4B_1^2+J^2}/{T}}|\psi^+\rangle\langle\psi^+|\big{]}.
\een
This $\rho$ has the Bloch representation \cite{hj08}

\ben\label{eqno11}
\rho=\frac{1}{4}[I\otimes I+\sum_{i=1}^3 c_i \sigma_i\otimes \sigma_i],
\een
where $\sigma_i~~(i=1, 2, 3)$ are the one-qubit Pauli operators.\\
The class of mixed states as in Eq. (\ref{eqno11}) have equal classical and quantum correlations (like bipartite pure states \cite{ll07}) provided
$$c_i=c_j>c_k$$ and

\ben\label{eqno12}
c_k=-c_i^2
\een
where $i\neq j \neq k \in \{1,2,3\}$ (see the Appendix to this chapter). Here $c_i, c_j, c_k$ are the diagonal elements of the correlation matrix defined by $c_{ij}=Tr(\rho \sigma_i\otimes \sigma_j).$
It is straightforward to check that the thermal state $\rho$ in Eq. (\ref{eqno10}) which has form of Eq. (\ref{eqno11}) satisfies conditions (12).
This explains the observations in Figs. (4.1), (4.2) and (4.3), that the two qubit thermal state $\rho$ for $B_1 = -B_2$ in Eq. (\ref{eqno10}) gives rise to equal $QD$ and $CC$ for all values of $B_1$ and temperature. In order to see why the common curve for $QD$ and $CC$ peaks at $B_1=0$ we can maximize the expression for $QD=CC$  with respect to $B_1$ and check that the maximum occurs at $B_1=0.$

\begin{figure}[!ht]
\begin{center}
\includegraphics[width=8cm,height=5cm]{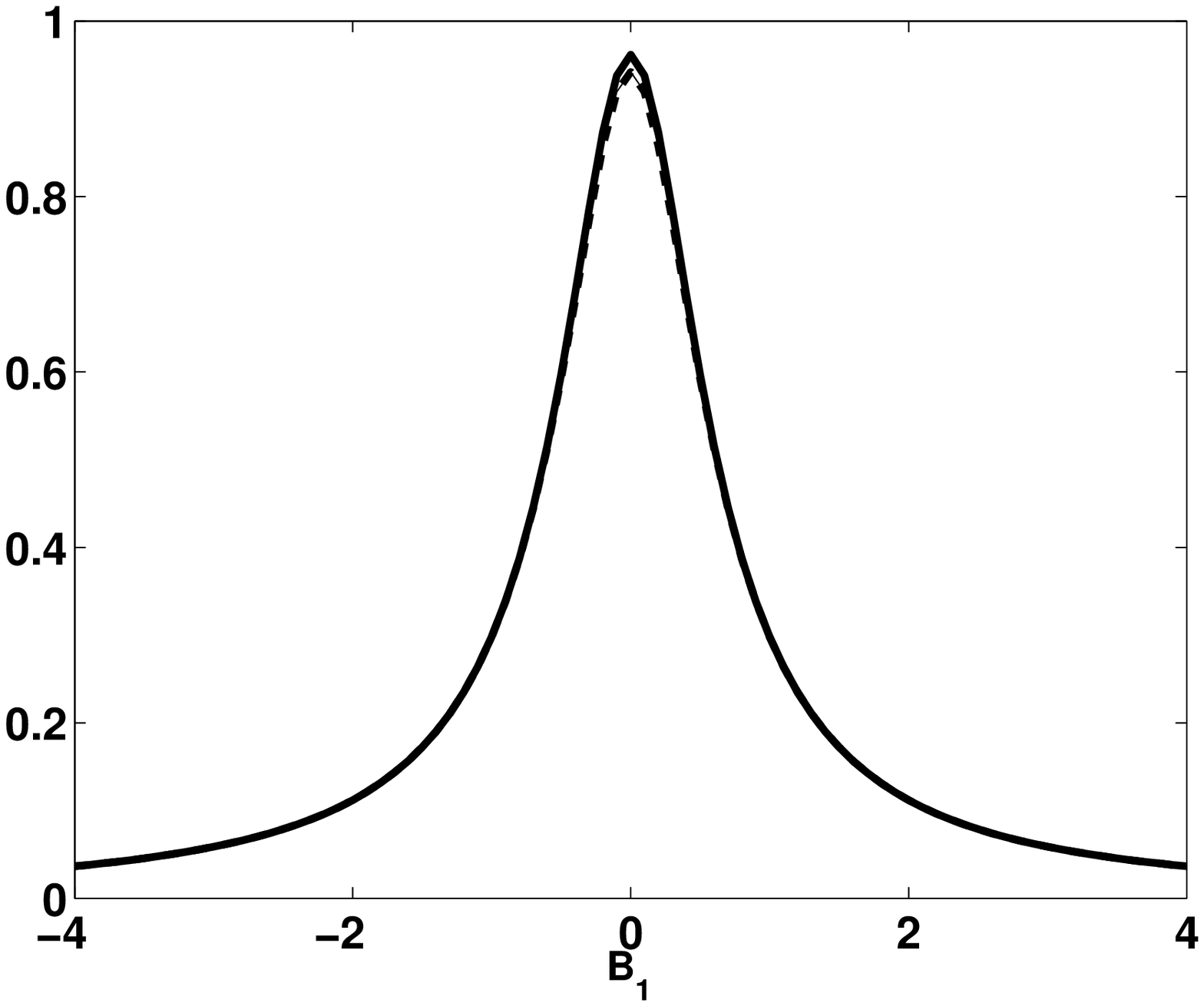}

Fig. (4.1) : $QD$ and $CC$ (dashed line) and $EN$ (solid line) as a function  of external magnetic field $B_1=-B_2$ at $T=0.2$
\end{center}
\end{figure}

\begin{figure}[!ht]
\begin{center}
\includegraphics[width=8cm,height=5cm]{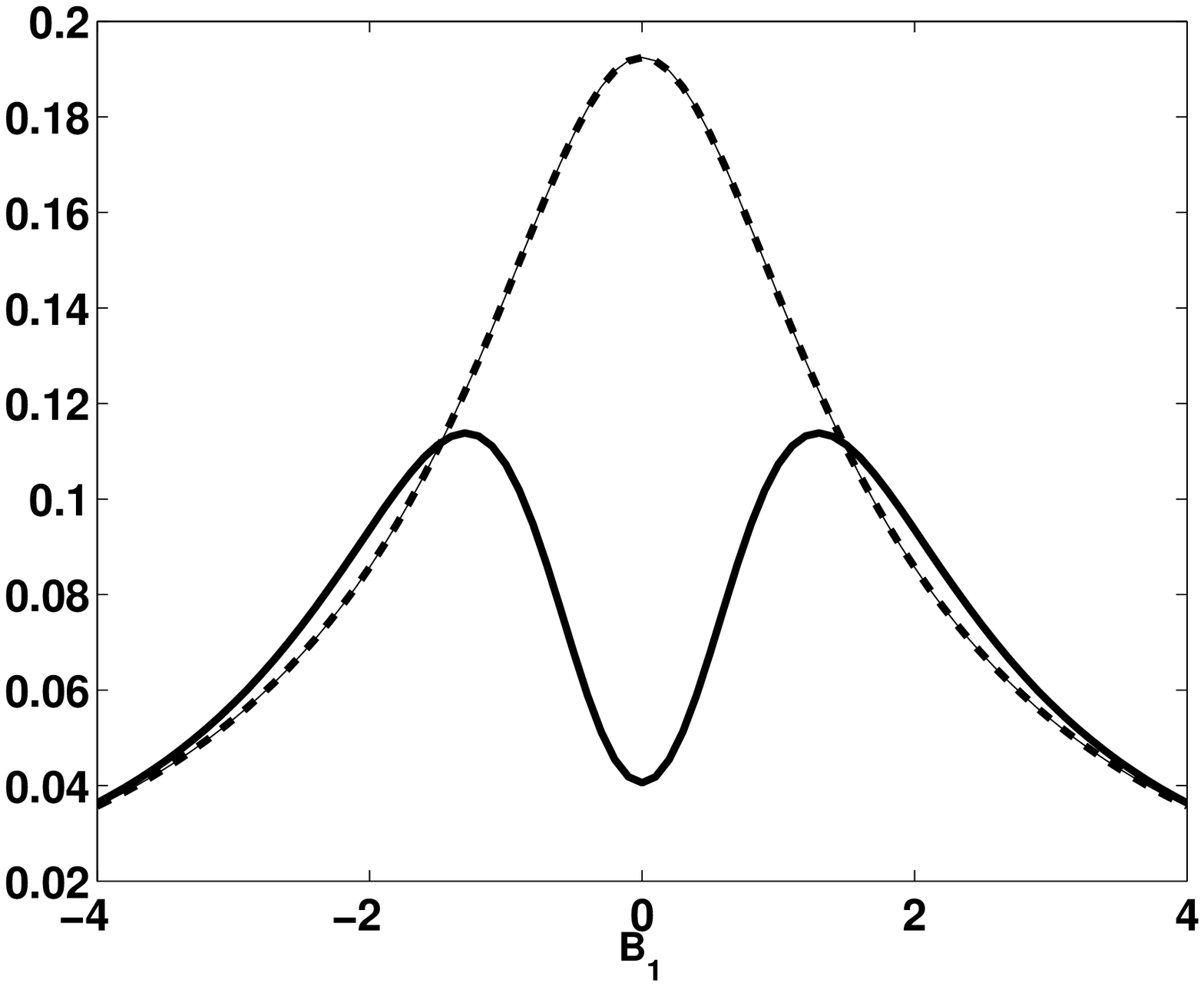}

Fig. (4.2) : $QD$ and $CC$ (dashed line) and $EN$ (solid line) as a function  of external magnetic field $B_1=-B_2$ at $T=0.9$
\end{center}
\end{figure}

\begin{figure}[!ht]
\begin{center}
\includegraphics[width=8cm,height=5cm]{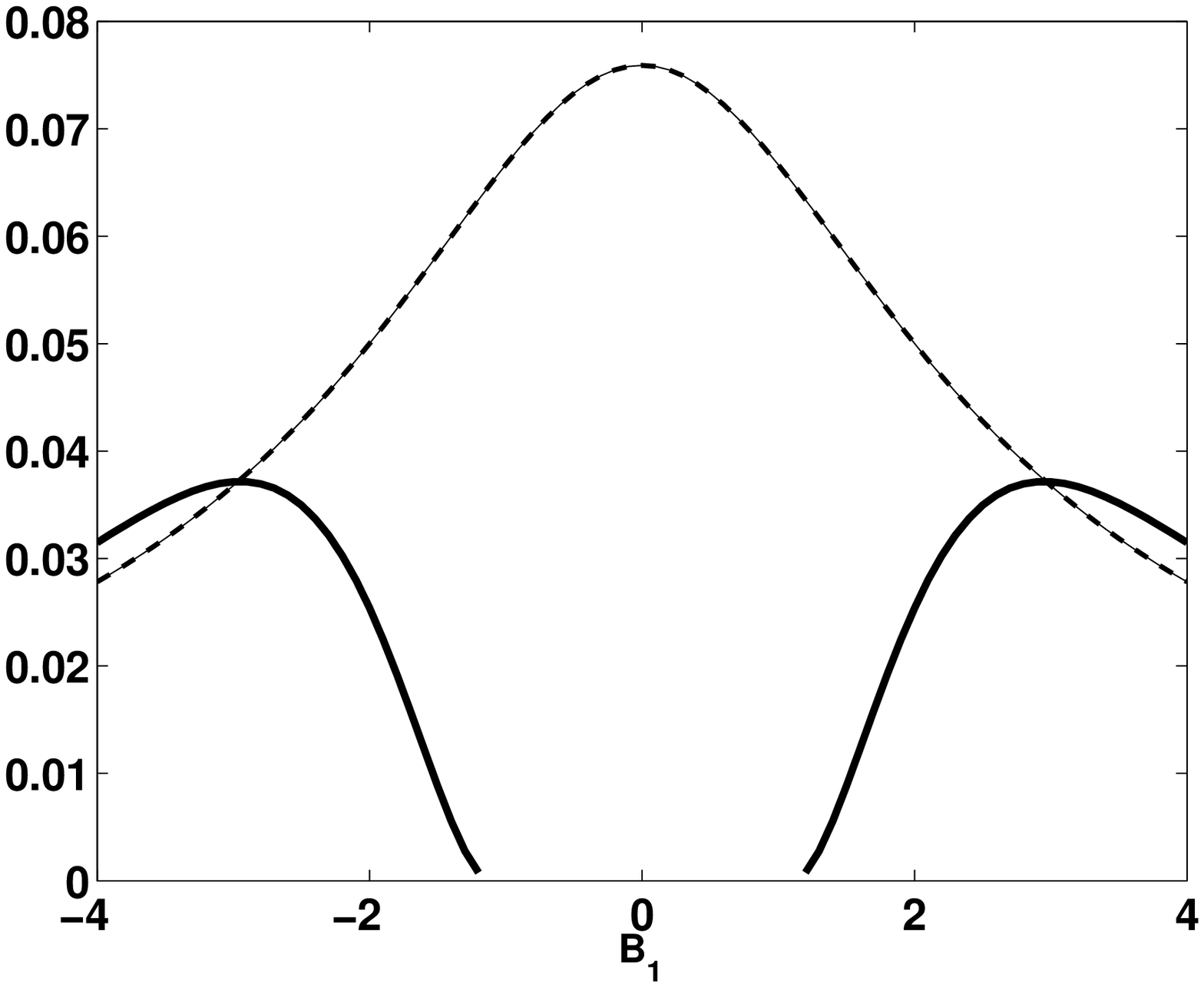}

Fig. (4.3) : $QD$ and $CC$ (dashed line) and $EN$ (solid line) as a function  of external magnetic field $B_1=-B_2$ at $T=1.5$
\end{center}
\end{figure}

From Figs. (4.1), (4.2) and (4.3),  we also see that $EN$ as a function of $B_1$ has a peak at $B_1=0$ for $T=0.2$, has a dip for $T=0.9$ and goes to zero over an interval symmetric about $B_1=0$ for $T=1.5$ \cite{scc03}. From Eq. (\ref{eqno10}) we see that the concurrence of the thermal state is governed by the admixture of the $|\psi^+\rangle$ and  $|\psi^-\rangle$ states. We expect concurrence to fall as the state  $|\psi^+\rangle$ classically mixes more and more with the ground state  $|\psi^-\rangle.$  For fixed $J=1$ and a fixed temperature T, this happens for $B_1=0$. That is why we got a dip in the $EN$ curve at $B_1=0$. The size of this dip increases with temperature. In fact the dip touches the $B_1$ axis when, at $B_1=0$ the concurrence is zero. To see this we note that the concurrence for $\rho$ in Eq. (\ref{eqno10}) is given by \cite{ow01}

\ben\label{eqno13}
C=\frac{2}{Z}\max\{|v|-\sqrt{u_1 u_2},0\}
\een
where $v, u_1, u_2$ are given in Eq. (\ref{eqno6}).
  Therefore, for $B_1=0$ and $J=1$, $C\geq0$ provided $\sinh\frac{1}{T} \geq 1$ or $T\leq 1.1346$. For $T=1.5$ and $J=1$, using the requirement $\sinh\frac{D}{T}=D$, we can find the range of $B_1$ around $B_1=0$ in which $C=0$. This is $-1.1456 \leq B_1 \leq 1.1456$. Figs. (4.1), (4.2) and (4.3) confirm the corresponding behavior of $EN.$

  Fig. (4.4) shows the variation of $EN, QD, CC$  with temperature at fixed values of $B_1=-B_2$. As expected we have $QD=CC$ for all temperatures. Both $EN$ and $QD~ (CC)$ curves have plateau at low temperatures corresponding to their ground state values, as at these temperatures, the ground state is not thermally connected to other exited states. Other interesting observation is  the vanishing of concurrence at a finite critical temperature $T_c$ which increases with $B_1$ value, while $QD$ and $CC$ asymptotically go to zero with temperature. The increase in $T_c$ with $B_1$ \cite{scc03} can be understood from the thermal state Eq. (\ref{eqno10}) which says that higher temperatures are required to get a given admixture of $|\psi^-\rangle$ and $|\psi^+\rangle$ for higher $B_1$ values.

\begin{figure}[!ht]
\begin{center}
\includegraphics[width=8cm,height=5cm]{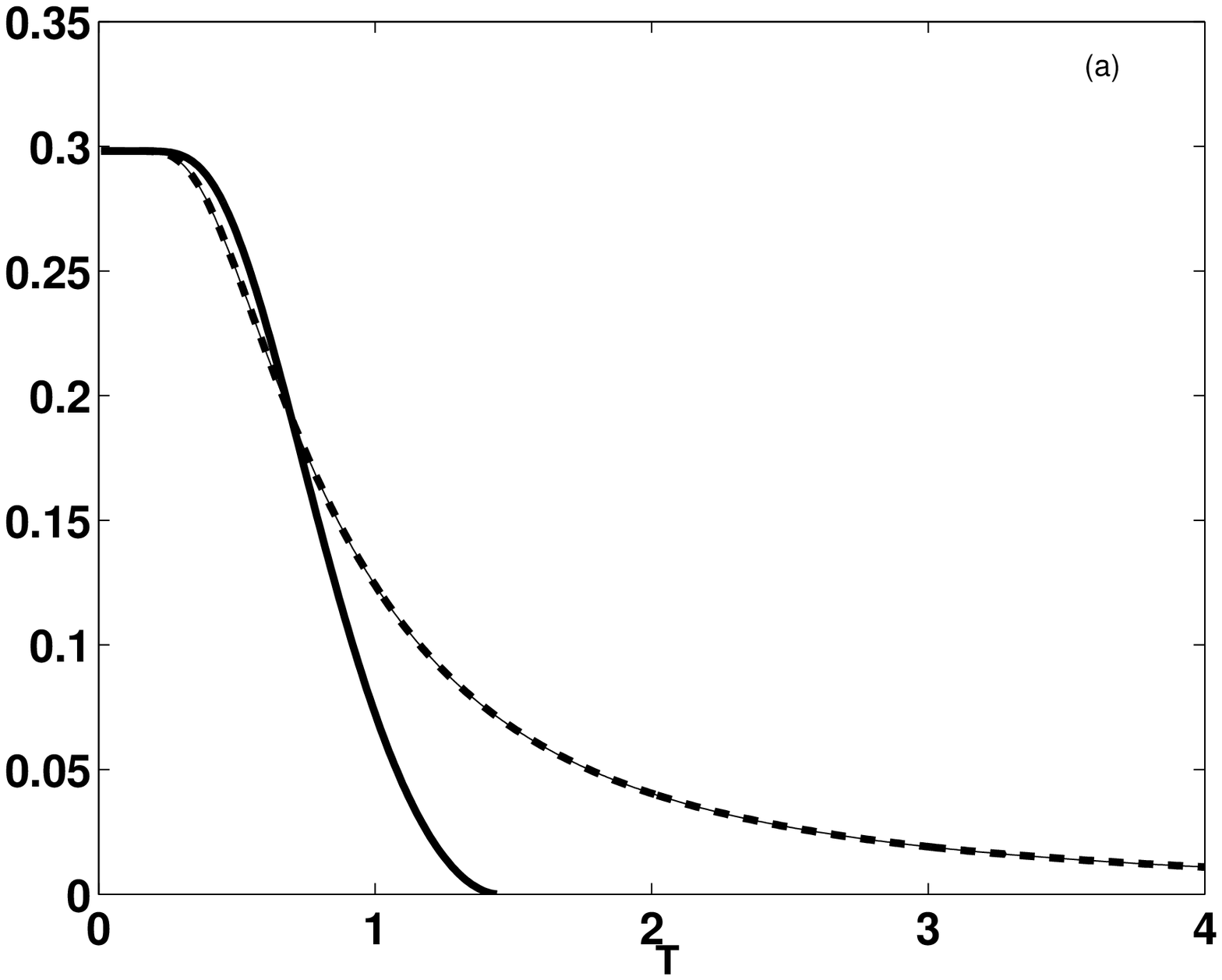}\\

\includegraphics[width=8cm,height=5cm]{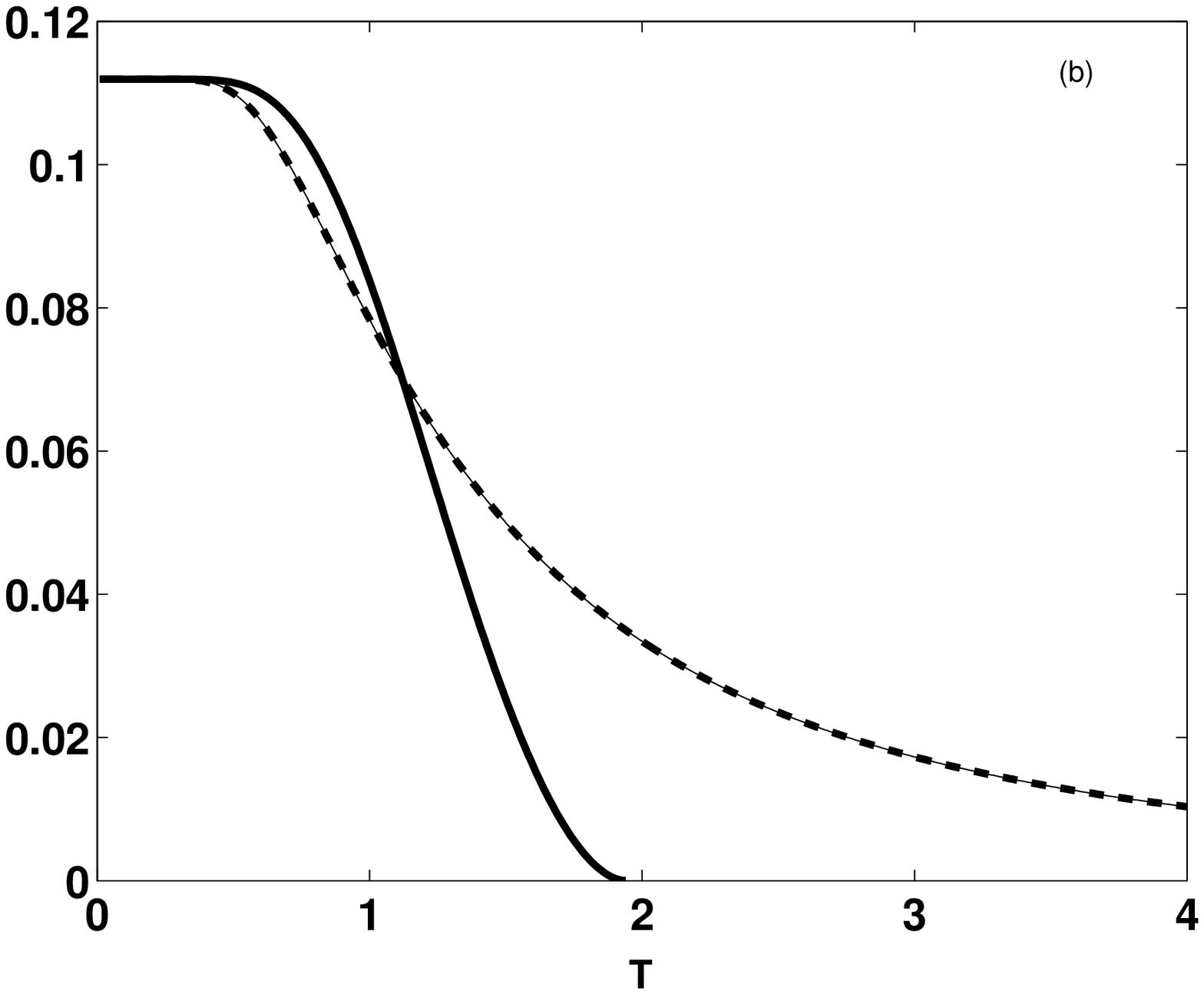}

Fig. (4.4) : $QD$ and $CC$ (dashed line) and $EN$ (solid line) as a function  of the absolute temperature $T$ for (a) $B_1=-B_2=1$
(b) $B_1=-B_2=2$
\end{center}
\end{figure}

\subsection{Case II: $B_2=-a B_1$ and $(J>0)$.}

 We now deal with the case $B_2=-a B_1$, $a\neq 1$ and positive. $B_1=0$ satisfies both, $B_2=-aB_1$ and $B_2=-B_1$, so that $QD=CC$ at $B_1=0$, for all temperatures T. For a fixed temperature T, it turns out that $QD > CC$ for $B_1 \neq 0$ if $a>1$ and $CC > QD$ for $B_1 \neq0$ if $0 < a <1$. This is depicted in Fig. (4.5), for $a=2$ and $a=1/2$ for $T=1.5$. The dominance of $QD$ over $CC$ (or vice versa) varies continuously with $a$. This observation gives us the key to control the contributions of $QD$ and $CC$ to a two qubit thermal state in Heisenberg model via the continuous variation of the applied magnetic field.
  The behavior of concurrence in this case can be analyzed in a way similar to the case $B_2=-B_1, (a=1)$. From Fig. (4.5), we see that for the same temperature, the range over which concurrence vanishes depends on $a$, this range decreases monotonically with $a$.  Also, the peak position of concurrence (or, $EN$) on the $B_1$ axis shifts monotonically towards $B_1=0$ as $a$ increases. Thus the main entanglement features can be controlled by varying external magnetic fields.

\begin{figure}[!ht]
\begin{center}
\includegraphics[width=8cm,height=5cm]{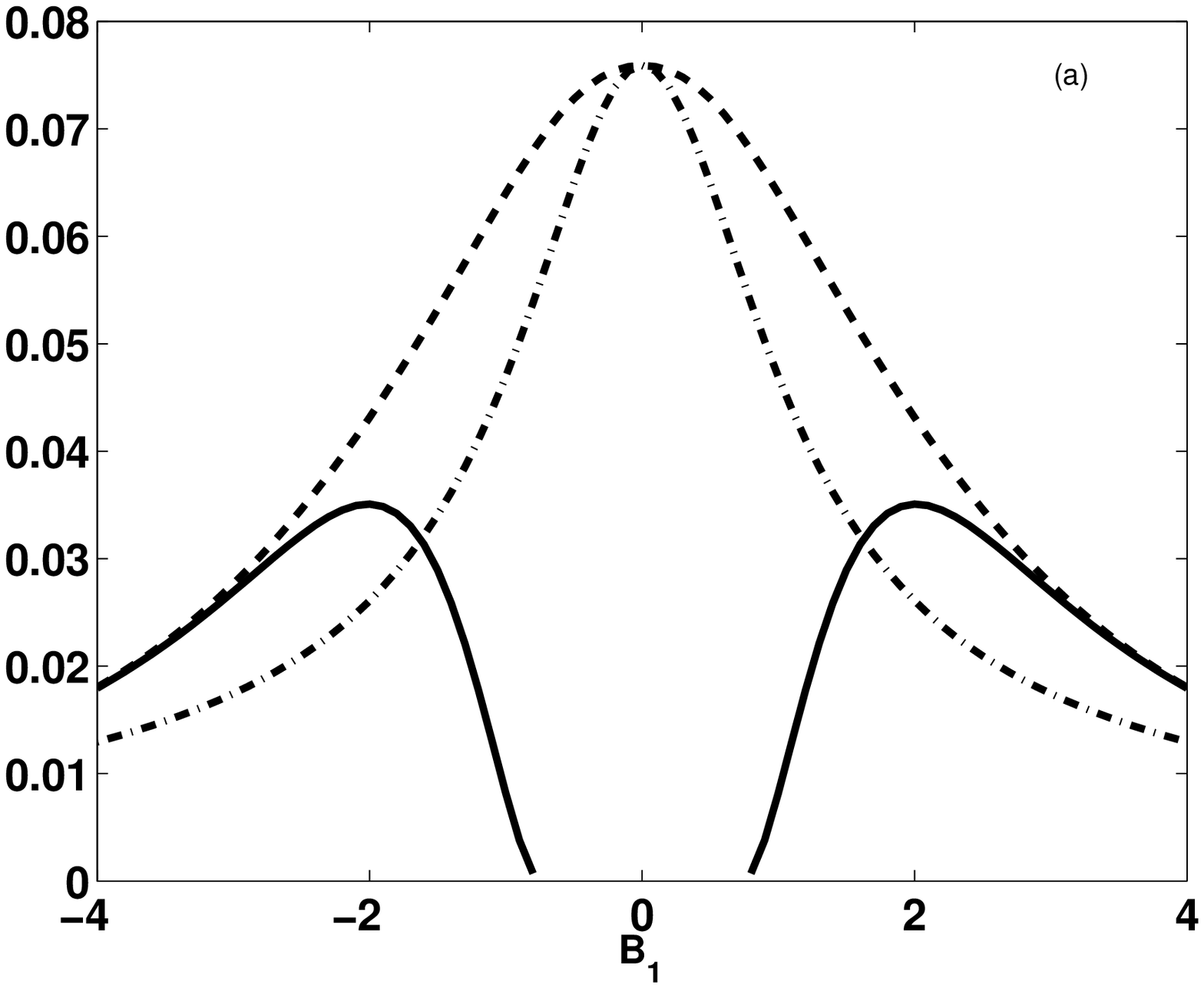}

\includegraphics[width=8cm,height=5cm]{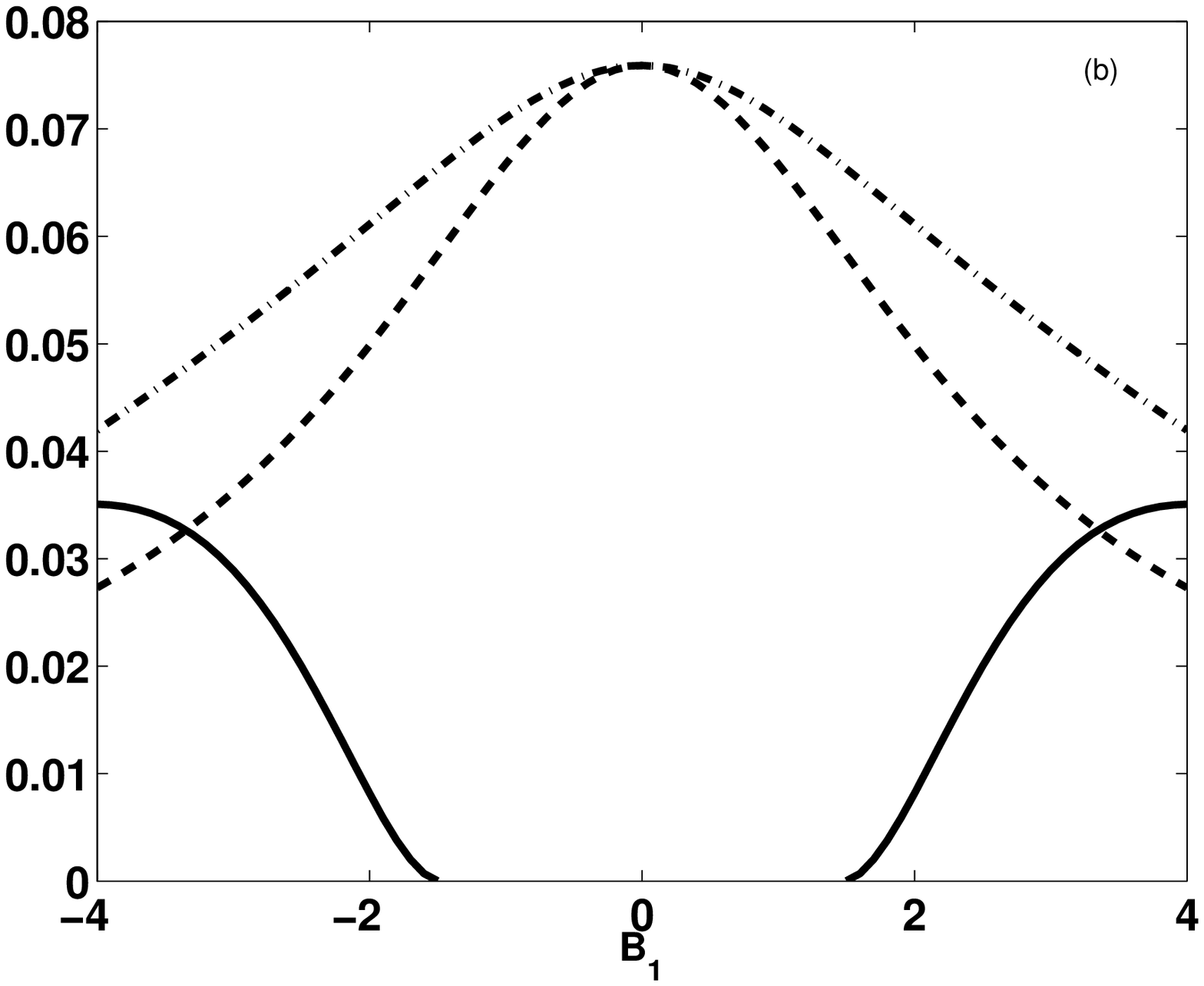}

Fig. (4.5) : $QD$ (dashed line), $CC$ (dash-dotted line) and $EN$ (solid line) as a function  of external magnetic field $B_1$ at $T=1.5$  (a) $B_2=-2B_1$
(b) $B_2=-B_1/2$.
\end{center}
\end{figure}

\subsection{Case III:  $B_1=B_2$. and $(J>0)$.}

 For uniform external magnetic field $B_1=B_2$, Figs. (4.6), (4.7) and (4.8), show the variation of $QD, CC$ and $EN$ with $B_1$ for temperatures $T=0.2, 0.9, 1.5$, respectively. We see that all three quantities are symmetric about $B_1=0$ where they have their maximum. Further, $QD > CC$, except at $B_1=0$, where $QD=CC$. For higher temperatures, the qualitative behavior of $QD$ and $CC$ remains the same, while $EN$ curve drops down below those of $QD$ and $CC.$ This can be qualitatively understood by looking at the thermal state given by
$$\rho=\frac{1}{Z}\big{[}e^{2B_1/{T}}|00\rangle\langle00|+e^{-2B_1/{T}}|11\rangle\langle11|$$
\ben\label{eqno14}
+e^{J/{T}}|\psi^-\rangle\langle\psi^-|+e^{-J/{T}}|\psi^+\rangle\langle\psi^+|\big{]}.
\een
For small temperatures, the entanglement of the thermal state is largely dictated by that of $|\psi^-\rangle$ and becomes dominant. At higher temperatures, admixture due to other states reduces the entanglement, so that $QD$ and $CC$ dominate. Such a complementary behavior of entanglement and discord can serve as a pointer to wards a possible connection between them.

\begin{figure}[!ht]
\begin{center}
\includegraphics[width=8cm,height=5cm]{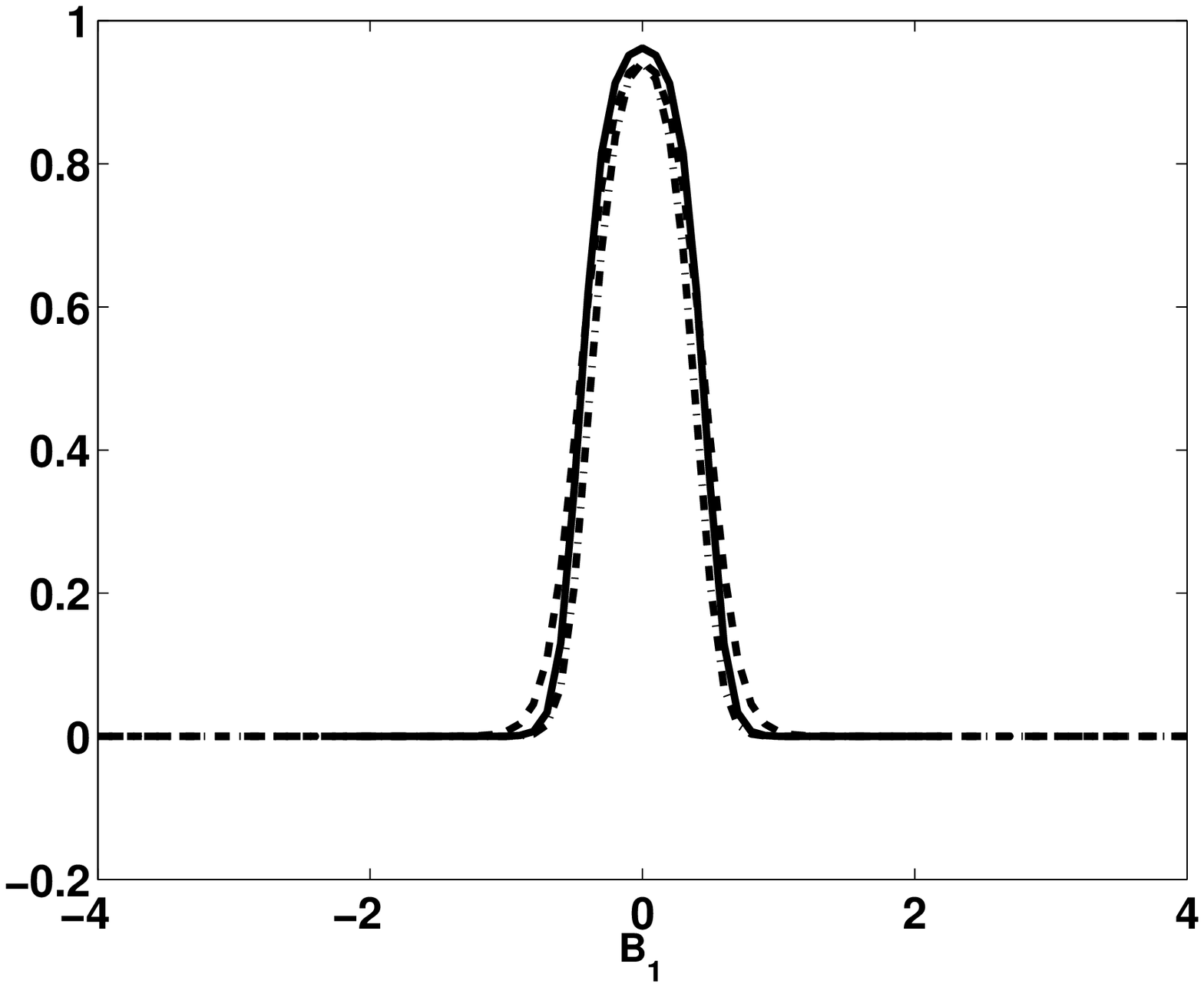}

Fig. (4.6) : $QD$ (dashed line), $CC$ (dash-dotted line) and $EN$ (solid line) as a function  of external magnetic field $B_1$ where $B_1=B_2$ at $T=0.2$
\end{center}
\end{figure}

\begin{figure}[!ht]
\begin{center}
\includegraphics[width=8cm,height=5cm]{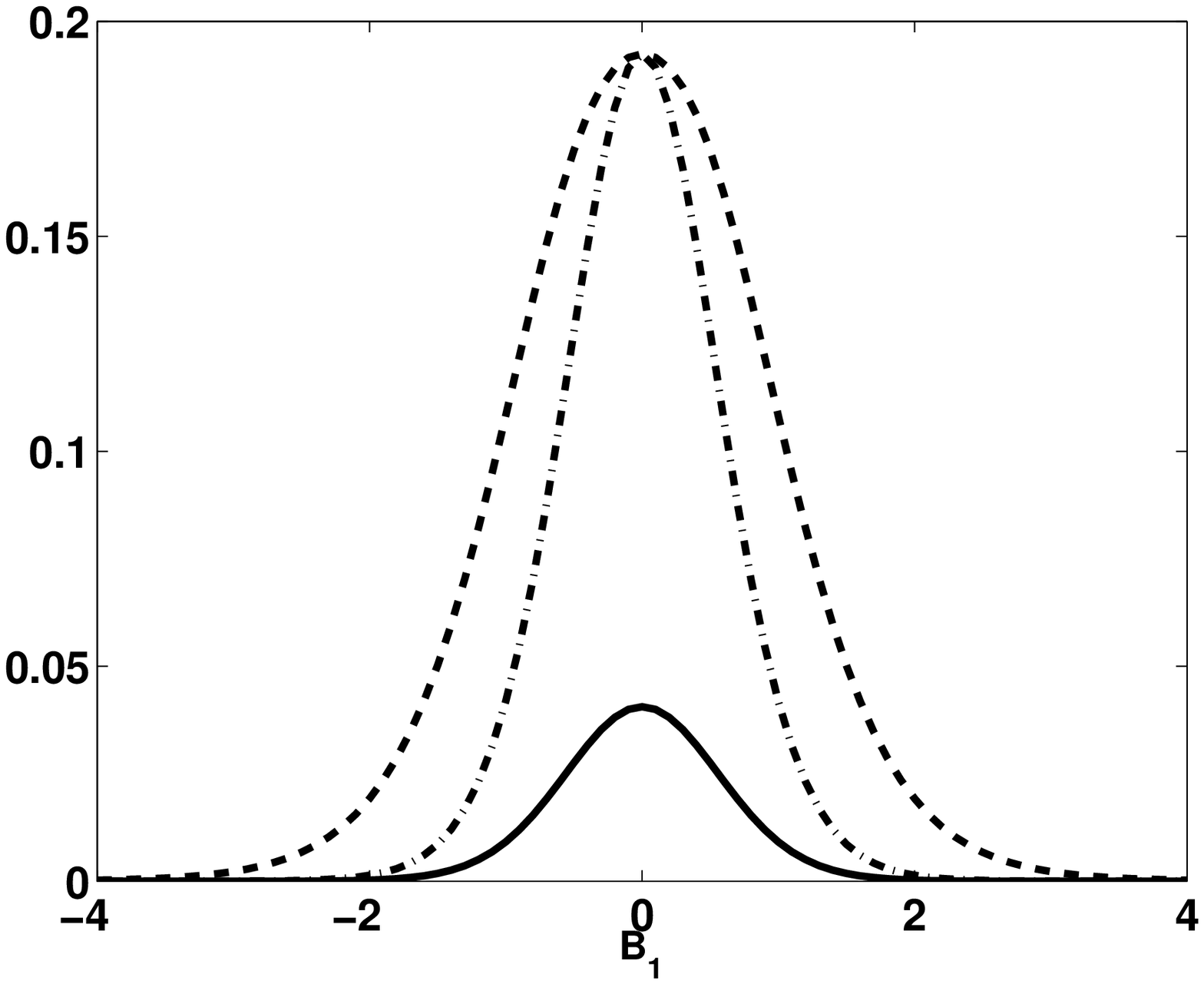}

Fig. (4.7): $QD$ (dashed line), $CC$ (dash-dotted line) and $EN$ (solid line) as a function  of external magnetic field $B_1$ where $B_1=B_2$ at $T=0.9$
\end{center}
\end{figure}

\begin{figure}[!ht]
\begin{center}
\includegraphics[width=8cm,height=5cm]{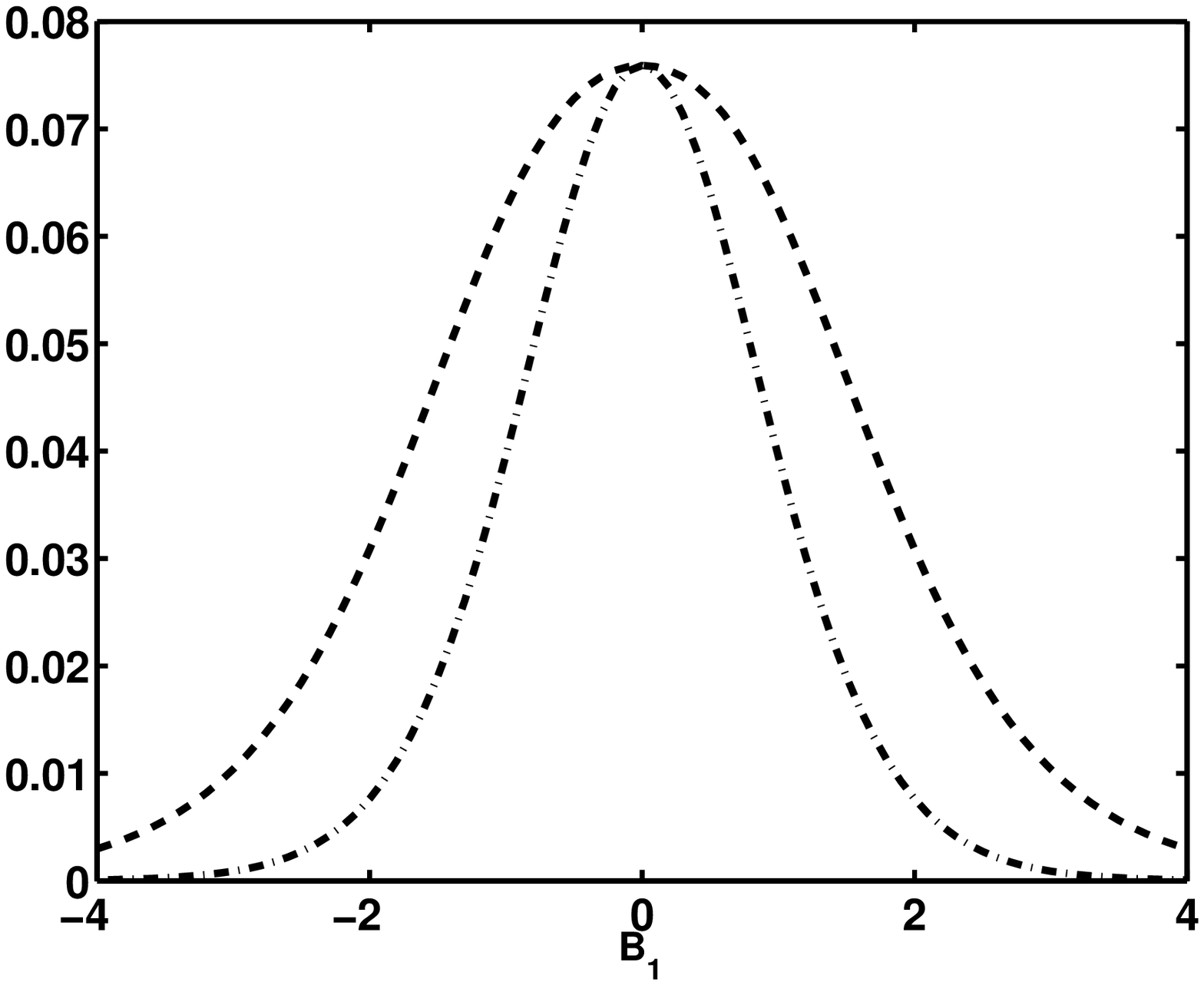}

Fig. (4.8) : $QD$ (dashed line), $CC$ (dash-dotted line) and $EN$ (solid line) as a function  of external magnetic field $B_1$ where $B_1=B_2$ at $T=1.5$
\end{center}
\end{figure}

Fig. (4.9), shows the variation of $QD, CC$ and $EN$ with temperature for $B_1=B_2=1$ and $B_1=B_2=2$. We see that for high temperatures, $QD$ hugely dominates $EN$, showing the robustness of $QD$ with temperature. As temperature becomes large $QD$ and $CC$ converge to wards each other. For larger values of $B_1$ this happens at higher temperatures. As the temperature increases, all the coefficients in the thermal mixture Eq. (\ref{eqno14}) tend to be equal and the thermal state approaches random mixture. Thus it seems that $QD$ and $CC$ approach each other as an arbitrary thermal state approaches a random mixture. Obviously, for random mixture $\rho=\frac{1}{4}(I\otimes I), ~ QD=CC=0.$ A quantitative analysis of the relative behaviors of $QD$ and $CC$ with temperature will be very interesting, but possibly have to wait for further developments in the theory.\\

\begin{figure}[!ht]
\begin{center}
\includegraphics[width=8cm,height=5cm]{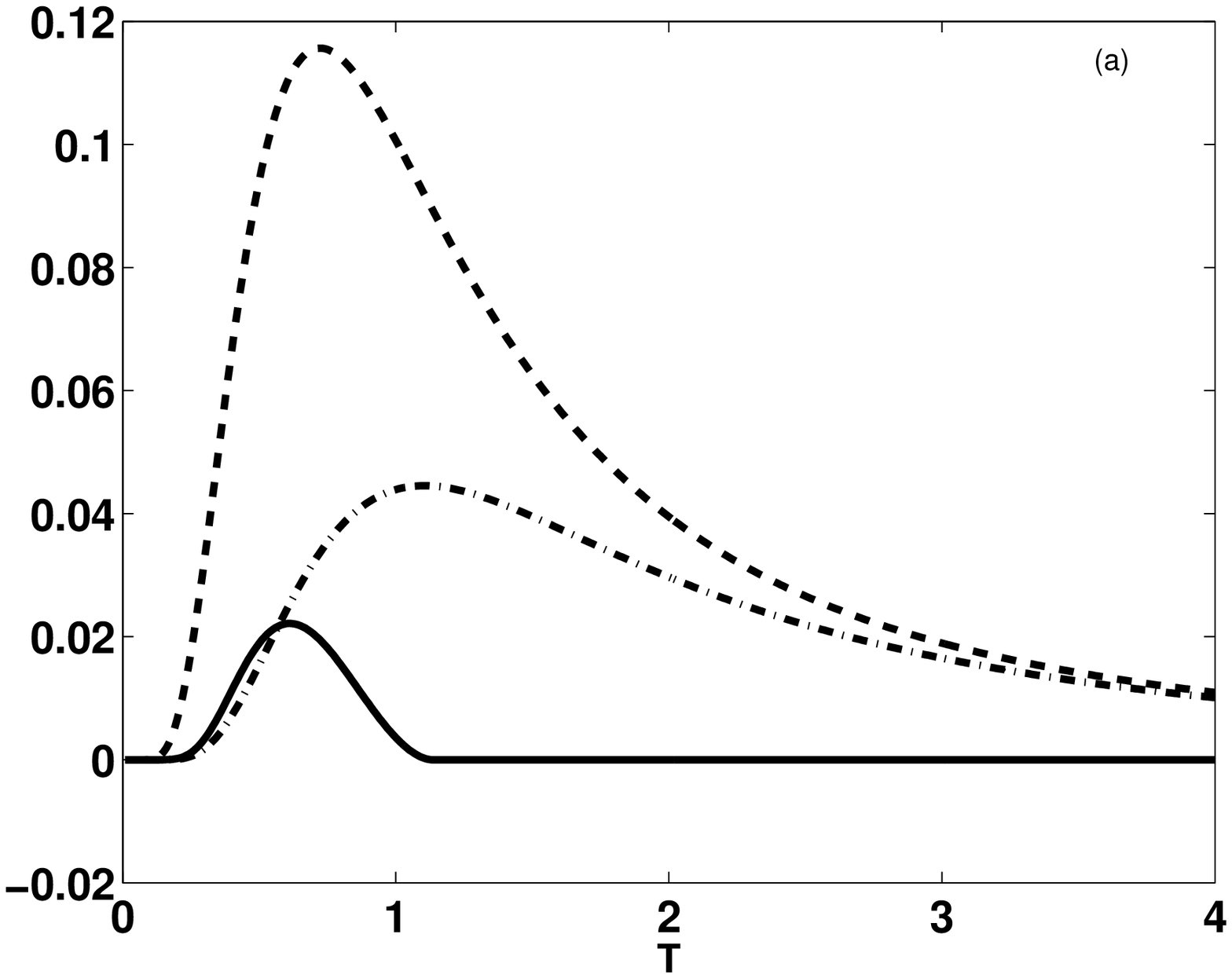}

\includegraphics[width=8cm,height=5cm]{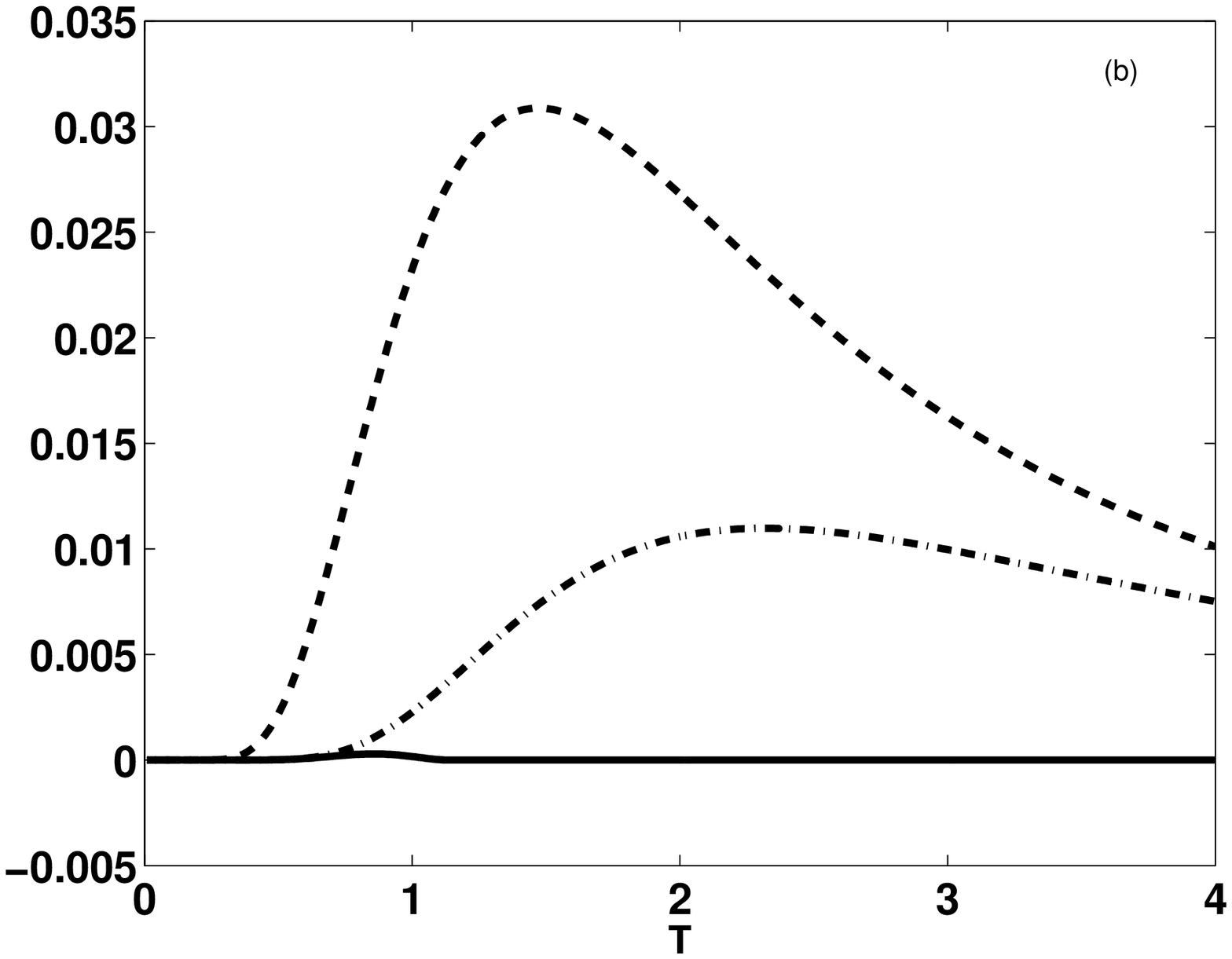}

Fig. (4.9) : $QD$ (dashed line), $CC$ (dash-dotted line) and $EN$ (solid line) as a function  of the absolute temperature $T$ for (a) $B_1=B_2=1$
(b) $B_1=B_2=2$
\end{center}
\end{figure}

\begin{figure}[!ht]
\begin{center}
\includegraphics[width=8cm,height=5cm]{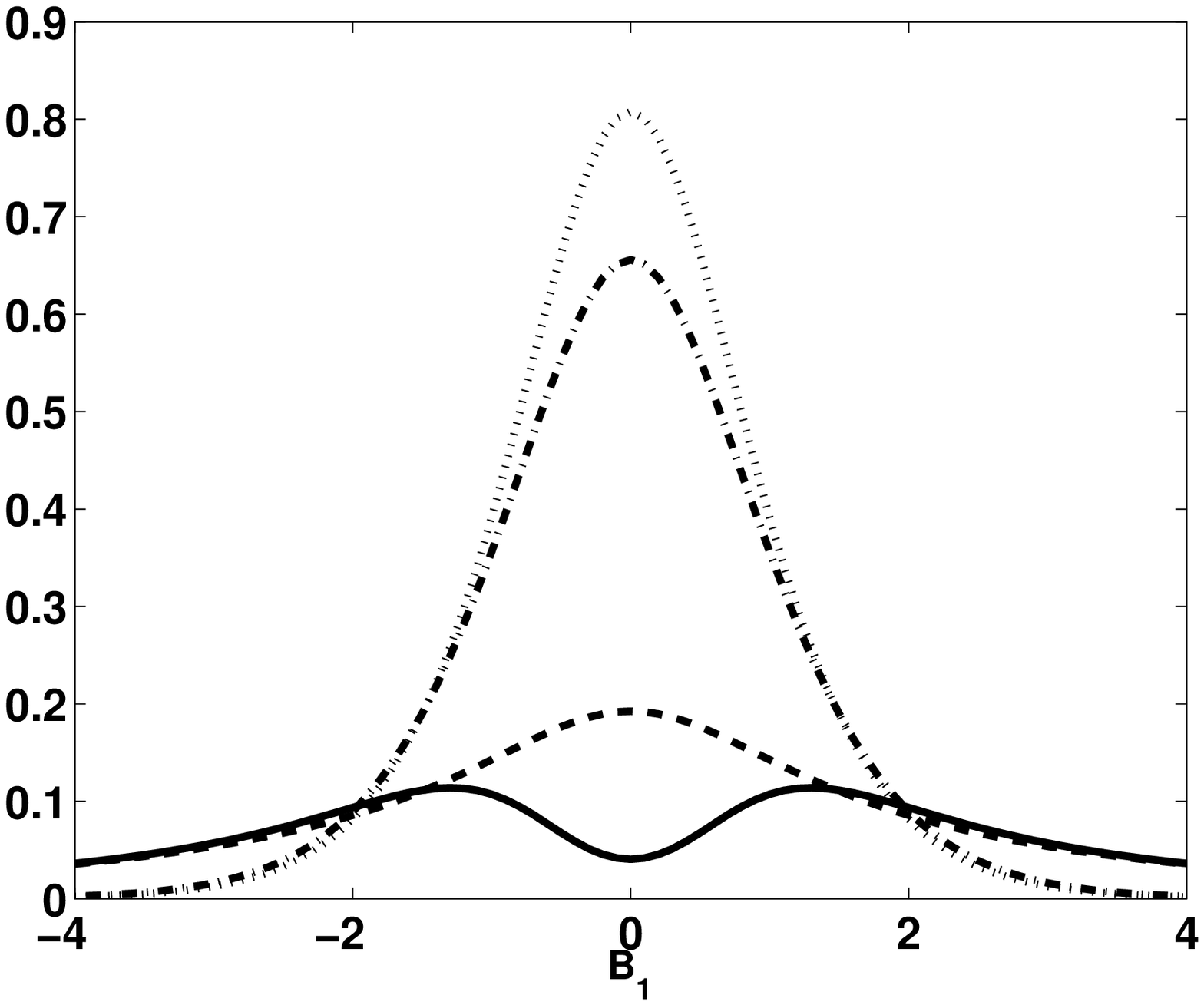}

Fig. (4.10) : $EN_{AB}$ (solid line), $QD_{\overleftarrow{AB}}$  (dashed line), $EN_{AE}$  (dot line) and $QD_{\overleftarrow{AE}}$  (dash-dotted line), as a function  of external magnetic field $B_1=-B_2$ at $T=0.9$
\end{center}
\end{figure}

\section{Monogamy relations between System and environment.}

It will be interesting to connect our results in previous section with the monogamy relations between the $EN$ and the classical correlation \cite{kw04} of two subsystems (qubits) and the environment

$$EN_{AB}+CC_{\overleftarrow{AE}}=S_A, $$

\ben\label{eqno15}
EN_{AE}+CC_{\overleftarrow{AB}}=S_A,
\een
and the relation between $EN$ and $QD$ \cite{fcoc10},

\ben\label{eqno16}
EN_{AB}+EN_{AE}=QD_{\overleftarrow{AB}}+QD_{\overleftarrow{AE}},
\een
showing us that $EN$ and $QD$ always exist is pairs. Here $A$,$B$ label the qubits and $E$ stands for the environment. We assume that environment (heat bath) comprises the universe  minus the qubits $A$ and $B$ so that the state $\rho_{ABE}$ is a pure state. Since the variation of all the quantities pertaining to the system $AB$ with $B_1$ and $T$ are obtained form the $XX$ model, we can use Eqs. ((\ref{eqno15}), (\ref{eqno16})) to find the corresponding dependence of $EN_{AE}$ and $QD_{\overleftarrow{AE}}$ on $B_1$ and $T$. Figs. (4.10) and (4.11), (for $B_1=-B_2$) show the variation of $EN_{AE}$ and $QD_{\overleftarrow{AE}}$ with $B_1$ and $T$.\\

The monogamic relations also help us establish a necessary and sufficient condition for $QD_{\overleftarrow{AB}}=CC_{\overleftarrow{AB}}$ when the environment is present.
This is : $QD_{\overleftarrow{AB}}=CC_{\overleftarrow{AB}}$ if and only if $\frac{1}{2}I_{AB}=EN_{AE}+EN_{AB}-QD_{\overleftarrow{AE}}.$ To prove the necessity we note that when $QD_{\overleftarrow{AB}}=CC_{\overleftarrow{AB}},$ (that is, $QD_{\overleftarrow{AB}}=\frac{1}{2}I_{AB}$), Eq. (\ref{eqno16}) can be written as

\ben\label{eqno17}
\frac{1}{2}I_{AB}=EN_{AE}+EN_{AB}-QD_{\overleftarrow{AE}}.
\een
Now suppose Eq. (\ref{eqno17}) is true. Then using Eq. (\ref{eqno16}) we have $$QD_{\overleftarrow{AB}}=\frac{1}{2}I_{AB}$$ which implies $QD_{\overleftarrow{AB}}=CC_{\overleftarrow{AB}}.$ Figs. (4.12) and (4.13), show the variation of both sides of Eq. (\ref{eqno17}) with $B_1$ and $T$ which establishes Eq. (\ref{eqno17}) for the XX model.

\begin{figure}[!ht]
\begin{center}
\includegraphics[width=8cm,height=5cm]{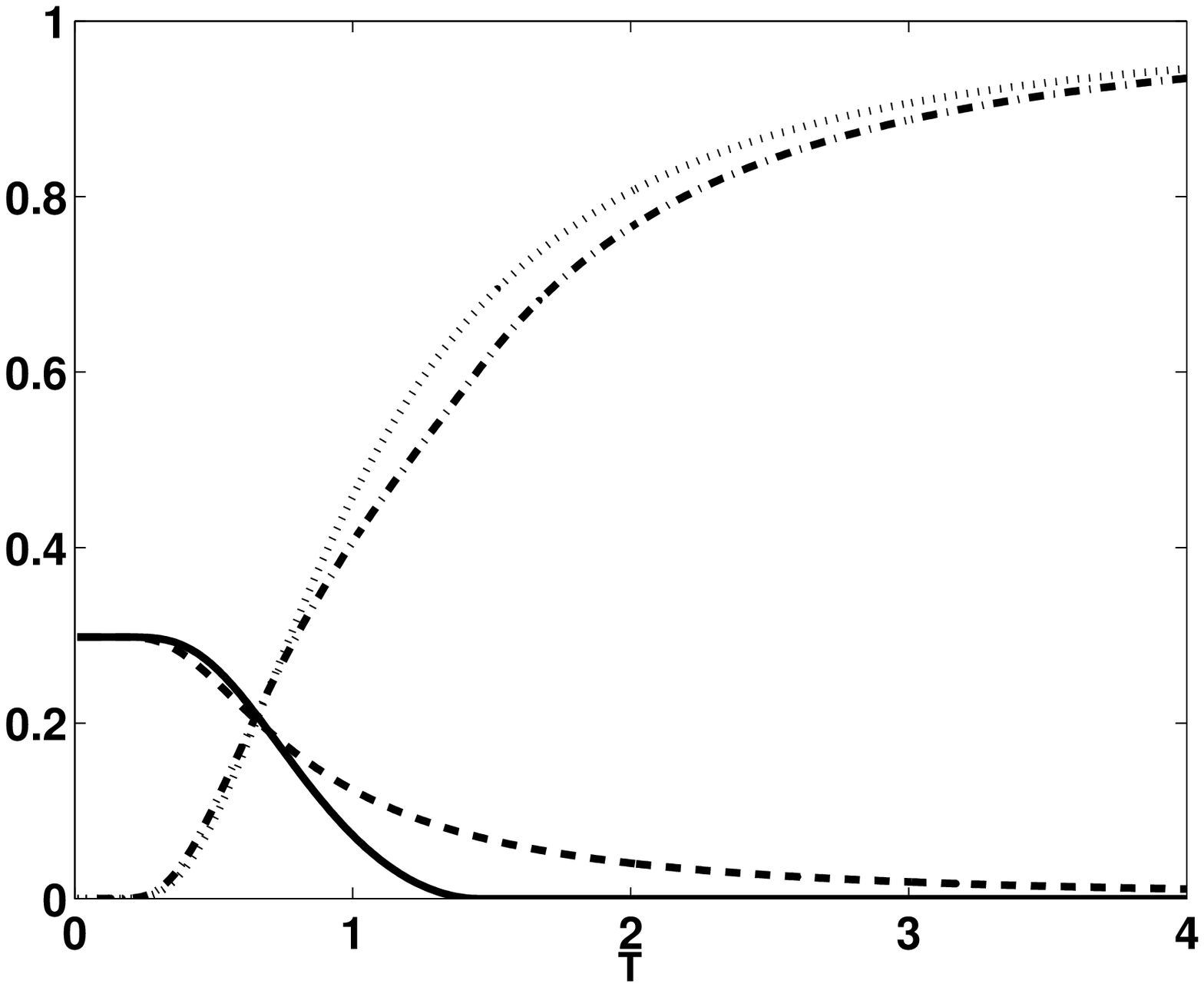}

Fig. (4.11) : $EN_{AB}$ (solid line), $QD_{\overleftarrow{AB}}$  (dashed line), $EN_{AE}$  (dot line) and $QD_{\overleftarrow{AE}}$  (dash-dotted line), as a function  of the absolute temperature $T$ for  $B_1=-B_2=1.$

\end{center}
\end{figure}

\begin{figure}[!ht]
\begin{center}

\includegraphics[width=8cm,height=5cm]{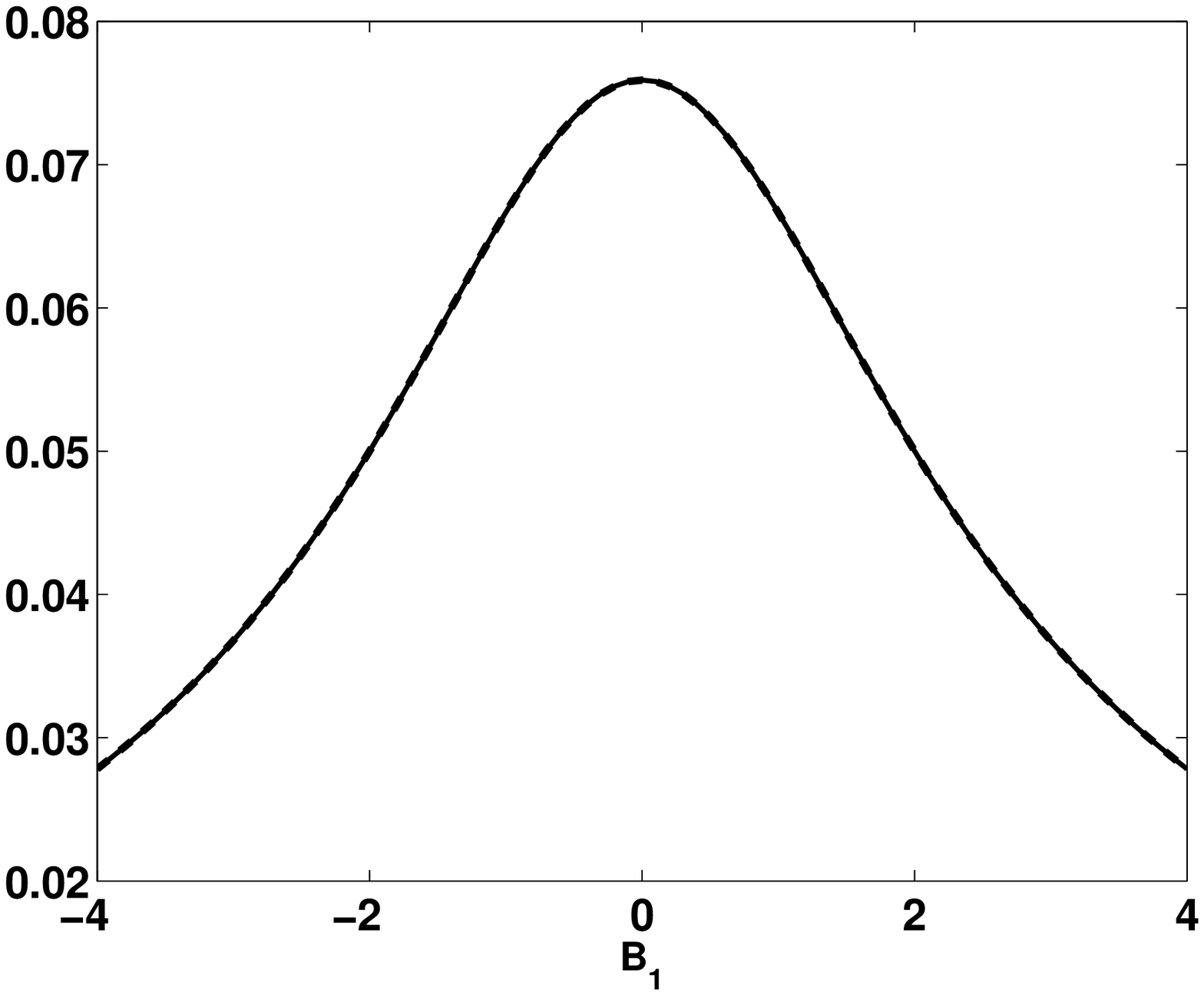}

  Fig. (4.12) : $\frac{1}{2}I_{AB}$ (solid line), $EN_{AE}+EN_{AB}-QD_{\overleftarrow{AE}}$  (dashed line), as a function of external magnetic field $B_2=-B_1$ at $T=1.5$

\end{center}
\end{figure}

\begin{figure}[!ht]
\begin{center}

\includegraphics[width=8cm,height=5cm]{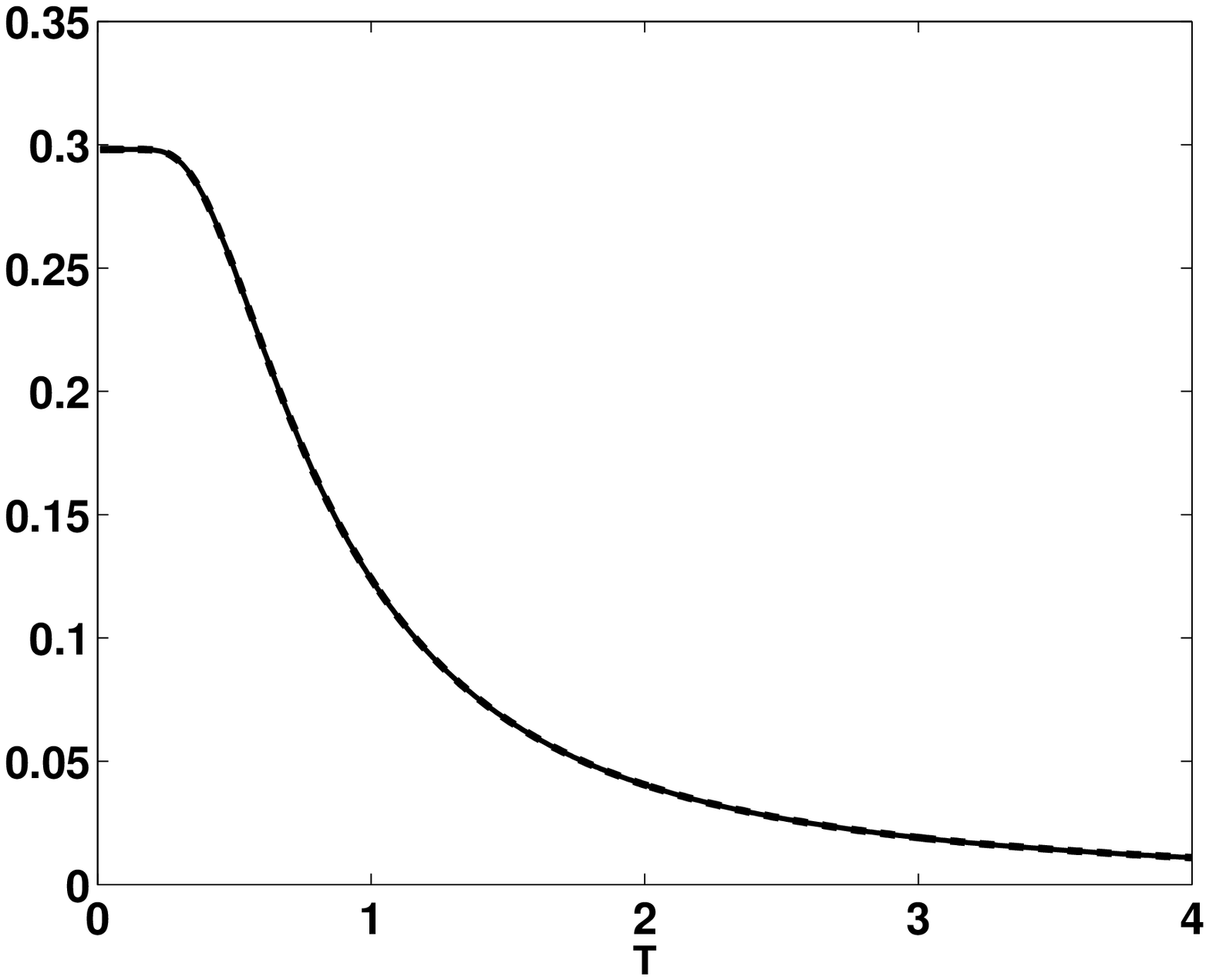}

Fig. (4.13) : $\frac{1}{2}I_{AB}$ (solid line), $EN_{AE}+EN_{AB}-QD_{\overleftarrow{AE}}$  (dashed line), as a function  of the temperature $T$ for $B_2=-B_1=-1.$

\end{center}
\end{figure}
\
\\
\
\\
\
\\
\
\\

\section{Summary and comments}

  In this chapter, we have studied the variation of $QD, CC$ and $EN$ in two qubit XX Heisenberg chain as functions of independently varied magnetic fields $B_1$ and $B_2$ on each qubit and also with temperature. We deal with two cases $B_1= - B_2$ (nonuniform field) and $B_1=B_2$ (uniform field). Our first observation is the complementary behavior of entanglement and $QD/CC$. For the nonuniform magnetic field, we get the interesting observation that $QD$ and $CC$ are equal for all $B_1= - B_2$ values as well as all temperatures. Surprisingly, this observation is explained quite simply, using the symmetric form of the thermal state. A very interesting observation is that the relative contributions of $QD$ and $CC$ can be tunably controlled by varying the applied magnetic field. Another interesting finding is that the equality of $QD$ and $CC$ of the subsystem (qubits) imposes a constraint on the distribution of $QD$ and $EN$ over the subsystem and its environment. Further investigation of general Heisenberg models like XXZ along these lines may turn out to be interesting and fruitful.\\

\large{\it Appendix: }

$Theorem$: If  the quantum state has  the Bloch representation \cite{hj08}
$$\rho=\frac{1}{4}[I\otimes I+\sum_{i=1}^3 c_i \sigma_i\otimes \sigma_i], $$
and $c_i=c_j>c_k$ and $c_k=-c_i^2$ where $i\neq j \neq k \in \{1,2,3\}$, then this state contains the same amount of quantum and classical correlation $(QD=CC)$.\\

\emph{Proof}: In Ref. \cite{lu08}  S. Luo evaluated analytically the quantum discord  for a large family of two-qubit states, which have the maximally mixed marginal and their Bloch representation is

  $$\rho=\frac{1}{4}[I\otimes I+\sum_{i=1}^3 c_i \sigma_i\otimes \sigma_i].$$

For this class of quantum states the quantum mutual information is\\
  $ \mathcal{I}(\rho)=\frac{1}{4}[(1-c_1-c_2-c_3) log_2(1-c_1-c_2-c_3)+(1-c_1+c_2+c_3) log_2(1-c_1+c_2+c_3)+(1+c_1-c_2+c_3) log_2(1+c_1-c_2+c_3)+(1+c_1+c_2-c_3) log_2(1+c_1+c_2-c_3)].$
We substitute the conditions above in the quantum mutual information. Puting $c=c_1=c_2 >c_3$ and $c_3=-c^2$, we get,\\
  $ \mathcal{I}(\rho)=\frac{1}{4}[(1-2c+c^2) log_2(1-2c+c^2)+(1-c^2) log_2(1-c^2)+(1-c^2) log_2(1-c^2)+(1+2c+c^2) log_2(1+2c+c^2)]$
After some algebraic simplification, we get\\
  $ \mathcal{I}(\rho)=(1-c) log_2(1-c)+(1+c) log_2(1+c)$\\
which equals $2 CC$ as in Ref. \cite{lu08}.
It is also easy to check that the above argument goes through when  $c=c_1=c_3 >c_2$ and $c_2=-c^2$ and when $c=c_3=c_2 >c_1$ and $c_1=-c^2$,
to get $ \mathcal{I}(\rho)=2CC$.
Thus, $$QD(\rho)=\mathcal{I}(\rho)-CC(\rho)=CC(\rho).$$

\chapter{Geometric measure of quantum discord for an arbitrary state of a bipartite quantum system.}

\begin{figure}[!ht]
\begin{center}
\includegraphics[width=8cm,height=1.25cm]{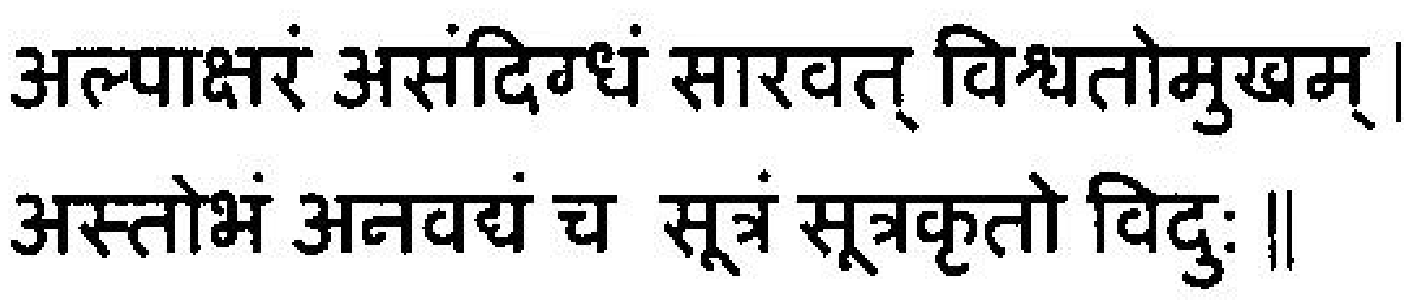}

Author unknown\\
\scriptsize\textsc{ A statement which has an essential message, which is short, simple and unambiguous in expression while generic and universal in application, is called a principle or a formula. }
\end{center}
\end{figure}

\section{Introduction}

  Quantum discord, (Olliver and Zurek \cite{oz01}), is a measure of the discrepancy between quantum versions of two classically equivalent expressions for mutual information and is found to be useful in quantification and application of quantum correlations in mixed states. It is viewed as a key resource present in certain quantum communication tasks and quantum computational models without containing much entanglement. A step toward the quantification of quantum discord in a quantum state was by Dakic, Vedral, and Brukner \cite{dvb10} who introduced a geometric measure of quantum discord and derived an explicit formula for any two-qubit state. Recently, Luo and Fu \cite{lf10} introduced a generic form of the geometric measure of quantum discord. Using these results we find an exact formula for the geometric measure of quantum discord for an arbitrary state of a $m\times n$ bipartite quantum system. This geometric measure is experimentally accessible.

Correlations in quantum states, with fundamental applications and implications for quantum information processing, are usually studied in the entanglement-versus-separability framework \cite{w89,hhhh09}. However, entanglement, while widely regarded as nonlocal quantum correlations, is not the only kind of correlation. An alternative classification for correlations based on quantum measurements has arisen in recent years and also plays an important role in quantum information theory \cite{phh08,lu08,lz08,ll08}. This is the quantum-versus-classical paradigm for correlations. In particular, the quantum discord as a measure of quantum correlations, initially introduced by Ollivier and Zurek \cite{oz01} and by Henderson and Vedral \cite{hv01}, is attracting increasing interest [143-164]. Recall that the quantum discord of a bipartite state $\rho$ on a system $H^a\otimes H^b$ with marginals $\rho^a$ and $\rho^b$ can be expressed as
 
\ben  \l{e1}Q(\rho) = \min_{\Pi^a}
\{I(\rho)-I(\Pi^a(\rho))\}.
\een
Here the minimum is over von Neumann measurements (one dimensional
orthogonal projectors summing to the identity) $\mf{\Pi}^a$ $=\{\Pi_k^a\}$ on subsystem $a$, and

\[\Pi^a(\rho) = \sum_{k} (\Pi_k^a\otimes I^b)\rho (\Pi_k^a\otimes I^b)\]
is the resulting state after the measurement. $I (\rho) = S(\rho^a) +
S(\rho^b)- S(\rho)$ is the quantum mutual information, $S(\rho) =
-tr\rho\ln\rho$ is the von Neumann entropy, and $I^b$ is the identity
operator on $H^b$.  Intuitively, quantum discord
may thus be interpreted as the minimal loss of correlations (as
measured by the quantum mutual information) due to measurement.
This formulation of quantum discord is equivalent
to the original definition of quantum discord by Ollivier and
Zurek \cite{oz01}. Recently, Dakic et al. introduced the following
geometric measure of quantum discord \cite{dvb10}:
\ben  \l{e2}
D(\rho)=\min_{\chi}||\rho-\chi||^2,
\een
where the minimum is over the set of zero-discord states [i.e.,
$Q(\chi) = 0$] and the geometric quantity
$$||\rho-\chi||^2 := tr(\rho-\chi)^2$$
is the square of Hilbert-Schmidt norm of Hermitian operators. A state $\chi$ on $H^a\otimes H^b$ is of zero discord if and only if it is a classical-quantum state \cite{kl98,oz01}, which can be represented as
\ben  \l{e3}
\chi = \sum_{k=1}^{m}p_k |k\ran\lan k|\otimes \rho_k ,
\een
where $\{p_k\}$ is a probability distribution, $\{|k\ran\}$ is an arbitrary orthonormal basis for $H^a$ and $\rho_k$ is a set of arbitrary states (density operators) on $H^b .$

Dakic et al \cite{dvb10} obtained an easily computable exact expression for the geometric measure of quantum discord for a two qubit system, which can be described as follows. Consider a two-qubit state $\rho$ expressed in its Bloch representation as (see below)
\begin{IEEEeqnarray}{rCl}
\rho & = & \frac{1}{4}\big{(} I^a\otimes I^b+\sum_{i=1}^3(x_i \sigma_i \otimes I^b+I^a \otimes y_i \sigma_i ) \nn \\
  && +\sum_{i,j=1}^3 t_{ij} \sigma_i \otimes \sigma_j \big{)},\nn
\end{IEEEeqnarray}
$\{\sigma_i\}$ being the Pauli spin matrices. Then its geometric measure
of quantum discord is given by \cite{dvb10}
\ben   \l{eq3}
D(\rho) = \frac{1}{4} (||x||^2 + ||T ||^2 - \lambda_{max}).
\een
Here $\vec{x} := (x_1,x_2,x_3)^t$ and $\vec{y} := (y_1,y_2,y_3)^t $ are coherent (column) vectors,
$T = (t_{ij} )$ is the correlation matrix, and $\lambda_{max}$ is the largest eigenvalue of the
matrix $\vec{x}\vec{x}^t + T T^t .$ The norms of vectors and matrices are the Euclidean norms, for example, $||x||^2 :=\sum_i x^2_i .$ Here and throughout this article, the superscript $t$ denotes transpose of vectors or matrices.

In order to obtain the exact formula for the quantum discord in an arbitrary bi-partite state, we set up the following scenario. Consider a bipartite system $H^a \otimes H^b$ with dim $H^a = m$ and dim $H^b = n$. Let $L(H^a)$ be the space consisting of all
linear operators on $H^a$. This is a Hilbert space with the Hilbert-Schmidt inner product
$$\langle X|Y \rangle = tr X^{\dagger} Y.$$
The Hilbert spaces $L(H^b)$ and $L(H^a\otimes H^b)$ are defined similarly.
Let $\{X_i : i = 1,2,\cdots,m^2\}$ and $\{Y_j : j = 1,2,\cdots,n^2\}$
be sets of Hermitian operators which constitute orthonormal bases for $L(H^a)$ and $L(H^b)$ respectively. Then
$$trX_i X_{i^{\prime}} = \delta_{ii^{\prime}} ,\;\; trY_jY_{j^{\prime}} = \delta_{jj^{\prime}}.$$
$\{X_i \otimes Y_j\}$ constitutes an orthonormal (product) basis for $L(H^a \otimes H^b)$ (linear operators on $H^a \otimes H^b$). In particular, any bipartite state $\rho$ on $H^a \otimes H^b$ can be expanded as
\ben   \l{e4}
\rho = \sum_{ij} c_{ij} X_i \otimes Y_j,
\een
with $c_{ij} = tr(\rho X_i \otimes Y_j).$

Quite recently, S. Luo and S. Fu introduced the following form of geometric measure of quantum discord \cite{lf10}
\ben   \l{e5}
D(\rho) = tr(CC^t )- \max_{A} tr(ACC^tA^t ),
\een
where $C = [c_{ij} ]$ (Eq.(\r{e4})) is an $m^2\ti n^2$ matrix and the maximum is taken over all $m\times m^2$-dimensional isometric (see below) matrices $A = [a_{ki} ]$ such that

\begin{IEEEeqnarray}{rCl} \l{e6}
a_{ki}  = tr(|k\rangle\langle k|X_i)=\lan k|X_i|k\ran,\;\; k = 1,2,\ldots,m\;;i = 1,2,\ldots,m^2   \nn  \\
\end{IEEEeqnarray}
and $\{|k\rangle\}$ is any orthonormal
basis in $H^a .$ We can expand the operator $|k\rangle\langle k|$ in the basis $\{X_i \}$ as

\begin{IEEEeqnarray}{rCl} \l{e7a}
|k\rangle\langle k| & = & \sum_i a_{ki} X_i,\;\; k = 1,2,\ldots,m .
\end{IEEEeqnarray}
$A=[a_{ki}]$ is an isometry in the sense that $AA^t=I^a$ and the row vectors $\vec{a}_k$ of the matrix $A$ satisfy
\ben  \l{e7}
||\vec{a}_k||^2=\sum_{i=1}^{m^2} a_{ki}^2=1.
\een
Further, using their definition it immediately follows that
\ben   \l{e8}
\sum_{k=1}^{m}a_{ki}=trX_i .
\een

We need the structure of the Bloch representation of density operators, which can be briefly described as follows \cite{mw95,bk03,hjq08}. Bloch representation of a density operator acting on the Hilbert space of a $d$-level quantum system $\mathbb{C}^d$ is given by
\ben  \l{e9}
 \rho = \frac{1}{d} (I_d + \sum_i s_i {\tilde{\lambda}}_i),
 \een
where the components of the coherent vector $\vec{s},$ defined via Eq.(\r{e9}), are given by $s_i=\frac{d}{2}tr(\rho \tilde{\lambda}_i).$
 Eq.(\r{e9}) is the expansion of $\rho$ in the Hilbert-Schmidt basis $\{I_d,\tilde{\lambda}_i; i=1,2,\dots,d^2-1\}$ where $\tilde{\lambda}_i$ are the traceless orthogonal hermitian generators of $SU(d)$ satisfying $tr(\tilde{\lambda}_i \tilde{\lambda}_j)=2\delta_{ij}$  \cite{mw95}. These generators are characterized by structure constants $f_{ijk}$ (completely antisymmetric tensor) and $g_{ijk}$ (completely symmetric tensor) of Lie algebra $su(d)$ via
\ben  \l{e01}
\tilde{\lambda}_i \tilde{\lambda}_j = \frac{2}{d}\del_{ij}I_{d} + i f_{ijk}\tilde{\lambda}_k + g_{ijk}\tilde{\lambda}_k
\een

A systematic construction of the generators of $SU(d)$ is known \cite{kim03} and are given by
\ben \l{e02}
\{\tilde{\lambda}_i\}_{i=1}^{d^2-1}=\{{\h{u}}_{jk},{\h{v}}_{jk},{\h{\om}}_{l}\},
\een
where
\benr \nn   \l{e03}
{\h{u}}_{jk}&=&|j\ran\lan k|+|k\ran\lan j|, {\h{v}}_{jk}=-i (|j\ran\lan k|-|k\ran\lan j|)\;\;1\leq j\leq k \leq d   \nn   \\
{\h{\om}}_{l}&=&\sqrt{\frac{2}{l(l+1)}} \sum_{j=1}^{l}(|j\ran\lan j|-l|l+1\ran\lan l+1|)\;\;1\leq l \leq d-1   \nn  \\
\eenr
with $\{|s\ran\}_{s=1}^{d}$ being some complete orthonormal basis in $H_{d}.$ In what follows, we take the generators of $SU(d)$ as defined via Eqs.(\r{e02},\r{e03}). 

We can represent the density operators acting on a bipartite system $H^a \otimes H^b$ with $dim (H^a) = m$
and $dim (H^b) = n$,  as

\begin{IEEEeqnarray}{rCl}  \l{e10}
\rho & = & \frac{1}{mn}(I_m\otimes I_n + \sum_i x_i \tilde{\lambda}_i\otimes I_n +\sum_j y_j I_m \otimes  \tilde{\lambda}_j \nn  \\
 && + \sum_{ij} t_{ij} \tilde{\lambda}_i \otimes \tilde{\lambda}_j),
\end{IEEEeqnarray}
where $\tilde{\lambda}_i , i=1,\ldots,m^2-1$ and $\tilde{\lambda}_j , j=1,\ldots,n^2-1$ are the generators of $SU(m)$ and $SU(n)$ respectively. Notice that $\vec{x} \in \mathbb{R}^{m^2-1}$ and $\vec{y}\in \mathbb{R}^{n^2-1}$ are the coherence vectors of the subsystems $A$ and $B$, so that they can be determined locally. These are given by \cite{bk03}
$$x_i=\frac{m}{2}tr(\rho \tilde{\lambda_i} \otimes I_n)=\frac{m}{2}tr(\rho_A \tilde{\lambda_i})$$
$$y_j=\frac{n}{2}tr(\rho I_m \otimes  \tilde{\lambda}_j )=\frac{n}{2}tr(\rho_B \tilde{\lambda_j}),$$
where $\rho_A=tr_B(\rho)$ and $\rho_B=tr_A(\rho)$ are the reduced density matrices. The correlation matrix $T=[t_{ij}]$ is given by

$$T=[t_{ij}]=\frac{m n}{4}[tr(\rho \tilde{\lambda_i} \otimes \tilde{\lambda_j})].$$

\section{Main result.}

We find the maximum in Eq.(\r{e5}) to obtain an exact analytic formula, as in the two-qubit case (Eq.(\r{eq3})) \cite{dvb10}.

\emph{Theorem 1}. Let $\rho$ be a bipartite state defined by Eq. (\r{e10}), then
\ben  \l{e11}
D(\rho)=\frac{2}{m^2n} \left[||\vec{x}||^2+\frac{2}{n} ||T||^2-\sum_{l=1}^{m-1} \eta_{(l+1)^2-1}\right],
\een
where $\eta_j,\;j=1,2,\cdots,m^2-1$ are the eigenvalues of the matrix $(\vec{x}\vec{x}^t+\frac{2TT^t}{n})$ arranged in increasing order (counting multiplicity).

We prove this theorem for arbitrary (finite) $m$ and $n$.

We choose the orthonormal bases $\{X_i\}$ and $\{Y_j\}$ in Eq.(\r{e4}) as the generators of $SU(m)$ and $SU(n)$ respectively  \cite{mw95}, that is,
\ben
X_1=\frac{1}{\sqrt{m}} I_m,  Y_1=\frac{1}{\sqrt{n}} I_n,                \nn     \\
\een
 and
 \ben
 X_i=\frac{1}{\sqrt{2}} \tilde{\lambda}_{i-1},\; i=2,3,\ldots,m^2           \nn      \\
 \een
 \ben
 Y_j=\frac{1}{\sqrt{2}} \tilde{\lambda}_{j-1},\; j=2,3,\ldots,n^2.           \nn       \\
 \een

 Since $tr \tilde{\lambda}_i=0;\;i=1,2,\cdots,m^2-1$, we have, via Eq.(\r{e8}),
 \ben
 \sum_{k=1}^{m} a_{ki}=tr X_i = tr \tilde{\lambda}_{i-1} = 0 , i=2,\ldots,m^2.     \nn    \\
 \een
 Therefore,
 \ben  \l{e12}
 a_{mi}=-\sum_{k=1}^{m-1} a_{ki},\;i=2,3,\cdots,m^2.
\een
We now proceed to construct the $m\times m^2$ matrix $A$ defined via Eq.(\r{e6}). We will use Eq.(\r{e12}). The row vectors of $A$ are
$$\vec{a}_k=(a_{k1},a_{k2},\cdots,a_{km^2}); k=1,2,\ldots,m.$$
Next we define
\ben   \l{e13}
\h{e}_k=\sqrt{\frac{m}{m-1}}(a_{k2},a_{k3},\ldots,a_{km^2}),\;k=1,2,\ldots,m-1
\een
and using Eq.(\r{e12}), we get
\ben   \l{e14}
\h{e}_m=-\sum_{k=1}^{m-1} \h{e}_k.
\een
We can prove
\ben  \l{e15}
||\h{e}_k||^2=1\;\;k=1,2,\ldots,m
\een
using the condition $||\vec{a}_k||^2=\sum_{i=1}^{9} a_{ki}^2=1$ (Eq.(\r{e7})) and using Eq.(\r{e8}) with $i=1,$ namely, $a_{k1}= tr(|k\rangle \langle k| X_1)=\frac{1}{\sqrt{m}}.$ 
  Further, isometry of the $A$ matrix ($AA^{t}=I$) implies
\ben  \l{e16}
\h{e}_i{\h{e}_j}^{t}=\frac{-1}{m-1}, j\ne i=1,2,\ldots,m.
\een
We can now construct the row vectors of $m\times m^2$ matrix $A$, using Eq.(\r{e13}) and Eq.(\r{e14}).
\ben
\vec{a}_k=\frac{1}{\sqrt{m}} (1, \sqrt{m-1}\h{e}_k),\; k=1,2,\cdots,m-1     \nn    \\
\een
\ben
\vec{a}_m=\frac{1}{\sqrt{m}} (1, - \sqrt{m-1} \sum_{k=1}^{m-1} \h{e}_k)      \nn       \\
\een
The matrix $A$ is, in terms of its row vectors defined above,
\begin{displaymath}
A=\frac{1}{\sqrt{m}}
\left(\begin{array}{cc}
1 & \sqrt{m-1}\h{e}_1\\
1 & \sqrt{m-1}\h{e}_2\\
1 & \sqrt{m-1}\h{e}_3\\
\vdots & \vdots\\
1 & -\sqrt{m-1}\sum_{k=1}^{m-1} \h{e}_k\\
\end{array}\right).
\end{displaymath}

We get the elements of $C=[c_{ij}]=[tr(\rho X_i\otimes Y_j)]$ using the definitions of the bases $\{X_i\}$ and $\{Y_j\}$ given above, in terms of the generators of $SU(m)$ and $SU(n)$. These are

$c_{11}=\frac{1}{\sqrt{mn}},$

$c_{1j}=\frac{\sqrt{2}}{n\sqrt{m}}y_j,\;j=2,\ldots,n^2$

$c_{i1}=\frac{\sqrt{2}}{m\sqrt{n}}x_i,\;i=2,\cdots,m^2$

$c_{ij}=\frac{2}{mn}t_{ij},\; i = 2,\ldots,m^2\;;\;j=2,\ldots,n^2$

where $x_i$ and $y_j$ are the components of the coherent vectors in Eq.(\r{e10}) and $t_{ij}$ are the elements of the correlation matrix $T.$ The matrix $C$ can then be written as

\begin{displaymath}
C =
\left(\begin{array}{cc}
\frac{1}{\sqrt{mn}} & \frac{\sqrt{2}}{n\sqrt{m}}\vec{y}^t \\
\frac{\sqrt{2}}{m\sqrt{n}}\vec{x} & \frac{2}{mn} T\\
\end{array}\right),
\end{displaymath}
 and
 \begin{IEEEeqnarray}{rCl} \l{e17}
 tr(CC^t) & = & \frac{1}{mn}+\frac{2}{n^2 m} ||\vec{y}||^2+\frac{2}{m^2n} ||\vec{x}||^2  \nn  \\
 &&+\frac{4}{n^2m^2} ||T||^2 .
\end{IEEEeqnarray}

Having constructed the matrices $A$ and $C,$ we get, for $tr(ACC^tA^t),$

\ben  \l{e18}
tr(ACC^tA^t)=\frac{1}{m}\left\{\frac{1}{n}+\frac{2}{n^2}||\vec{y}||^2 +\frac{2(m-1)}{m^2n}\left[\sum_{j=1}^{m-1}\h{e}_jG\h{e}_j^t +
\sum_{i=1}^{m-1}\sum_{j=1}^{m-1}\h{e}_i G\h{e}_j^t \right]\right\}.
\een

where
\ben  \l{e19}
G=\vec{x}\vec{x}^t + \frac{2TT^t}{n}
\een
is the $(m^2-1)\ti (m^2-1)$ real symmetric matrix. For $m=n=2,$ as in \cite{dvb10},

\ben  \l{e20}
tr(ACC^tA^t)=\frac{1}{4}\left[1+||\vec{y}||^2 +\h{e}_1G\h{e}_1^t\right],
\een

which attains the maximum of $tr(ACC^tA^t)$, when $\h{e}_1^t$ is an eigenvector of matrix $G$ for largest eigenvalue, which give us Eq.(\r{eq3}).
The eigenvectors of $G$ span $\mb{R}^{m^2-1}$ and form a orthonormal basis of $\mb{R}^{m^2-1}.$ Let $\eta_1 ,\eta_2 ,\ldots,\eta_{m^2-1}$ be the eigenvalues of $G$ arranged in increasing order (counting multiplicity). Let $(|\h{f}_1\rangle ,|\h{f}_2\rangle ,\ldots,|\h{f}_{m^2-1}\rangle)$ be the corresponding orthonormal eigenvectors of $G.$ Here and in what follows we denote the column vectors in a orthonormal basis in $\mb{R}^{m^2-1}$ by ket and the corresponding row vectors by bra. Whether a ket (bra) vector belongs to $\mb{R}^{m^2-1}$ or $H^a$ can be understood with reference to context.  

The last constraint on vectors $\{\h{e}_j\}$  making up matrix $A$ is that $\{\h{e}_j\},\;j=1,\ldots, m$ must be the coherent vectors of $m$ pure states comprising an orthonormal basis of $H^{a}$ \cite{bk03}. We can span all such sets of coherent vectors (each making up an $A$ matrix) using Lemma 1 proved in the Appendix. As seen from Lemma 1, in order to span all the sets of coherent vectors making up matrix $A,$ we can start with the set of coherent vectors $\{\h{n}_j\},j=1,\ldots,m$ corresponding to some orthonormal basis in $H^{a}$ and then obtain all possible other sets of coherent vectors composing matrix $A$ via Eq.(A1) corresponding to the unitary transformations $U$ taking the initial orthonormal basis to other orthonormal bases in $H^{a}.$ We choose the initial orthonormal basis with coherent vectors $\{\h{n}_j\}$ to be the standard (computational) basis $\{|k\ran \}\;k=1,\ldots, m .$ We can write
\ben  \l{e21}
\h{n}_j= \sum_{i=1}^{m^2-1} \epsilon_{ij}|i\rangle \;\; j=1,\ldots, m
\een
where now the set $\{|i\ran \}\;i=1,\ldots, m^2-1$ is the orthonormal basis in $\mb{R}^{m^2-1} ,$ the $i$th basis vector consisting of $1$ at the $i$th place and zero at all other places. $\epsilon=[\epsilon_{ij}]$ is a $m^2\ti m$ matrix with its $j$th column being vector $\h{n}_j$ expressed in $\{|i\ran \}$ basis. Obviously, $$\epsilon_{kj}=\sqrt{\frac{m}{2(m-1)}}\langle j|\lambda_k|j \rangle .$$
Substituting Eq.(A1) and Eq.(\r{e21}) in Eq.(\r{e18}) we get,

\begin{IEEEeqnarray}{rCl}\l{e51}
tr(ACC^tA^t) & = & \frac{1}{m}\left\{\frac{1}{n}+\frac{2}{n^2}||\vec{y}||^2+
 \frac{2(m-1)}{m^2n}\left[\sum_{j=1}^{m-1}\sum_{k=1}^{m^2-1}\sum_{p=1}^{m^2-1}\epsilon_{kj}\epsilon_{pj}\langle k|O G O^t |p\rangle \right.\right. \nn \\
&& +\: \left. \left.\sum_{i=1}^{m-1}\sum_{j=1}^{m-1} \sum_{k=1}^{m^2-1}\sum_{p=1}^{m^2-1}\epsilon_{ki}\epsilon_{pj}\langle k|O G  O^t|p\rangle\right]\right\}.
\end{IEEEeqnarray}

We write $G=\sum_{q=1}^{m^2-1}\eta_q |\h{f}_q\ran \lan \h{f}_q |,$ with its eigenvalues arranged in non-decreasing order, giving

\begin{IEEEeqnarray}{rCl}\l{e22}
tr(ACC^tA^t) & = & \frac{1}{m}\left\{\frac{1}{n}+\frac{2}{n^2}||\vec{y}||^2+\frac{2(m-1)}{m^2n}\sum_{k=1}^{m^2-1}\sum_{p=1}^{m^2-1}\sum_{q=1}^{m^2-1}\eta_q\langle k|O |\h{f}_q\rangle \langle \h{f}_q|O^t|p\rangle \right.\nn \\
&& \left.\left[ \sum_{j=1}^{m-1}\epsilon_{kj}\epsilon_{pj}+\sum_{i=1}^{m-1}\sum_{j=1}^{m-1} \epsilon_{ki}\epsilon_{pj}\right]\right\}.
\end{IEEEeqnarray}
We substitute $\epsilon_{kj}=\sqrt{\frac{m}{2(m-1)}}\langle j|\lambda_k|j \rangle$ in Eq.(\r{e22}), to get

\begin{IEEEeqnarray}{rCl}\l{e23}
tr(ACC^tA^t) & = &\frac{1}{m}\left\{\frac{1}{n}+\frac{2}{n^2}||\vec{y}||^2+ \frac{1}{m n}\sum_{k=1}^{m^2-1}\sum_{p=1}^{m^2-1}\sum_{q=1}^{m^2-1}\eta_q\langle k|O |\h{f}_q\rangle \langle \h{f}_q|O^t|p\rangle \right. \nn \\
&& \left. \left[ \sum_{j=1}^{m-1}\langle j|\tilde{\lambda}_k|j \rangle\langle j|\tilde{\lambda}_p|j \rangle +\sum_{i=1}^{m-1}\sum_{j=1}^{m-1} \langle i|\tilde{\lambda}_k|i \rangle\langle j|\tilde{\lambda}_p|j \rangle\right]\right\}.
\end{IEEEeqnarray}

To evaluate the sums in Eq.(\r{e23}) we need the average values of $\tilde{\lambda}_k$ operators in the standard (computational) basis states in $H^{a} .$ Using the definition of $\tilde{\lambda}_k,k=1,\ldots, m^2-1$ operators in Eqs.(\r{e02},\r{e03}) \cite{kim03} and the fact that $|i\ran,|j\ran$ in Eq.(\r{e23}) correspond to the standard basis in $H^{a} ,$ we see that $\langle i|\tilde{\lambda}_k|i\rangle \neq 0,\; k=1,2,\cdots,m^2-1$ only for $k=(l+1)^2-1,\;\;1 \leq l \leq m-1 .$ Also, $ \langle j|\tilde{\lambda}_p|j\rangle \neq 0,\; p=1,2,\cdots,m^2-1$ only for $p=(s+1)^2-1,\;\;1 \leq s \leq m-1 .$ For $\tilde{\lambda}_{(l+1)^2-1}$ or $\tilde{\lambda}_{(s+1)^2-1}$ we have,   

$$\tilde{\lambda}_{(l(s)+1)^2-1}=\sqrt{\frac{2}{l(s)(l(s)+1)}}(\sum_{j=1}^{l(s)} |j\rangle \langle j|- l(s)|l(s)+1\rangle\langle l(s)+1|).$$ 
where $$ 1 \leq l(s) \leq m-1 .$$
%\begin{widetext}
%\ben
%\begin{IEEEeqnarray}{rCl}\l{e24}
%tr(ACC^tA^t) & = &\frac{1}{m}\left\{\frac{1}{n}+\frac{2}{n^2}||\vec{y}||^2+
 %\frac{1}{m n}\sum_{l=1}^{m-1}\sum_{s=1}^{m-1}\sum_{q=1}^{m^2-1}\eta_q\langle (l+1)^2-1|O |\h{f}_q\rangle \langle \h{f}_q|O^t|(s+1)^2-1\rangle
  %\right. \nn \\ 
  %&&\: \left. \left[\sum_{j=1}^{m-1}\langle j|\lambda_{(s+1)^2-1}|j \rangle\langle j|\lambda_{(s+1)^2-1}|j \rangle +\sum_{i=1}^{m-1}\sum_{j=1}^{m-1} \langle i|\lambda_{(s+1)^2-1}|i \rangle\langle j|\lambda_{(s+1)^2-1}|j \rangle\right]\right\}.
%\end{IEEEeqnarray}
%\een
%\end{widetext}
Using the definition of $\tilde{\lambda}_{(l(s)+1)^2-1}$ above, we can easily prove

\begin{displaymath}
\sum_{j=1}^{m-1} \langle j|\tilde{\lambda}_{(l+1)^2-1}|j\rangle \langle j|\tilde{\lambda}_{(s+1)^2-1}|j\rangle = \left\{ \begin{array}{ll}
2 & \textrm{if $1\leq l=s\leq m-2$}\\
\frac{2}{m} & \textrm{if $l=s=m-1$}\\
0 & \textrm{otherwise}
\end{array} \right.
\end{displaymath}

\begin{displaymath}
\sum_{i=1}^{m-1} \langle i|\tilde{\lambda}_{(l+1)^2-1}|i\rangle \sum_{j=1}^{m-1} \langle j|\lambda_{(s+1)^2-1}|j\rangle = \left\{ \begin{array}{ll}
\frac{2(m-1)}{m} & \textrm{if $l=s=m-1$}\\
0 & \textrm{otherwise}
\end{array} \right.
\end{displaymath}
Following the para after Eq.(\r{e23}) and the above equations, Eq.(\r{e23}) can be reduced to

%\begin{widetext}
\ben   \l{e25}
%\begin{IEEEeqnarray}{rCl}
tr(ACC^tA^t)  = \frac{1}{m}\left\{\frac{1}{n}+\frac{2}{n^2}||\vec{y}||^2+\frac{2}{m n}\sum_{q=1}^{m^2-1} \eta_q\left[\sum_{l=1}^{m-1}|\langle (l+1)^2-1|O |\h{f}_q\rangle|^2 \right]\right\}.  \\
%\end{IEEEeqnarray}
\een
%\end{widetext}

To get the maximum of $tr(ACC^tA^t)$ over $A$ matrices we use Cauchy-Schwarz inequality $$|\langle (l+1)^2-1|O |\h{f}_{q}\rangle|^2 \leq \langle (l+1)^2-1|(l+1)^2-1\rangle \langle \h{f}_q|O^t O |\h{f}_q\rangle = 1 ,$$ which tells us that the desired maximum is obtained by choosing $O$ in Eq.(\r{e25}) to be that orthogonal matrix  which takes the eigenbasis of $G$ matrix to the standard basis in $\mb{R}^{m^2-1} .$ We denote the inverse of this orthogonal matrix by $O^{t}_{max}.$ Thus we get,  

\begin{IEEEeqnarray}{rCl}\l{e28}
\max_A tr(ACC^tA^t) & = &\frac{1}{m}\left\{\frac{1}{n}+\frac{2}{n^2}||\vec{y}||^2+
 \frac{2}{m n}\left[\sum_{l=1}^{m-1} \eta_{(l+1)^2-1}\right]\right\}.
 \end{IEEEeqnarray}

 Finally, Eq.(\r{e17}), Eq.(\r{e28}) and  Eq.(\r{e5}) together imply

\begin{IEEEeqnarray}{rCl}\l{e29}
D(\rho)=\frac{2}{m^2 n}\left[||\vec{x}||^2+\frac{2}{n}||T||^2-\sum_{l=1}^{m-1}\eta_{(l+1)^2-1}\right].
\end{IEEEeqnarray}

To complete the proof we have to show that the vectors $${\h{e}}_{j}^{t}=O^{t}_{max}{\h{n}}_{j}\;\; ; j=1,\ldots,m^2-1$$ are the coherent vectors of some pure states in $H^a$ given that the vectors $\{{\h{n}}_{j}\}$ are the coherent vectors of the computational basis states in $H^a .$ A necessary and sufficient condition for the vectors $\{{\h{e}}_{j}^{t}\}$ to be the coherent vectors of some pure states in $H^a$ is
\ben  \l{e04}
{\h{e}}_{j}^{t}\ast {\h{e}}_{j}^{t} = {\h{e}}_{j}^{t},\;\;j=1,\ldots ,m
\een
where the star product of vectors $\vec{a}$ and $\vec{b}$ is defined as $$(\vec{a}\ast\vec{b})_k = \sqrt{\frac{m(m-1)}{2}}\frac{1}{m-2}\sum_{i,j}g_{ijk}a_i b_j ,$$  $g_{ijk}$ being the elements of the completely symmetric tensor for $SU(m),$ defined as in Eq.(\r{e01}). The orthogonal operator $O^{t}_{max}$ maximizing $tr(ACC^tA^t)$ is given by $$O^{t}_{max}= \sum_{k}|{\h{f}}_k\ran \lan\h{k}| ,$$ so that the vectors ${\h{e}}_{j}^{t}$ are, using Eq.(\r{e21})  $${\h{e}}_{j}^{t} = O^{t}_{max}{\h{n}}_{j} = \sum_{i=1}^{m^2-1} \epsilon_{ij}|{\h{f}}_i\rangle \;\; j=1,\ldots, m .$$ Thus vectors ${\h{e}}_{j}^{t}$ have the same components with respect to the rotated basis $\{|{\h{f}}_i\ran \}$ as the components of ${\h{n}}_{j}$ with respect to the standard basis. (That is, both the standard basis and the vectors ${\h{n}}_{j}$ undergo the same rigid rotation.) It follows that ${\h{e}}_{j}^{t}$ satisfy Eq.(\r{e04}) if vectors ${\h{n}}_{j}$ do. This completes the proof.   

{\emph Remark 1} : 

The choice of $\{\tilde{\lambda}_k\}$ operators, as defined in Eqs.(\r{e02},\r{e03}) \cite{kim03} is not a restriction causing loss of generality. An arbitrary permutation of $\{\tilde{\lambda}_k\}$ by a permutation matrix $P$ replaces $G$ by $PGP^t = \sum_{q=1}^{m^2-1}\eta_{P(q)} |\h{f}_{P(q)}\ran \lan \h{f}_{P(q)} |$ causing appropriate re-arrangement of eigenvectors $|\h{f}_q \ran $ and the corresponding eigenvalues $\eta_q .$ This leaves Eq.(\r{e28}) invariant under such a permutation.

{\emph Remark 2} :

We expect $D(\rho)$ to vanish if $\rho$ is a classical quantum state as in Eq.(\r{e3}). This follows trivially from the fact that the generators of the $SU(m)$ group are traceless, that is, $$tr(\tilde{\lambda}_i) = \sum_{k=1}^{m}\lan k|\tilde{\lambda}_i|k\ran = 0 \; i=1,2,\ldots,m^2-1,$$
where $\{|k \ran\};k=1,2,\ldots,m$ is an arbitrary orthonormal basis in $H^a .$ 

We emphasize that the geometric measure of quantum discord in Eq.(\r{e11}) is experimentally accessible, as all the quantities involved depend on the average values of $\{\tilde{\lambda}_i\}$ and $\tilde{\lambda}_i \ot \tilde{\lambda}_j$ (generators of $SU(m)$ and $SU(n)$ respectively) which are hermitian operators and hence are in principle measurable for any $m$ and $n.$ However, it turns out that feasible experimental methods to measure  $\tilde{\lambda}_i , \tilde{\lambda}_j$ are available only for qubits and qutrits \cite{bk08,m06} in which case they are simply related to spin angular momentum components of a spin-$\frac{1}{2}$ and spin $1$ particle respectively. Thus we can experimentally determine the quantum discord in a bipartite system comprising qutrits and qubits. It is important to note that this experimental determination of the quantum discord does not require a detailed knowledge of the state of the system and can apply to any stage of quantum information process in which the state of the quatum system may not be known. In general, determination of the unknown state of a quantum system is a formidable task.   

For $m=n=3$, we have

\begin{IEEEeqnarray}{rCl}\l{e30}
D(\rho)=\frac{2}{27}\left[||\vec{x}||^2+\frac{2}{3}||T||^2-(\eta_3+\eta_8)\right].
\end{IEEEeqnarray}

We now give some examples using Eq.(\r{e11}).

\emph{Example 1.} We consider the $m\times m$-dimensional Werner state
$$\rho=\frac{m-z}{m^3-m} I+\frac{mz-1}{m^3-m} F,\; z\in[-1,1]$$
with $F=\sum_{kl} |k\rangle\langle l|\otimes |l\rangle\langle k|.$
First, we calculate the coherent vector $\vec{x}.$ We have
\begin{IEEEeqnarray}{rCl}
x_i  = \frac{m}{2} tr(\rho \tilde{\lambda}_i\otimes I) &=&\frac{m}{2}\left[\frac{m-z}{m^3-m} tr(\tilde{\lambda}_i\otimes I)\right. \nn  \\
&&+\left.\frac{mz-1}{m^3-m} tr(F (\tilde{\lambda}_i\otimes I))\right].$$
\end{IEEEeqnarray}
Since $tr(\tilde{\lambda}_i)=0,\; (i=1,2,\ldots,m^2-1)$, the first term in the above equation is zero. This gives

$$x_i  = \frac{m}{2} \left[\frac{mz-1}{m^3-m}\sum_{l} \langle l|\tilde{\lambda}_i|l\rangle \right]=\frac{m}{2}\left[\frac{mz-1}{m^3-m} tr(\tilde{\lambda}_i)\right]=0,$$
so that $\vec{x}=0.$

Now, we calculate the elements of the correlation matrix
\begin{IEEEeqnarray}{rCl}
t_{ij} & = &\frac{m^2}{4} tr(\rho \tilde{\lambda}_i \otimes \tilde{\lambda}_j) = \frac{m^2}{4}\left[\frac{m-z}{m^3-m} tr(\tilde{\lambda}_i\otimes \tilde{\lambda}_j)\right. \nn \\
&&+\left.\frac{mz-1}{m^3-m} tr(F (\tilde{\lambda}_i\otimes \tilde{\lambda}_j))\right].
\end{IEEEeqnarray}
The first term in the above equation is zero, so that
$$t_{ij}=\frac{m^2}{4}\left[\frac{mz-1}{m^3-m} \sum_l \langle l| \tilde{\lambda}_i \tilde{\lambda}_j|l\rangle \right]=\frac{m^2}{4}\left[\frac{mz-1}{m^3-m} tr( \tilde{\lambda}_i \tilde{\lambda}_j)\right].$$
Using the orthonormality of the generators $\{\tilde{\lambda}_i\}$ we get
$$t_{ij}=\frac{m^2}{4}\left[\frac{mz-1}{m^3-m} (2 \delta_{ij})\right]$$ or, $$ t_{ii}=\frac{m^2}{2} \frac{(mz-1)}{(m^3-m)} .$$
Thus $T$ matrix is diagonal and all diagonal elements are equal. This gives,
$$\frac{2||T||^2}{m}=\frac{2}{m}\sum_{i=1}^{m^2-1} t_{ii}^2=\frac{2}{m} \left[\frac{m^4}{4}\frac{(mz-1)^2}{(m^3-m)^2} (m^2-1)\right]$$
$$=\frac{m(mz-1)^2}{2(m^2-1)} ,$$
 the eigenvalues of $\frac{2}{m}TT^t$ are all equal :  $$\eta_i=\frac{m^3(mz-1)^2}{2(m^3-m)^2},$$
Substituting $||\vec{x}||^2=0$ and the expressions for $\frac{2||T||^2}{m}$ and $\{\eta_i\}$  as above in Eq.(\r{e11}) we get, after some algebra, for the quantum discord of the $m\times m$-dimensional Werner state
$$D(\rho)=\frac{(mz-1)^2}{m(m-1)(m+1)^2}.$$

A Werner state is separable if and only if $z \in [0,1],$ but its
geometric measure of quantum discord vanishes if and only if
$z = 1/m.$

\emph{Example 2.} We consider the particular case $m = n = 3$ and the bipartite pure state
$$|\psi\rangle = \frac{1}{2} |11\rangle+\frac{1}{2} |22\rangle+\frac{1}{\sqrt{2}} |33\rangle,$$
then by straightforward evaluation based on the original definition, its geometric measure of
quantum discord is $D(|\psi\rangle\langle \psi|) = \frac{5}{8} .$
We show that Eq.(\r{e11}) yields the same result. 

The detailed calculation is as follows.
$$|\psi\rangle\langle \psi|= \frac{1}{9}\{I\otimes I-\frac{\sqrt{3}}{4} I\otimes \lambda_8 -\frac{\sqrt{3}}{4}\lambda_8 \otimes I+\frac{9}{8} \lambda_1\otimes \lambda_1+\frac{9}{8} \lambda_2\otimes \lambda_2+\frac{9}{8} \lambda_3\otimes \lambda_3$$
$$+\frac{9}{4\sqrt{2}} \lambda_4\otimes \lambda_4-\frac{9}{4\sqrt{2}} \lambda_5\otimes \lambda_5+\frac{9}{4\sqrt{2}} \lambda_6\otimes \lambda_6-\frac{9}{4\sqrt{2}} \lambda_7\otimes \lambda_7+\frac{15}{8} \lambda_8\otimes \lambda_8\}$$
Using this expression for $|\psi\ran\lan\psi|$, we get,
\begin{displaymath}
G=\vec{x}\vec{x}^t + \frac{2}{3}TT^t
=\left(\begin{array}{cccccccc}
\frac{27}{32} & 0 & 0 & 0 & 0 & 0 & 0 & 0\\
0 & \frac{27}{32} & 0 & 0 & 0 & 0 & 0 & 0\\
0 & 0 & \frac{27}{32} & 0 & 0 & 0 & 0 & 0\\
0 & 0 & 0 & \frac{27}{16} & 0 & 0 & 0 & 0\\
0 & 0 & 0 & 0 & \frac{27}{16} & 0 & 0 & 0\\
0 & 0 & 0 & 0 & 0 & \frac{27}{16} & 0 & 0\\
0 & 0 & 0 & 0 & 0 & 0 & \frac{27}{16} & 0\\
0 & 0 & 0 & 0 & 0 & 0 & 0 & \frac{81}{32}\\
\end{array}\right).
\end{displaymath}
and from Eq.(\r{e30}) we get, using the values of $||\vec{x}||^2 ,$ $\frac{2}{3}||T||^2$ and $\eta_3 + \eta_8 ,$
\benn
D(\rho)=\frac{2}{27}\left[\frac{3}{16}+\frac{2}{3}\times\frac{279}{16}-\left(\frac{27}{32}+\frac{81}{32}\right)\right]=\frac{5}{8}.
\eenn

\emph{Example 3.} We consider the two qutrit state
\ben \l{e48}
\rho=p |e\rangle\langle e|+ (1-p)\frac{I}{9} \\
\een
where $|e\rangle =\frac{1}{\sqrt{6}}(|2\rangle\otimes |2\rangle +|3\rangle\otimes |3\rangle +|2\rangle\otimes |1\rangle +|1\rangle\otimes |2\rangle +|1\rangle\otimes |3\rangle +|3\rangle\otimes |1\rangle) $, $I$ is the identity operator and $\{|i\rangle; i=1,2,3\}$ is the standard basis in $C^3$.
Fig.(5.1) shows the variation of $D(\rho)$ (Eq.(\r{e11})) and the lower bound on $D(\rho),$ as given in \cite{lf10}, namely, $tr(CC^{t})-\sum_{i=1}^{m}\lambda_{i}$ (where $\{\lambda_{i}\}$ are the eigenvalues of $CC^{t}$ listed in the decreasing order, counting multiplicity), with $p.$ We see that $D(\rho)$ (Eq.(\r{e11})) dominates this lower bound for $p > 0.2.$
 
\begin{figure}[!ht]
\begin{center}
\includegraphics[width=9cm,height=6cm]{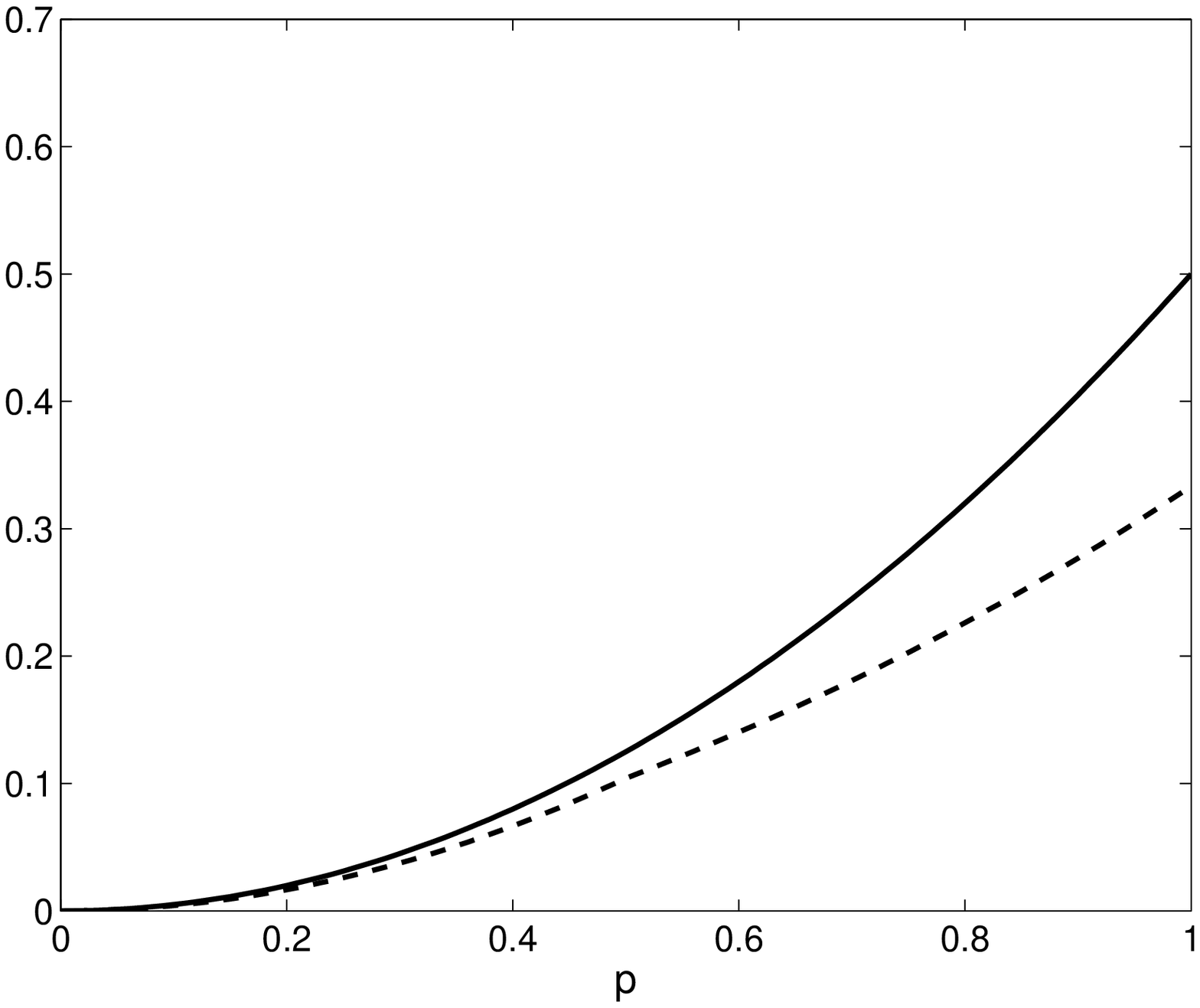}\\
Fig.(5.1) : Quantum discord (Eq.(\r{e11})) (solid line) and its lower bound (Eq.(5.6) in \cite{lf10}) (dashed line) as a function of $p$ (Eq.(\r{e48})).

\end{center}
\end{figure}

\emph{Example 4.} We consider the two qutrit state
\ben \l{e49}
\rho=p |e_1\rangle\langle e_1|+ (1-p)|e_2\rangle\langle e_2| \\
\een

where  $|e_1\rangle =\frac{1}{2} |11\rangle+\frac{1}{2} |22\rangle+\frac{1}{\sqrt{2}} |33\rangle$ and $|e_2\rangle =\frac{1}{\sqrt{6}}(|2\rangle\otimes |2\rangle +|3\rangle\otimes |3\rangle +|2\rangle\otimes |1\rangle +|1\rangle\otimes |2\rangle +|1\rangle\otimes |3\rangle +|3\rangle\otimes |1\rangle).$
Fig.(5.2) shows the variation of $D(\rho)$ (Eq.(\r{e11})) and the lower bound on $D(\rho),$ as in \cite{lf10}, namely, $tr(CC^{t})-\sum_{i=1}^{m}\lambda_{i}$ (where $\{\lambda_{i}\}$ are the eigenvalues of $CC^{t}$ listed in the decreasing order, counting multiplicity), with $p.$ We see that $D(\rho)$ (Eq.(\r{e11})) dominates this lower bound.
 
\begin{figure}[!ht]
\begin{center}
\includegraphics[width=9cm,height=6cm]{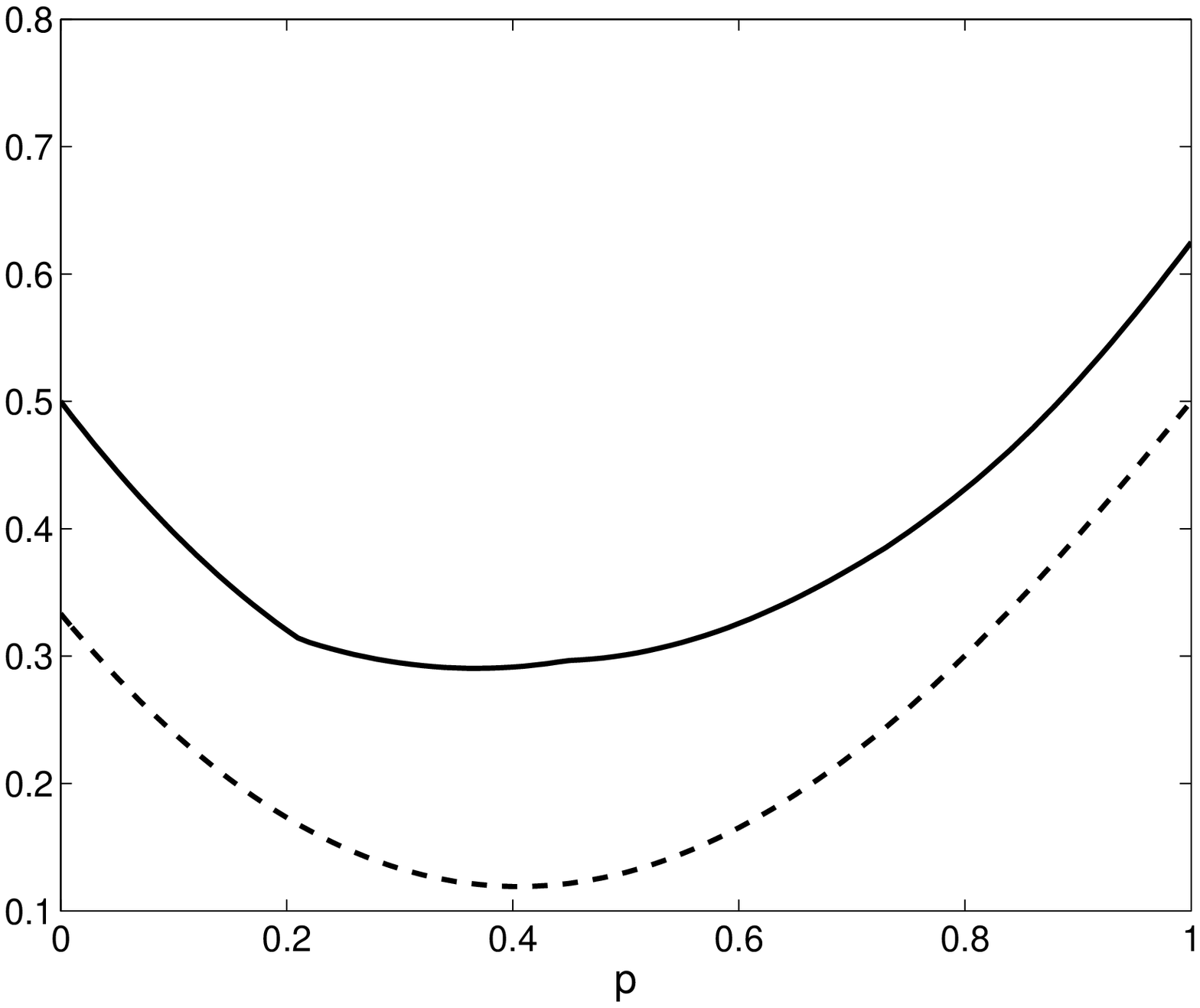}\\
Fig.(5.2) : Quantum discord (Eq.(\r{e11})) (solid line) and its lower bound (Eq.(5.6) in \cite{lf10}) (dashed line) as a function of $p$ (Eq.(\r{e49})).

\end{center}
\end{figure}

\
\\
\
\\
\
\\
\
\\
\
\\
\
\\
\
\\
\
\\
\
\\

\newpage
\large{\textbf{Appendix to the chapter 5}}
\
\\
\begin{center}
\scriptsize\textsc{ A child of five would understand this. Send someone to fetch a child of five.}\\{ Groucho Marx}
\end{center}

In this appendix we prove the lemma 1. 
For a given pure state $|i\ran\lan i|$ and an unitary operator $U$ acting on $H^{a} ,$ there exists an orthogonal operator $O=[O_{\al\b}]$ acting on $\mb{R}^{m^2-1}$ such that   

$$\h{e}_j = \h{n}^t_j O  \eqno{(A1)}$$

where $\h{n}_j$ is the  coherence (column) vector of the state $|i\ran\lan i|$ and $\h{e}_j$ is the  coherence (row) vector of the state 
$U|i\ran\lan i|U^{\dagger} .$ 

{\it Proof}: We use the following fact which is easily proved.

$$U{\tilde{\la}}_{\al}U^{\dagger} = \sum_{\b}O_{\al\b}{\tilde{\la}}_{\b}\;\;\al=1,\ldots,m^2-1 . \eqno{(A2)}$$

for some orthogonal operator $O=[O_{\al\b}] .$ 

We now have, using Eq.(A2)

$$\sum_{\b}(\h{e}_j)_{\b}{\tilde{\la}}_{\b}=U|i\ran\lan i|U^{\dagger}=\sum_{\al}(\h{n}_j)_{\al}U{\tilde{\la}}_{\al}U^{\dagger}=
\sum_{\b}\left(\sum_{\al}(\h{n}_j)_{\al}O_{\al\b}\right){\tilde{\la}}_{\b}$$
$$=\sum_{\b}(\h{n}^t_j O)_{\b}{\tilde{\la}}_{\b} $$  

which proves the Lemma 1.\\

\chapter{Summary and Future Directions.}

%\begin{center}
%\scriptsize \textsc{This is not the end. Nor is this a beginning of the end.\\ This may at most be the end of a beginning.\\
%-{\it Sir Winston Churchill.}\footnote{From his last address to the British parliament as the prime minister of U.K. ( 3rd june 1946)}.}
%\end{center}

\begin{figure}[!ht]
\begin{center}
\includegraphics[width=12cm,height=3cm]{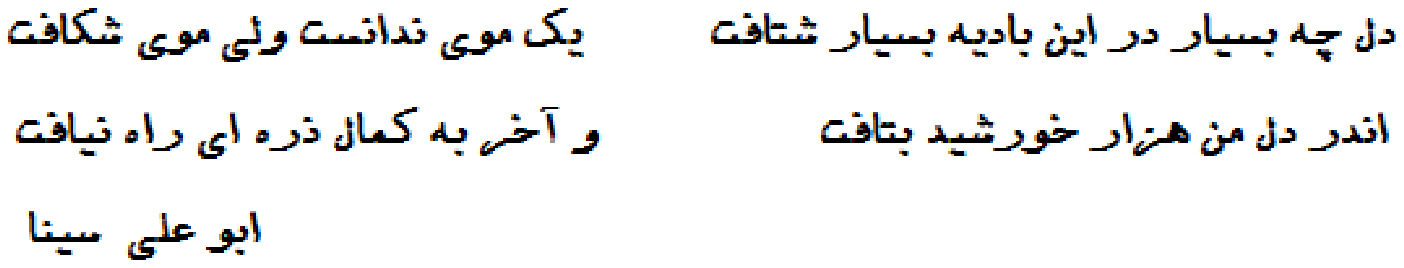}
\end{center}
\end{figure}

The first chapter is introductory and does not present any new work. Therefore, we start summarizing from the second chapter.

In chapter 2, we deal with the problem of controlled production of entanglement that is, producing the multipartite quantum systems in required entangled states. The motivation of this work is as follows. It is now well known that entanglement is a resource for accomplishing various kinds of quantum information processing tasks and quantum communication protocols, thus future technology may require uninterrupted supply of entangled quantum systems just as the present day technology requires uninterrupted supply of energy. Therefore, it is worth searching for the general methods of preparing entangled quantum states of multipartite quantum systems.
This can be achieved, at least in principle, because different interaction Hamiltonians can be expressed in a single generic form. We have implemented some protocols for entanglement generation for two qubits, two qutrits and three qubits using the geometric measure of entanglement suggested by us \cite{hj08,hj09}.

A possible question that can be addressed is the effect of ancillas in entanglement production. In ref. \cite{dvclp01} ancillas are proved to improve efficiency of entanglement production in the two qubit case.

A possible analytical approach to find the entanglement capacity for two qutrits and three qubits is another challenge.

In chapter 3, we have given a geometric measure of entanglement for a multipartite fermionic system. We construct a isomorphism which enables us to view a fermionic (Fock) state as multiqubit state to which we can apply the geometric the geometric entanglement measure invented by us \cite{hj08,hj09}. We think that our way of defining the local operations in the case of fermionic systems has removed the ambiguities prevailing before. We have tested our measure on the generically important systems as Hubbard dimer and trimer.

The basic open question in this area is the possible role of entanglement in the wide range of cooperative phenomena in many body systems, in particular quantum phase transition. We can use our measure to quantitatively study the dependence of entanglement on various interaction parameters in the model Hamiltonian, which in turn are connected to the physical behavior of systems. One of the immediate problems is the fate of entanglement when the Coulomb interactions between different sites are different (in Hubbard trimer).

In chapter 4, We investigate how thermal quantum discord $(QD)$ and classical correlations $(CC)$ of  a two qubit one-dimensional XX Heisenberg chain in thermal equilibrium depend on temperature of the bath as well as on nonuniform external magnetic fields applied to two qubits and varied separately. We show that the behavior of $QD$ differs in many unexpected ways from thermal entanglement $(EOF)$. For the nonuniform case, $(B_1= - B_2)$ we find that $QD$ and $CC$ are equal for all values of $(B_1=-B_2)$ and for different temperatures. We show that, in this case, the thermal states of the system belong to a class of mixed states and satisfy certain conditions under which $QD$ and $CC$ are equal. The specification of this class and the corresponding conditions are completely general and apply to any quantum system in a state in this class and satisfying these conditions. We further find that the relative contributions of $QD$ and $CC$ can be controlled easily by changing the relative magnitudes of $B_1$ and $B_2$. Finally, we connect our results with the monogamy relations between the EOF, classical correlations and the quantum discord of two qubits and the environment.

Our work reported in chapter 5 solves the problem of finding a geometric measure of $QD$ in a bipartite state.

As for the future work with quantum discord, we think that finding a viable relation between quantum discord and entanglement may be very useful for mixed states. Such relations involving inequalities have been obtained \cite{fcoc10}. Quantifying $QD$ for many particle systems and relating it to various physical properties (like order parameters for the case of quantum phase transition) is an area we find interesting. We feel that in this area $QD$ may be more useful than entanglement because $QD$ measures total quantum correlations while in most cases entanglement measures quantum correlations breaking Bell inequalities.

\end{document}